\documentclass[a4paper,fleqn]{cas-dc}

\usepackage[numbers]{natbib}

\usepackage[utf8]{inputenc}
\usepackage[english]{babel}
\usepackage[ruled,linesnumbered]{algorithm2e} 
\usepackage[inline]{enumitem}
\usepackage[acronym]{glossaries}
\usepackage{lipsum}
\usepackage{nth}
\usepackage{ifthen}
\usepackage{hyphenat}
\usepackage{lipsum}
\usepackage{tikz}
\usepackage[draft]{tikzpeople}
\usepackage{pgf-umlcd}
\usepackage{fvextra}
\usepackage{forest}
\usetikzlibrary{angles}
\definecolor{drawColor}{RGB}{128 128 128}
\newcommand{\circleSize}{0.25em}
\newcommand{\angleSize}{0.8em}

\usepackage{csquotes}
\usepackage{outlines}
\usepackage{printlen}
\usepackage{tabularx}
\usepackage{booktabs}
\usepackage{multirow}
\usepackage{graphicx}
\usepackage{rotating}
\usepackage{listings}
\usepackage{footmisc} 
\usepackage[frozencache=true,cachedir=minted-cache]{minted}
\usepackage{todonotes}
\usepackage{subcaption}
\usepackage[export]{adjustbox}

\newcommand{\defineReviewCommands}[2]{
	\expandafter\def\csname#1\endcsname##1{\todo[color=#2]{##1 -#1}}\expandafter\def\csname#1i\endcsname##1{\todo[inline,caption={},color=#2]{##1 -#1}}}

\DeclareUnicodeCharacter{042F}{\russianr}

\usetikzlibrary{arrows,arrows.meta,positioning,shapes.arrows,shapes.geometric,shapes.symbols,angles,backgrounds,decorations.pathreplacing,matrix,angles,patterns,intersections,calc}

\newacronym{cps}{CPS}{Cyber-Physical System}
\newacronym{cpps}{CPPS}{Cyber-Physical Production System}
\newacronym{ppr}{PPR}{Product-Process-Resource}
\newacronym{fpd}{FPD}{Formalized Process Description}
\newacronym{spl}{SPL}{Software Product Line}
\newacronym{sple}{SPLE}{Software Product Line Engineering}
\newacronym{foda}{FODA}{Feature-Oriented Domain Analysis}
\newacronym{ovm}{OVM}{Orthogonal Variability Modeling}
\newacronym{cpr}{CPR}{Construction Primitives}
\newacronym{uml}{UML}{Unified Modeling Language}
\newacronym{dsl}{DSL}{Domain-Specific Language}
\newacronym{slr}{SLR}{Systematic Literature Review}
\newacronym[longplural={product comparison matrices}]{pcm}{PCM}{product comparison matrix}
\newacronym{pg}{PG}{Precedence Graph}
\newacronym{vert}{VERT}{Variability Evolution Roundtrip Transformation}
\newacronym[longplural={type comparison matrices}]{tcm}{TCM}{type comparison matrix}
\newacronym{vme}{VME}{variability modeling expert}
\newacronym{vm}{VM}{Variability Modeling}
\newacronym{fm}{FM}{Feature Model}
\newacronym{dm}{DM}{Decision Model}
\newacronym{ipse}{IPSE}{Iterative Process Sequence Exploration}
\newacronym{eipse}{eIPSE}{Extended Iterative Process Sequence Exploration}
\newacronym{ide}{IDE}{integrated development environment}
\newacronym{pprdsl}{PPR--DSL}{Product-Process-Resource Domain-Specific Language}
\newacronym{v4rdiac}{V4rdiac}{Variability for 4diac}
\newacronym{pp2dm}{PP2DM}{Product-and-Product to Decision Model}
\newacronym{iot}{IoT}{Internet of Things}
\newacronym{cnf}{CNF}{Conjunctive normal form}
\newacronym{cvl}{CVL}{Common Variability Language}
\newacronym{uvl}{UVL}{Universal Variability Language}
\newacronym{pla}{PLA}{Polyactic Acid}
\newacronym{cdc}{CDC}{Cross-Discipline Constraint}
\newacronym{cli}{CLI}{Command Line Interface}
 
\def\tsc#1{\csdef{#1}{\textsc{\lowercase{#1}}\xspace}}
\tsc{WGM}
\tsc{QE}
\tsc{EP}
\tsc{PMS}
\tsc{BEC}
\tsc{DE}

\clearpage{}
\definecolor{darkCyan}{RGB}{0,145,207}
\definecolor{lightCyan}{RGB}{73,186,235}

\definecolor{darkBlue}{RGB}{69,92,221}
\definecolor{lightBlue}{RGB}{108,129,253}

\definecolor{darkViolet}{RGB}{118,68,211}
\definecolor{lightViolet}{RGB}{149,109,224}

\definecolor{darkRed}{RGB}{195,37,37}
\definecolor{lightRed}{RGB}{229,95,69}

\definecolor{darkOrange}{RGB}{233,105,20}
\definecolor{lightOrange}{RGB}{255,143,43}

\definecolor{darkBrown}{RGB}{167,88,35}
\definecolor{lightBrown}{RGB}{198,124,69}

\definecolor{darkGreen}{RGB}{92,147,21}
\definecolor{lightGreen}{RGB}{115,192,17}

\definecolor{darkTeal}{RGB}{76,131,114}
\definecolor{lightTeal}{RGB}{114,173,155}

\definecolor{darkYellow}{RGB}{255,225,6}
\definecolor{lightYellow}{RGB}{255,246,128}

\definecolor{darkGrey}{RGB}{141,156,163}
\definecolor{lightGrey}{RGB}{178,194,203} 

\definecolor{myLighterGray}{gray}{0.97}
\definecolor{myDarkerGray}{gray}{0.84}

\newcommand{\code}[1]{\small\textsf{#1}\normalsize}

\renewcommand{\subsectionautorefname}{Section}

\newcommand{\travart}{{\fontsize{10}{12}\selectfont T}RA{\fontsize{10}{12}\selectfont V}AR{\fontsize{10}{12}\selectfont T}\xspace}

\ifthenelse{0 > -1}{
    \newcommand{\rr}[1]{\textcolor{orange}{[RR] #1}}
    \newcommand{\km}[1]{\textcolor{purple}{[KM] #1}}
    \newcommand{\hsf}[1]{\textcolor{darkViolet}{[HSF] #1}}
    \newcommand{\kf}[1]{\textcolor{darkGreen}{[KF] #1}}
    \newcommand{\sg}[1]{\textcolor{darkBlue}{[SG] #1}} 
    
}{
    \newcommand{\hsf}[1]{}
    \newcommand{\sbi}[1]{}
    \newcommand{\rr}[1]{}
    \newcommand{\km}[1]{}
    \newcommand{\kf}[1]{}
    
    \newcommand{\sg}[1]{}
}

\defineReviewCommands{sandra}{cyan}
\defineReviewCommands{sayyid}{darkViolet}
\defineReviewCommands{kevin}{darkGreen}
\defineReviewCommands{kristof}{purple}
\defineReviewCommands{rick}{orange}

\newcommand{\productFM}{Product~\gls{fm}}
\newcommand{\processDM}{Process~\gls{dm}}
\newcommand{\resourceFM}{Resource~\gls{fm}}
\clearpage{}

\begin{document}
\let\WriteBookmarks\relax
\def\floatpagepagefraction{1}
\def\textpagefraction{.001}
\def\subsectionautorefname{Sec.}
\def\figureautorefname{Fig.}
\def\listingautorefname{Lst.}
\def\tableautorefname{Tab.}

\setlength{\intextsep}{0.1cm}
\setlength{\textfloatsep}{0.1cm}

\shorttitle{Variability Modeling of Products, Processes, and Resources in Cyber-Physical Production Systems Engineering}

\shortauthors{Kristof Meixner et~al.}

\title [mode = title]{Variability Modeling of Products, Processes, and Resources in Cyber-Physical Production Systems Engineering}

\author[1,2]{Kristof Meixner}[
    orcid=0000-0001-7286-1393
]
\cormark[1]
\ead{kristof.meixner@tuwien.ac.at}
\credit{Conceptualization, Data curation, Formal analysis, Investigation, Methodology, Project administration, Software, Validation, Visualization, Writing - Original draft preparation, Writing - review \& editing}

\author[3]{Kevin Feichtinger}[orcid=0000-0003-1182-5377
]
\ead{kevin.feichtinger@kit.edu}
\credit{Conceptualization, Data curation, Formal analysis, Investigation, Methodology, Project administration, Software, Validation, Visualization, Writing - Original draft preparation, Writing - review \& editing}

\author[4]{Hafiyyan Sayyid Fadhlillah}[orcid=0000-0001-8361-6190
]
\ead{hafiyyan.fadhlillah@jku.at}
\credit{Conceptualization, Data curation, Formal analysis, Investigation, Methodology, Software, Validation, Visualization, Writing - Original draft preparation, Writing - review \& editing}

\author[5]{Sandra Greiner}[orcid=0000-0001-8950-0092
]
\ead{sandra.greiner@unibe.ch}
\credit{Conceptualization, Formal analysis, Investigation, Methodology, Writing - Original draft preparation, Validation, Writing - Original draft preparation, Writing - review \& editing}

\author[1]{Hannes Marcher}[]
\ead{hannes.marcher@tuwien.ac.at}
\credit{Software, Writing - Original draft preparation}

\author[4,6]{Rick Rabiser}[
    orcid=0000-0003-3862-1112
]
\ead{rick.rabiser@jku.at}

\credit{Conceptualization, Funding acquisition, Resources, Writing - Original draft preparation, Writing - review \& editing}

\author[2]{Stefan Biffl} [
    orcid=0000-0002-3413-7780
]
\ead{stefan.biffl@tuwien.ac.at}

\credit{Conceptualization, Funding acquisition, Resources}

\affiliation[1]{organization={Christian Doppler Laboratory SQI, TU Wien},
    addressline={Favoritenstraße 9-11}, 
    city={Vienna},
postcode={1040}, 
country={Austria}}

\affiliation[2]{organization={Information Systems Engineering, TU Wien},
    addressline={Favoritenstraße 9-11}, 
    city={Vienna},
postcode={1040}, 
country={Austria}}
    
\affiliation[3]{organization={CRC 1608, KASTEL -- Dependability of Software-intensive Systems, Karlsruhe Institute of Technology},
    addressline={Am Fasanengarten 5}, 
    city={Karlsruhe},
postcode={76131}, 
country={Germany}}

\affiliation[4]{organization={Christian Doppler Laboratory VaSiCS, Johannes Kepler University Linz},
    addressline={Altenberger Straße 69}, 
    city={Linz},
postcode={4040}, 
country={Austria}}

\affiliation[5]{organization={Software Engineering Group, Institute of Computer Science, University of Bern},
    addressline={Neubrückstraße 10}, 
    city={Bern},
postcode={3012}, 
country={Switzerland}}

\affiliation[6]{organization={LIT CPS Lab, Johannes Kepler University Linz},
    addressline={Altenberger Straße 69}, 
    city={Linz},
postcode={4040}, 
country={Austria}}

\cortext[cor1]{Corresponding author}

\begin{abstract}
~\glspl{cpps}, such as \emph{automated car manufacturing plants}, execute a \emph{configurable} sequence of production steps to manufacture products from a product portfolio.
In \gls{cpps} engineering, domain experts start with \emph{manually} determining feasible production step sequences and resources based on \emph{implicit knowledge}. 
This process is hard to reproduce and highly inefficient.
In this paper, we present the \gls{eipse} approach to derive variability models for products, processes, and resources from a domain-specific description. 
To automate the integrated exploration and configuration process for a \gls{cpps}, we provide a toolchain which automatically reduces the configuration space and allows to generate \gls{cpps} artifacts, such as control code for resources. 
We evaluate the approach with four real-world use cases, including the generation of control code artifacts, and an observational user study to collect feedback from engineers with different backgrounds. 
The results confirm the usefulness of the \gls{eipse} approach and accompanying prototype to straightforwardly configure a desired \gls{cpps}.
\end{abstract}

\begin{keywords}
Variability Modeling \sep 
Feature Modeling \sep
Decision Modeling \sep
Production Process Variability \sep
Cyber-Physical Production System.
\end{keywords}

\maketitle

\glsresetall

\section{Introduction}
\label{sec:introduction}
\glspl{spl} are ``a set of software-intensive systems that share a common, managed set of features satisfying the specific needs of a particular market segment or mission developed from a common set of core assets in a prescribed way~\citep{clements2002software}.'' 
\emph{Variability modeling} is a crucial activity of \gls{spl} engineering. 
It captures the commonalities and differences of these software-intensive systems explicitly and manifests them in a variability model, such as a \gls{fm}~\citep{kang1990feature} or \gls{dm}~\citep{schmid2011comparison}. 
\glspl{fm} focus on configuring products by respecting tree and cross-tree constraints, whereas \glspl{dm} also allow configuring sequences of options, which, in particular, is useful to model process variability.

\glspl{cpps}, such as \emph{automated car manufacturing plants}, use the latest information and communication technology to manufacture customized products with modern production techniques~\citep{MONOSTORI20149}. 
Each \gls{cpps} is built by assembling various hardware components (e.g., sensors and actuators) that are controlled by software to deliver one or more production system functionalities. 
Such systems execute a \emph{configurable and flexible} sequence of production steps to manufacture products from a product portfolio, provoking variation in how to realize the \gls{cpps}~\citep{Jaervenpaeae2019,MONOSTORI20149}. 
Hence, variability modeling for \glspl{cpps} faces the challenge of capturing multiple aspects~\citep{Fang2013,galster2013variability} that reach from the variability of products to the sequence of production steps and employed production resources. Furthermore, engineers of different disciplines, such as mechanics, electrics, and software engineering, with different views on the planned \gls{cpps} collaborate to build it iteratively~\citep{biffl2017introduction}.
Due to this multidisciplinarity, \textit{separating the concerns is essential}, which is also true for variability modeling~\citep{Ananieva2016}.

To build a \gls{cpps}, typically, \gls{cpps} engineers start with determining a feasible sequence of production process steps \emph{manually} for the products in the portfolio~\citep{lee1989disassembly}.
Then, they define which production resources can execute these production process steps, examine the performance characteristics, and estimate the \gls{cpps} variant's construction cost. 
The production process steps and production resources are later used for designing and implementing \gls{cpps} artifacts, e.g., the control software. 
As the result originates from a completely manual process founded in \emph{implicit domain knowledge}, resulting primarily \emph{from experience and undocumented dependencies}, it is \textit{hard and, most of the time, impossible to reproduce}.
However, repetition of the planning may be necessary when a new product variant is introduced, ``a frequent scenario in \gls{cpps} engineering''~\citep{Tolio2010}.
The manual configuration is also highly inefficient and tedious due to the large number of possible production sequences, challenging engineers to find practically feasible ones.
Thus, this laborious, time-consuming, and error-prone activity calls for a methodology to automate and document the derivation of production sequence and resource configurations from a product configuration.

To address some of these challenges, we developed different automation facilities in previous work.
We developed 
\begin{enumerate*}[label=(\roman*)]
    \item the \gls{pprdsl}~\citep{MeixnerETFA2021PPRDSL} 
to model products with the required production processes and resources systematically; 
\item the \travart{} framework~\citep{FeichtingerTraVarT2021} 
to transform engineering artifacts containing variability information into state-of-the-art variability models automatically;
\item the \gls{ipse}~\citep{MeixnerVamos2022ProcessExploration} approach that utilized these approaches to handle \gls{cpps} variability through \gls{spl} techniques and transformations of the \gls{pprdsl} into a \gls{fm} and a \gls{dm}.
\end{enumerate*}
However, the \gls{ipse} approach had \textit{no automation and tool support to explore and configure process sequences}, and \textit{did not include production resource modeling and configuration} and \textit{artifact generation}.

In this paper, we extend the semi-automated \gls{ipse} process and aim to answer the following research questions: 
\begin{enumerate}[nosep, label={\upshape\bfseries RQ\arabic*}]
    \item \emph{How can \gls{cpps} engineers be supported in modeling, exploring, and configuring the combined variability of products, production processes, and production resources, to generate corresponding \gls{cpps} artifacts?}     The \gls{eipse} approach establishes a process that includes these artifacts and incorporates feedback loops that reflect the iterative development.
    \item \emph{How and to what extent can \gls{cpps} design be automated using variability modeling and \gls{cpps} concepts?}     We provide the \gls{eipse} tool architecture, which starts with the manually defined \gls{pprdsl} and derives the remaining artifacts automatically.
\end{enumerate}

In its first state~\citep{MeixnerVamos2022ProcessExploration}, the \gls{ipse} converted a given \gls{pprdsl} into a \gls{fm} to capture the variety of products and their parts, and a \gls{dm}, to represent the production process sequences, using \travart{}~\citep{FeichtingerTraVarT2021}. 
We extend \gls{ipse} to derive a second kind of \gls{fm} to capture the variability in production resources. 
Additionally, we utilize \glspl{cdc}~\citep{FadhlillahSPLC22,Fadhlillah22} to express dependencies between variation points from those three variability models. 
Furthermore, we reduce the effort of manually configuring the resulting \gls{dm} by offering an editor, which decreases the number of configuration options with each decision taken, and we empirically examine the gained automation .
Thus, on top of our previous work, this paper contributes: 
\begin{itemize} [nosep]
    \item The \gls{eipse} approach with additional steps for production resource definition and configuration, artifact generation, and feedback loops.
    \item An extended prototype to assess the feasibility of the \gls{eipse} approach, complemented with
    \begin{itemize}   [noitemsep]
        \item support for transforming the \gls{pprdsl} into a \resourceFM{} and \glspl{cdc} to elicit resource variability and dependencies among the different variability models
        \item a novel \gls{dm} editor for modeling and configuring DOPLER \glspl{dm}, used in this context to represent and configure production process sequences 
        \item an automated reduction of the possible \processDM{} configuration based on the product configuration to lower the complexity of production process sequence configuration
        \item support for resource configuration and control code artifact generation through integration with \gls{v4rdiac}~\citep{FadhlillahSPLC22,Fadhlillah22} to automate the creation of control software.
    \end{itemize}
\item An empirical evaluation of \begin{itemize}
        \item the feasibility of selecting an adequate \gls{cpps} variant in four \emph{real-world case studies}~\citep{Meixner2021} 
        \item applying the \gls{eipse} approach in a new case study performed by engineers with heterogeneous backgrounds inexperienced in the approach, and thereby 
        \item the first exploration of the joint usage of feature and decision models for configuring \glspl{cpps} and creating \gls{cpps} artifacts.
    \end{itemize}
    \item A report on the gathered insights from exploring production sequences in this highly automated way.
\end{itemize}
We postulate that the \gls{eipse} approach, compared to the baseline of the traditional manual and hard-to-reproduce approach,  
\begin{enumerate*}[label=(\roman*)]
    \item helps to externalize the implicit knowledge of engineers,
    \item reduces the effort of \gls{cpps} configuration, including the exploration of feasible production process sequences and 
    \item separates the concerns of different engineering disciplines and, further, 
    \item benefits the reproducibility of the configuration process.
\end{enumerate*}
With these aspects, the approach directly addresses criteria of the VDI~3695 guideline for optimizing industrial plant engineering~\citep{vdi_3695,Jazdi2010}.
In particular, configuration and knowledge management, description languages, re-use, and the integration of disciplines.
Additional material to this paper, such as the model artifacts and variability models, can be found online.\footnote{\label{material}Additional material to \gls{eipse}: \url{https://github.com/tuw-qse/eipse}}

The results of our empirical evaluation demonstrate that we can automate the subsequent configuration process of a \gls{cpps} by using variability models; a clear benefit compared to the manual assembly and exploration process. 
Our subjects spend the most time defining the \gls{pprdsl}, while configuring the \gls{cpps} boils down to generating and configuring the variability models.
However, this shows how much effort it is to externalize their implicit knowledge, which, on the other hand, supports engineers downstream the engineering process.
Particularly, configuring the production process sequences through the \gls{dm} benefits from our prototype, which displays only the decisions that can be made in the respective configuration step. 
Furthermore, the separation of concerns of splitting and linking the variability models for products, production processes, and production resources allows their configuration by respective experts. 
Consequently, the \gls{eipse} process with tool support meets the expectations of automating and eliciting the configuration process of \glspl{cpps}, improving its reproducibility. 

The remainder of this work is structured as follows:
\autoref{sec:background} summarizes the background on \glspl{cpps} engineering and variability modeling. \autoref{sec:use_case} presents an illustrative use case. 
\autoref{sec:methodology} describes our research methodology. 
\autoref{sec:approach} presents the \gls{eipse} approach and corresponding tooling.
\autoref{sec:evaluation} describes the evaluation of the \gls{eipse} approach.
\autoref{sec:discussion} discusses the results and implications of the approach.
\autoref{sec:related_work} summarizes related work that aims at achieving similar goals and
\autoref{sec:conclusion} concludes this work by providing an outlook on future work.
 
\section{Background}
\label{sec:background}

This section provides background information on \gls{cpps} engineering and variability modeling.

\subsection{CPPS Engineering}
\label{sec:cpps}

\glspl{cpps} are next-generation production systems that interact autonomously with their environment, aiming for flexible production of customized products that build a product family~\citep{Meixner2020,MONOSTORI20149}. 
\glspl{cpps} are \glspl{cps}~\citep{gunes2014survey} with the purpose of manufacturing goods.
\gls{cpps} engineering resides in a multidisciplinary environment that involves engineers from diverse disciplines, such as mechanical, electrical, and computer sciences~\citep{biffl2017introduction}.

The engineers collaborate to create various artifacts representing aspects of the \gls{cpps}. 
For instance, they create technical documents such as text documents or spreadsheets for defining requirements or \textit{bills of materials}. 
The engineers also create engineering artifacts, e.g., CAD drawings or AutomationML artifacts~\citep{62714} to represent the \gls{cpps}'s physical and functional design. 
Additionally, the engineers use domain-specific language defined by industrial standards, such as IEC~61499~\citep{61499}, to implement the control software. 
Furthermore, several of these artifacts, e.g., \glspl{tcm}, contain information on variability in the \gls{cpps}, such as product types or production processes~\citep{FeichtingerVariVolution2020}. 

One prominent concept in \gls{cpps} engineering is the \gls{ppr} approach~\citep{schleipen2015requirements}, which describes
\begin{enumerate*}[label=(\roman*)]
  \item \textbf{p}roducts and their parts,
  \item production \textbf{p}rocesses required to manufacture the products, and
  \item production \textbf{r}esources that execute the processes.
\end{enumerate*}
The \gls{fpd}~\citep{vdi_3682} allows for modeling these \gls{ppr} aspects in a visual (and partly formalized) model (cf.~\autoref{fig:ppr-model}).
Complementary, we contributed the \gls{pprdsl}~\citep{Meixner2021}, which we use in this work, as a machine-readable \gls{ppr} format that represents the variability and constraints among the \gls{ppr} aspects.

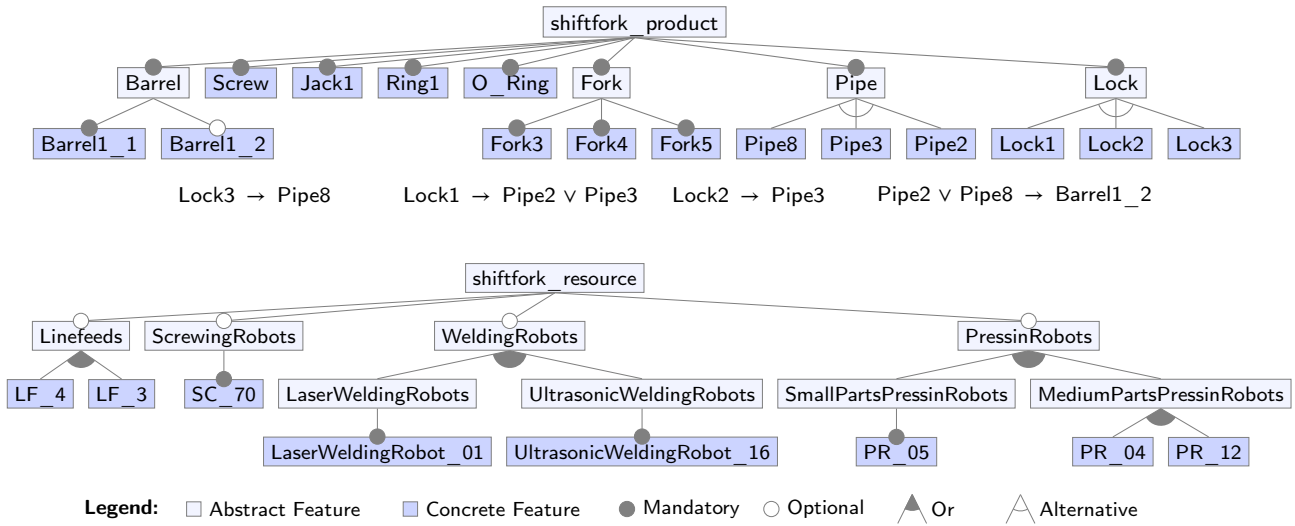
\begin{figure*}[ht]
    \centering
    \subfloat[][]{\forestset{
	/tikz/mandatory/.style={
		circle,fill=drawColor,
		draw=drawColor,
		inner sep=\circleSize
	},
	/tikz/optional/.style={
		circle,
		fill=white,
		draw=drawColor,
		inner sep=\circleSize
	},
	featureDiagram/.style={
		for tree={
			text depth = 0,
			parent anchor = south,
			child anchor = north,
			draw = drawColor,
			edge = {draw=drawColor},
		}
	},
	/tikz/abstract/.style={
		fill = blue!85!cyan!5,
		draw = drawColor
	},
	/tikz/concrete/.style={
		fill = blue!85!cyan!20,
		draw = drawColor
	},
	mandatory/.style={
		edge label={node [mandatory] {} }
	},
	optional/.style={
		edge label={node [optional] {} }
	},
	or/.style={
		tikz+={
			\path (.parent) coordinate (A) -- (!u.children) coordinate (B) -- (!ul.parent) coordinate (C) pic[fill=drawColor, angle radius=\angleSize]{angle};
		}	
	},
	/tikz/or/.style={
	},
	alternative/.style={
		tikz+={
			\path (.parent) coordinate (A) -- (!u.children) coordinate (B) -- (!ul.parent) coordinate (C) pic[draw=drawColor, angle radius=\angleSize]{angle};
		}	
	},
	/tikz/alternative/.style={
	},
	/tikz/placeholder/.style={
	},
	collapsed/.style={
		rounded corners,
		no edge,
		for tree={
			fill opacity=0,
			draw opacity=0,
			l = 0em,
		}
	},
	/tikz/hiddenNodes/.style={
		midway,
		rounded corners,
		draw=drawColor,
		fill=white,
		minimum size = 1.2em,
		minimum width = 0.8em,
		scale=0.9
	},
}

\footnotesize
\begin{forest}
	featureDiagram
	[shiftfork\_product,abstract[Barrel,abstract,mandatory[Barrel1\_1,concrete,mandatory][Barrel1\_2,concrete,optional]][Screw,concrete,mandatory][Jack1,concrete,mandatory][Ring1,concrete,mandatory][O\_Ring,concrete,mandatory][Fork,abstract,mandatory[Fork3,concrete,mandatory][Fork4,concrete,mandatory][Fork5,concrete,mandatory]][Pipe,abstract,mandatory[Pipe8,concrete,alternative][Pipe3,concrete][Pipe2,concrete]][Lock,abstract,mandatory[Lock1,concrete,alternative][Lock2,concrete][Lock3,concrete]]]	
	\matrix [anchor=north west] at (current bounding box.north east) {
		\node [placeholder] {}; \\
	};
\node () at (-1.5,-2.2) {\footnotesize Lock1  $\,\to\,$ Pipe2 $\lor$ Pipe3};
\node () at (1.5,-2.2) {\footnotesize Lock2 $\,\to\,$ Pipe3};
\node () at (-5,-2.2) {\footnotesize Lock3 $\,\to\,$ Pipe8};
\node () at (5,-2.2) {\footnotesize Pipe2 $\lor$ Pipe8 $\,\to\,$ Barrel1\_2};
\end{forest}
 \label{fig:fm_shift_product}} \\
    \subfloat[][]{\forestset{
	/tikz/mandatory/.style={
		circle,fill=drawColor,
		draw=drawColor,
		inner sep=\circleSize
	},
	/tikz/optional/.style={
		circle,
		fill=white,
		draw=drawColor,
		inner sep=\circleSize
	},
	featureDiagram/.style={
		for tree={
			text depth = 0,
			parent anchor = south,
			child anchor = north,
			draw = drawColor,
			edge = {draw=drawColor},
		}
	},
	/tikz/abstract/.style={
		fill = blue!85!cyan!5,
		draw = drawColor
	},
	/tikz/concrete/.style={
		fill = blue!85!cyan!20,
		draw = drawColor
	},
	mandatory/.style={
		edge label={node [mandatory] {} }
	},
	optional/.style={
		edge label={node [optional] {} }
	},
	or/.style={
		tikz+={
			\path (.parent) coordinate (A) -- (!u.children) coordinate (B) -- (!ul.parent) coordinate (C) pic[fill=drawColor, angle radius=\angleSize]{angle};
		}	
	},
	/tikz/or/.style={
	},
	alternative/.style={
		tikz+={
			\path (.parent) coordinate (A) -- (!u.children) coordinate (B) -- (!ul.parent) coordinate (C) pic[draw=drawColor, angle radius=\angleSize]{angle};
		}	
	},
	/tikz/alternative/.style={
	},
	/tikz/placeholder/.style={
	},
	collapsed/.style={
		rounded corners,
		no edge,
		for tree={
			fill opacity=0,
			draw opacity=0,
			l = 0em,
		}
	},
	/tikz/hiddenNodes/.style={
		midway,
		rounded corners,
		draw=drawColor,
		fill=white,
		minimum size = 1.2em,
		minimum width = 0.8em,
		scale=0.9
	},
}

\footnotesize
\scalebox{0.95}{
\begin{forest}
	featureDiagram
	[shiftfork\_resource,abstract[Linefeeds,abstract,optional[LF\_4,concrete,or][LF\_3,concrete]][ScrewingRobots,abstract,optional[SC\_70,concrete,mandatory]][WeldingRobots,abstract,optional[LaserWeldingRobots,abstract,or[LaserWeldingRobot\_01,concrete,mandatory]][UltrasonicWeldingRobots,abstract[UltrasonicWeldingRobot\_16,concrete,mandatory]]][PressinRobots,abstract,optional[SmallPartsPressinRobots,abstract,or[PR\_05,concrete,mandatory]][MediumPartsPressinRobots,abstract[PR\_04,concrete,or][PR\_12,concrete]]]]	
	\matrix [anchor=north west] at (current bounding box.north east) {
		\node [placeholder] {}; \\
	};
    \node [label=center:\textbf{Legend:}] () at (-6,-3.1) {};
    \node [abstract,label=right:Abstract Feature] () at (-5,-3.1) {};
    \node [concrete,label=right:Concrete Feature] () at (-2,-3.1) {};
    \node [optional,label=right:Optional] () at (3,-3.1) {};
    \node [mandatory,label=right:Mandatory] () at (1,-3.1) {};
\coordinate (a) at (4.95,-2.9);
        \coordinate (b) at (-0.2, -0.4);
        \coordinate (c) at (0.2,-0.4);
        \filldraw[drawColor] (a) - +(4.85,-3.1) - +(5.05,-3.1)- +(a);
		\draw[drawColor] (a) -- +(b);
		\draw[drawColor] (a) -- +(c);
		\fill[drawColor] (4.85,-3.1) arc (240:300:0.2);
    \node [or,label=right:Or] () at (5,-3.1) {};
        \draw[drawColor] (6.45,-2.9) -- +(b);
		\draw[drawColor] (6.45,-2.9) -- +(c);
		\draw[drawColor] (6.35,-3.1) arc (240:300:0.2);
    \node [alternative,label=right:Alternative] () at (6.5,-3.1) {};
\end{forest}
} \label{fig:fm_shift_resource}}
     \caption{FeatureIDE \glspl{fm}~\citep{FeatureIDE} of product (top) and production resource (bottom) variability of the \emph{shift fork} case study.
     }
    \label{fig:fm_shift}
\end{figure*}

\subsection{Variability Modeling}
\label{sec:cpps_vm}
Modeling variability explicitly is crucial for (software) product line engineering.  
In this work, two dimensions of variability play a role: 
\glspl{cpps} manufacture a product line of goods, such as families of cars, whereas \glspl{cpps} can also be configured in various ways. 
Furthermore, the sequence of the production steps involves dependencies between product parts and production resources and may vary in the estimated cost.
The multidisciplinary nature of \gls{cpps} engineering calls for different views on the \gls{cpps} involving different variability models~\citep{Meixner2020DocSym,Meixner2019}: feature models to capture structural variability of product parts and production resources, and decision models to capture the behavioral variability of production process sequences. 

\textbf{Feature models} \citep{kang1990feature} elicit commonalities and differences in a feature tree and allow for defining cross-tree constraints, e.g., that one feature requires or excludes another. 
Given this model, we can perform a configuration by selecting and deselecting features while conforming to the expressed constraints.
For instance, the \gls{fm} on the top of~\autoref{fig:fm_shift} captures commonalities and differences of a shift fork (c.f.,~\autoref{sec:use_case}). 
The model consists of \emph{mandatory} features representing product parts required in all variants, such as a \emph{Screw} and the three types of forks.
The model also contains \emph{optional} features, such as \emph{Barrel1\_2}, and feature groups, such as the alternative \emph{Pipe} group that allows selecting only one type of pipe.
The \gls{fm} on the bottom of~\autoref{fig:fm_shift} captures commonalities and differences of potential production resources, such as two \textit{Linefeeds} that can be used to feed material into the \gls{cpps}.

\begin{table*}[ht!]
\centering
\caption{Excerpt of the generated DOPLER \gls{dm}~\citep{DOPLER} representing the process variability of the \emph{shift fork} case study in tabular notation~\citep{Schmid2004}.}
\label{tab:dm_fork1}
\resizebox{\textwidth}{!}{\begin{tabular}{lllll}\toprule
ID &
  Question &
Range &
Visible/Relevant if &
  Constraint/Rule \\ \midrule 
  \rowcolor{myLighterGray}
Pipe &
  Which Pipe types? &
Pipe2 | Pipe 3 | Pipe8 &
false &
   \\
\rowcolor{myDarkerGray}
Barrel1\_2 &
  Install Barrel1\_2? &
true | false &
false &
   \\
  \rowcolor{myLighterGray}
Lock &
  Which Lock types? &
Lock1 | Lock2 | Lock3 &
false &
  \begin{tabular}[c]{@{}l@{}}Lock1 $\Longrightarrow$ Pipe = Pipe2 $\vee$ \\ ~~~~~~~~~~~~~~Pipe = Pipe3 \\ Lock2 $\Longrightarrow$ Pipe = Pipe3   \\ Lock3 $\Longrightarrow$ Pipe = Pipe 8\end{tabular} \\
\rowcolor{myDarkerGray}
InsertPipe &
  Install InsertPipe? &
true | false &
false &
   \\
\rowcolor{myLighterGray}
InsertPipe2 &
  Install InsertPipe2? &
true | false &
Pipe == Pipe2 &
  InsertPipe2 $\Longrightarrow$ InsertPipe \\
\rowcolor{myDarkerGray}
InsertLock &
  Install InsertLock? &
true | false &
false &
   \\
\rowcolor{myLighterGray}
InsertLock1 &
  Install InsertLock1? &
true | false &
Lock == Lock1 &
  InsertLock1 $\Longrightarrow$ InsertLock \\
\rowcolor{myDarkerGray}
InsertLock2 &
  Install InsertLock2? &
true | false &
Lock == Lock2 &
  InsertLock2 $\Longrightarrow$ InsertLock \\
\rowcolor{myLighterGray}
InsertBarrel1\_2 &
  Install InsertBarrel1\_2? &
true | false &
Barrel1\_2 &
   \\
\rowcolor{myDarkerGray}
  PressBarrel1\_2 &
  Install PressBarrel1\_2? &
true | false &
Barrel1\_2 \& InsertBarrel1\_2 \& InsertPipe &
  \\
\dots & \dots & \dots & \dots & \dots \\
  \bottomrule
\end{tabular}}
\end{table*}
 
\textbf{Decision models} \citep{schmid2011comparison} are rooted in the Synthesis meth\-od~\citep{campbell1990introduction}, which supports the reuse of processes and the necessary assets for configuring an application engineering process.
\glspl{dm} include only varying features and their constraints. 
For instance, the DOPLER approach \citep{DOPLER} comprises \glspl{dm} and asset models for defining variability in the problem space and reusable elements and their dependencies in the solution space, respectively. 
It maps assets onto the decisions, but decisions are unaware of the assets.

\autoref{tab:dm_fork1} represents an exemplary \gls{dm} in tabluar representation. 
A decision in a \gls{dm} consists of a unique \texttt{ID} and a text describing the decision (usually a \texttt{Question}).
These decisions are configured based on the specified \texttt{Range}.
For instance, only one of the three locks can be selected in the enumeration decision \emph{Lock}. 
\texttt{Constraint/Rule} and \texttt{Visible/Relevant if} relationships between decisions specify \emph{(post-)conditions} and hierarchical or logical \emph{(pre-)conditions}.
The \texttt{Visible/Relevant if} relationship defines preconditions that need to be satisfied for the decision to be selectable.
For instance, \emph{InsertLock1} can only be selected if \emph{Lock1} is selected. 
A visibility condition \emph{false} (e.g., \emph{InsertLock}) entails that the decision can not be selected by developers but rather by selecting another decision that relies on this decision. 
Similarly, abstract features in \glspl{fm} are automatically selected if one of their child features is selected. 
For instance, the selection of \emph{InsertLock1} triggers the selection of \emph{InsertLock}. 
These explicit dependencies among selected decisions reflect a configuration order, which models behavioral variability.

\textbf{\glspl{cdc}}~\citep{Fadhlillah22,FadhlillahSPLC22} model the relationships among different types of variability models, likely employed by different (engineering) disciplines, as shown in \autoref{lst:cdc-fork}.
A \gls{cdc} identifies the involved variability model using the variability type and the unique id.
For instance, \emph{CDC1} in \autoref{lst:cdc-fork} refers to the relation of a \productFM{} feature (\texttt{shiftfork\_product\#Pipe2}), which concerns product engineers, to the \processDM{} decision (\texttt{shiftfork\_process\#Pipe2}), which concerns production process engineers, to model dependent processes in the decision model.
Similarly, \textit{CDC3} models the relation of the \textit{WeldLock} process to the \textit{WeldingRobot} production resource that concerns production resource engineers.

\begin{listing}[t]
\begin{minted}[linenos,tabsize=2,breaklines,fontsize=\footnotesize,xleftmargin=0.6cm]{sql}
CDC1) shiftfork_product#Pipe2 => shiftfork_process#Pipe2
CDC2) shiftfork_product#Pipe2 => shiftfork_process#InsertPipe2
CDC3) shiftfork_process#WeldLock => 
     shiftfork_resource#WeldingRobots
CDC4) shiftfork_product#Lock1 => shiftfork_product#Pipe2 || 
     shiftfork_product#Pipe3;
\end{minted}
\vspace{-1.25em}
\caption{Excerpt of \glspl{cdc}~\citep{Fadhlillah22} of the \emph{shift fork} case study.}
    \label{lst:cdc-fork}
\end{listing}

\section{Motivating Case Study}
\label{sec:use_case}

This section presents the shift fork case study~\citep{Meixner2021}, one example of a \gls{cpps}, based on which we illustrate our contribution in the following sections.
Shift forks are part of a manual transmission in a car.\footnote{Exemplary figure of \emph{shift forks}: \url{https://w.wiki/3DCf}}  
They shift the cuffs with two forks along the pipes to their correct position to connect the transmission's gears.
Typically, a single shift fork moves a particular cuff for two gears, e.g., the first and the third gear.
This design results in a product portfolio of four shift fork variants required for a particular type of car transmission.

In the first \gls{cpps} engineering phase, so-called \textit{basic engineers} analyze the product portfolio that the \gls{cpps} should produce~\citep{Meixner2021}, which represents a classical (mechanical) product line with common and varying features.
The engineers examine various artifacts, such as CAD drawings or product prototypes, to identify the commonalities and differences of the product line~\citep{Meixner2021}.
\begin{figure}[t]
    \centering
    \includegraphics[width=0.99\columnwidth]{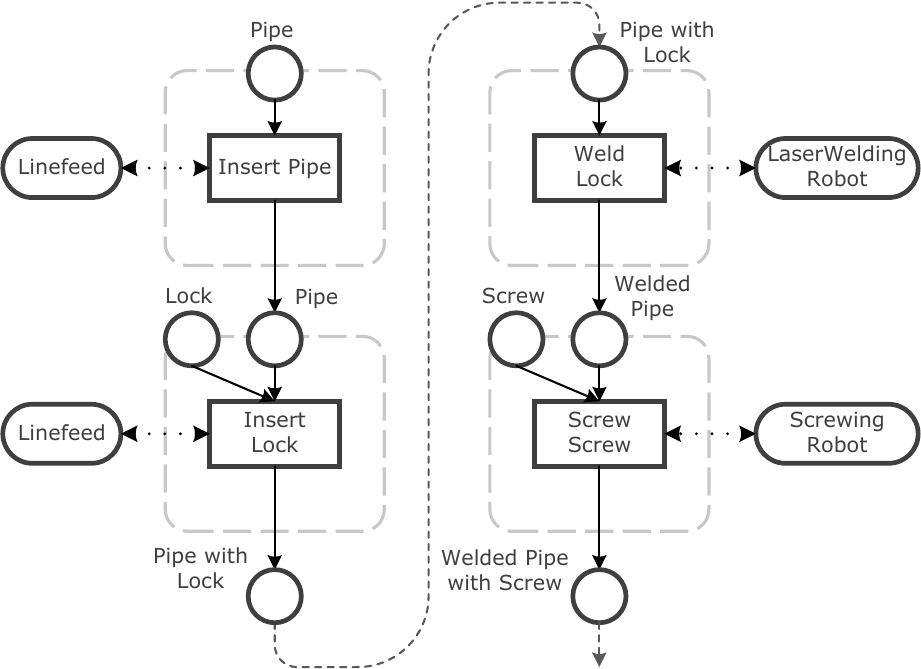}
    \caption{Excerpt of a \gls{ppr} model for the production steps of a shift fork in VDI~3682 notation~\citep{vdi_3682}.}
    \label{fig:ppr-model}
\end{figure}
Furthermore, the engineers examine the required partial production steps, such as joining the fork and the pipe of the shift fork by welding them together~\citep{Meixner2021}.
\autoref{fig:ppr-model} shows a section of such a production step sequence for a single product in VDI~3682 notation~\citep{vdi_3682} (cf.~\autoref{sec:background}), where, in the first step, a \textit{Pipe} (product) \textit{is inserted} (process) into the \gls{cpps} using a \textit{Linefeed} (resource).\footnote{In contrast to the figure, the engineers first plan the single process steps, e.g., \emph{insert pipe} in an isolated way.}
In this step, they often disassemble the prototypes and put them ``back together''~\citep{lee1989disassembly}.
The engineers aim to design feasible production processes for the requested product line.
They complement these designs with potential basic \gls{cpps} designs and corresponding rough cost estimates~\citep{Meixner2021}.\footnote{
If customers accept a \gls{cpps} design and cost estimate, the artifacts of \textit{basic engineering} are handed over to the different disciplines of \textit{detail engineering}.
As the time of a cost estimate and its acceptance can be wide apart, the engineers in \textit{detail engineering} need to reiterate over those designs, ideally, reuse, and detail them.}

Some product parts require a specific production sequence. 
For instance, the pipe must be available to mount forks onto the pipe.
Still, several degrees of freedom to assemble a particular product remain, which must be resolved either in engineering or operation.
Traditionally, the engineers design feasible process sequences, e.g., in tools like DelMia,\footnote{\label{delmia} DelMia: \url{https://www.3ds.com/products-services/delmia/}} based on implicit knowledge, by using heterogeneous and partial data representations. 
The engineers use the sequences to reason about process characteristics, such as the production duration.
However, as the decisions are typically undocumented, a change to the \gls{cpps}, which occurs frequently, may cause redesigning the entire \gls{cpps} from scratch \citep{Paetzold2017,Meixner2021}.
Therefore, making this knowledge more explicit is a decade-old challenge in \gls{cpps} engineering~\citep{Drath2008,drath2009datenaustausch}.

\begin{listing}[t!]
\begin{minted}[linenos,tabsize=2,breaklines,fontsize=\footnotesize,xleftmargin=0.6cm]{bash}
Product "Pipe": { name: "Abstract Pipe", isAbstract: true }
Product "Pipe2": { name: "Pipe 2",
  implements: ["Pipe"], 
  excludes: [ "Pipe3", "Pipe8" ]
}

Product "Lock": { name: "Abstract Lock", isAbstract: true , 
  requires: ["Pipe"]
}
Product "Lock1": { name: "Lock 3",
  implements: ["Lock"], excludes: [ "Lock2", "Lock3" ]
}

Process "InsertPipe": { name: "InsertPipe", isAbstract: true }
Process "InsertPipe2": { name: "InsertPipe2",
  implements: ["InsertPipe"],
  inputs: [ {productId: "Pipe2"} ],
  outputs: [ {OP1: {productId: "Pipe2"}} ],
  resources: [ { resourceId: "Linefeeds" } ]
}
Process "WeldLock": { name: "WeldLock", isAbstract: true ,
  requires: [ "InsertLock", "InsertPipe", ... ],
  inputs: [ {productId: "Lock"}, {productId: "Pipe"} ],
  outputs: [ {OP2: {productId: "ForkProduct"}}],
  resources: [ {resourceId: "WeldingRobot"} ]
}
Process "WeldLock1": { name: "WeldLock1",
  implements: [ "WeldLock" ], inputs: [ "Lock1" ]
}

Resource "WeldingRobot": { name: "WeldingRobot", 
  isAbstract: true }
Resource "LaserWeldingRobot_01":{ name: "LaserWeldingRobot_01",
  implements: [ "LaserWeldingRobots" ]
}

Constraint "C1": {
  definition: "Lock1,Pipe2,Pipe3 -> Lock1 implies Pipe2 OR Pipe3" 
}
\end{minted}
    \vspace{-1.25em}
    \caption{Excerpt of a \gls{ppr} model for the \emph{shift fork} case study in \gls{pprdsl}~\citep{MeixnerETFA2021PPRDSL}.}
    \label{lst:fork}
\end{listing}

To elicit their domain knowledge, \gls{cpps} engineers can use the \gls{pprdsl}~\citep{MeixnerETFA2021PPRDSL} to represent \gls{ppr} aspects and their constraints. 
\autoref{lst:fork} shows an excerpt of a \gls{pprdsl} file that defines \emph{shift fork} product parts (lines 1-12), the atomic production process steps and variants (lines 14-29), the production resources (lines 31-35), and the constraints between these \gls{ppr} aspects (lines 37-39).
The entire \gls{pprdsl} file is available online.\footref{material}

For instance, Line~1 defines the abstract \code{Product} \emph{Pipe} that builds the central part of a \emph{shift fork}. 
The \code{Product} \emph{Pipe2}, defined in the following line, represents a concrete pipe and excludes another variant of a pipe (i.e., \emph{Pipe3} and \emph{Pipe8}). 
Similarly, \emph{InsertPipe} in Line 14 represents an abstract \code{Process} which is implemented, for instance, by the \code{Process} \emph{InsertPipe2}. 
The \gls{pprdsl} further enumerates its input and output products, i.e., \emph{Pipe2}, and required production resources. 
Examples of production resources, which can be abstract or concrete, are stated in Lines 31-33. 
The last lines of the excerpt, Lines 37 to 39, add the \code{Constraint} \emph{C1} which defines that the \textit{Lock1} \code{implies} the presence of either \textit{Pipe2} \code{or} \textit{Pipe3}.

Thus, the \gls{pprdsl} describes the products and their parts, the required production process steps, and the production resources of the shift fork case study precisely.
Particularly, it elicits the variability and dependencies among the three aspects. 
 
\section{Methodology}
\label{sec:methodology}

\begin{figure}
    \centering
    \includegraphics[width=\columnwidth]{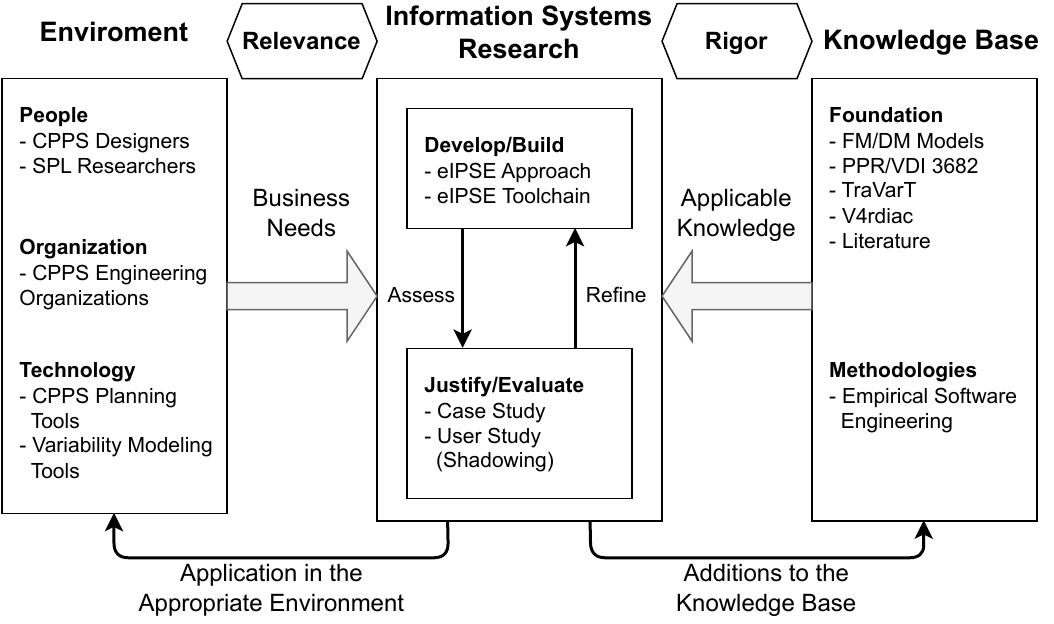}
    \caption{Design Science methodology \citep{hevner2007three,hevner2008design} for this work.}
    \label{fig:designscience}
\end{figure}

To develop the \gls{eipse} approach, we applied the Design Science methodology~\citep{hevner2007three,hevner2008design}.
This methodology aims to solve problems by covering the design and investigation of artifacts that define, above others, methods and technical solutions in a problem context to improve something in that context~\citep{hevner2008design,wieringa2014design}.
In this work, the method is the iterative design and investigation of the \gls{eipse} approach in the context of \gls{cpps} engineering.
Fig.~\ref{fig:designscience} shows the Design Science methodology adapted to this work, with the \emph{relevance} cycle on the left, the \emph{design} cycle in the middle, and the \emph{rigor} cycle on the right.

Our previous work~\citep{FeichtingerVariVolution2020, FeichtingerTraVarT2021, MeixnerVamos2022ProcessExploration, MeixnerETFA2021PPRDSL} builds the \emph{grounding} in the \emph{rigor cycle}.
Furthermore, we aim to address literature gaps, in particular, the integration of different variability models of \glspl{cpps}~\citep{Kruger:2017:BSP:3106195.3106217} including their behavior~\citep{Fang2013,galster2013variability} and the separation of concerns between engineers of different disciplines~\citep{Ananieva2016}.
Based on discussions with stakeholders of our industry partners and our previous work, we obtain additional requirements for an integrated \gls{cpps} variability modeling approach (cf.~\autoref{sec:requirements}) for the \emph{relevance cycle}.

We iteratively perform the following tasks for the \emph{design cycle} and the \emph{develop/build activity} (cf. \autoref{sec:approach}):
First, to address the identified requirements, we extend our former approach with respective steps, resulting in the \gls{eipse} approach, and investigate which steps can be further automated (cf., ~\autoref{subsec:approach_extendedIPSE}). 
We lay out the architecture of the \gls{eipse} toolchain to illustrate a prototype implementation (cf.,~\autoref{subsec:architecture}).
We integrate a DOPLER~\gls{dm} into the state-of-the-art technology and develop the remaining transformation operations for converting the \gls{pprdsl} into three variability models.
We develop a \emph{\gls{dm} Editor} as a prototype for modeling and configuring the \processDM{} and design the automated reduction of the \processDM{} configuration based on the configuration of the \productFM{}. 
Then, we adapt \gls{v4rdiac} to use the distinct configurations of the \productFM{} and \processDM{} for generating artifacts. 
In this way, we establish the \gls{eipse} toolchain to support the steps of the \gls{eipse} approach (cf.~\autoref{sec:approach}).
Furthermore, we aim to separate the concerns of \textit{basic engineers} and the disciplines involved in \textit{detail engineering}, particularly product design, production process engineering, and production resource engineering.

For the \emph{evaluate activity} of the \emph{design cycle}, i.e., the evaluation of the approach and prototype, we rely on a three-fold approach.
First, we applied the \gls{eipse} approach to a set of previously published case studies~\citep{Meixner2021} from the \gls{cpps} domain to investigate the feasibility of the approach (cf. \textbf{EQ1} and \autoref{sec:usecase_evaluation}).
Second, we conducted an observational user study~\citep{wohlinExperimentationInSoftwareEngineering2012,runeson2012case,singer2008software} with six engineers to provide evidence on the perceived usability of the \gls{eipse} approach and learn about its usage.
Therefore, we requested study subjects to perform the \gls{eipse} process on a newly introduced case study (cf. \textbf{EQ1} and \autoref{sec:user_evaluation}).
A detailed description of the user study can be found in the appendix, cf. Section~\ref{sec:guideline}.
Third, we investigate the configuration space's reduction for the production sequences of one particular case study using the \gls{eipse} approach (cf.~\textbf{EQ2} and \autoref{sec:reduction_evaluation}).
Finally, we examined the generation of parameterized \gls{cpps} artifacts for a particular \gls{cpps} configuration for each of the four previously mentioned case studies (cf. \textbf{EQ3} and \autoref{subsec:eval_controlSWgeneration}).
These measures complete the \emph{design cycle}.
 
We discuss our findings and lessons learned and accompany them with additional material\footref{material} and a demonstration video\footnote{\label{demonstration}eIPSE demonstration video: \url{https://youtu.be/eoNNDOusXKA}} as additions to the knowledge base.
Furthermore, we briefly describe measures to assess the practical impact of the application in \gls{cpps} engineering.
These measures aim to reach a broad audience in academia and industry and complete the \emph{rigor} and \emph{relevance cycle} (cf.~\autoref{sec:discussion}).

\section{Adopting Variability Modeling for CPPS Process Exploration and Configuration}
\label{sec:approach}
This section details employing variability modeling for exploring configuration options in deriving a functional \gls{cpps} design.
Firstly, \autoref{sec:requirements} states the requirements for a \gls{cpps} production process exploration and production resource configuration approach as elicited from industrial needs in previous work. 
Inferred from these requirements, \autoref{subsec:approach_extendedIPSE} describes increasing automation in CPPS planning and configuring through the \glsfirst{eipse} approach. 
To assess the feasibility of the \gls{eipse} approach and to follow the Design Science methodology, 
\autoref{subsec:architecture} presents our prototype implementation, respectively. 

\subsection{Requirements for eIPSE}
\label{sec:requirements}

To gather requirements for automating the configuration of \glspl{cpps}, we conducted interviews with industrial partners in previous work~\citep{MeixnerVamos2022ProcessExploration} on which we build in this article. 
The resulting requirements \textbf{R1}-\textbf{R3} mainly motivate knowledge representation and tool support as follows:

\textbf{R1. Production Variability Exploration.} The approach shall collect variability knowledge from \gls{cpps} engineering (artifacts) that is required to explore production process sequence options efficiently.
The approach must incorporate new knowledge from product line evolution, such as changes of the products that the \gls{cpps} manufactures or process simulation and optimization iteratively.

\textbf{R2. Product \& Process Variability Representation.} 
The knowledge representation shall describe
\begin{enumerate}[label=(\roman*), nosep]
    \item the products, which form a product line,
    \item the atomic production process steps, which form a process line, both with their dependencies and \glspl{cdc} in an industrial variability artifact, e.g., using the \gls{pprdsl}~\citep{MeixnerETFA2021PPRDSL}, and
    \item variability models of the product and process line in state-of-the-art variability models (e.g., \glspl{fm} or \glspl{dm}).
\end{enumerate}

\textbf{R3. Variability Transformation \& Configuration.} 
The \gls{ipse} method and tool support shall 
\begin{enumerate*}[label=(\roman*)]
    \item automate transformations of the variability represented in the industrial artifacts into state-of-the-art variability models and
    \item provide guidance through product and process variability model exploration and configuration supported by tools, such as the FeatureIDE~\citep{FeatureIDE} or DOPLER tool~\citep{DOPLER} to better support multidisciplinary engineering. 
\end{enumerate*}

However, requirements R1-R3 only focus on exploring and configuring products and processes without considering production resources. 
Thus, based on previous work and discussions with stakeholders, we defined two additional requirements (\textbf{R4} and \textbf{R5}) that address the necessity to include the production resources in the \gls{ipse}. 

\textbf{R4. Resource Variability Representation.} 
After exploring feasible production sequences, the engineers goal is to ``search'' for suitable production resources, e.g., welding robots, that are able to execute the production steps (cf. \autoref{sec:use_case}).
Therefore, the \gls{eipse} knowledge representation shall model a product line of potential production resources that can execute respective production process steps.
The product line should be represented in an industrial variability artifact, e.g., a variability model. 
Additionally, the \gls{eipse} method shall respect the production resource variability model when being transformed and configured (\textbf{R3}). 
The production resource variability model shall be transformed into state-of-the-art variability models and adequate tool support shall guide configuration.

\textbf{R5. Cross-Discipline Constraint Representation.} 
The \gls{cpps} engineering process expects engineers to link the developed concepts to a system design, e.g., that a production process requires a particular type of welding robot.
To complement the overall \gls{eipse} knowledge representation, we must be able to describe dependencies (include/exclude relations) between variability knowledge across different \gls{ppr} concepts.

These requirements reflect the need of engineers to continuously incorporate additional product requirements and subsequently assign production processes and resources to the \gls{cpps} design.
Furthermore, they represent the engineers' demand to obtain a holistic overview of a \gls{cpps}'s variability in their preferred engineering artifacts and models that can be better computed, e.g., for satisfiability, such as state-of-the-art variability models.
Beyond that, the requirements build the foundation for the automation of an integrated \gls{cpps} variability modeling approach.

\subsection{The Extended IPSE Approach}
\label{subsec:approach_extendedIPSE}
In prior work~\citep{MeixnerVamos2022ProcessExploration}, we introduced the linear \gls{ipse} approach, which utilizes state-of-the-art variability models to enable the systematic and reproducible exploration of potential production process sequence and resource configurations based on a product configuration.
These variability models need to support \textit{structural} variability to represent the hierarchical structure of products and \textit{behavioral} variability to represent the potential sequences of process steps.
The tool support for the original \gls{ipse} approach allows for \textit{exploring process sequences} manually, but \textit{does not} include the modeling and configuration of \textit{production resources} and the \textit{artifact generation}.
To address these issues and satisfy the additional requirements \textbf{R4} and \textbf{R5}, we utilize a \gls{fm} to represent resource variability and \glspl{cdc}~\citep{Fadhlillah22} to represent one or more dependencies across \gls{ppr} concepts of different disciplines.
Here, each \gls{cdc} is a propositional logic constraint based on the variation points defined in \productFM{}, \processDM{}, and \resourceFM{} (cf.~\autoref{lst:cdc-fork}). 
To this end, we extend our previous work with the \gls{eipse} approach and toolchain.

\begin{figure*}[h!t]
  \centering
  \includegraphics[width=0.99\textwidth]{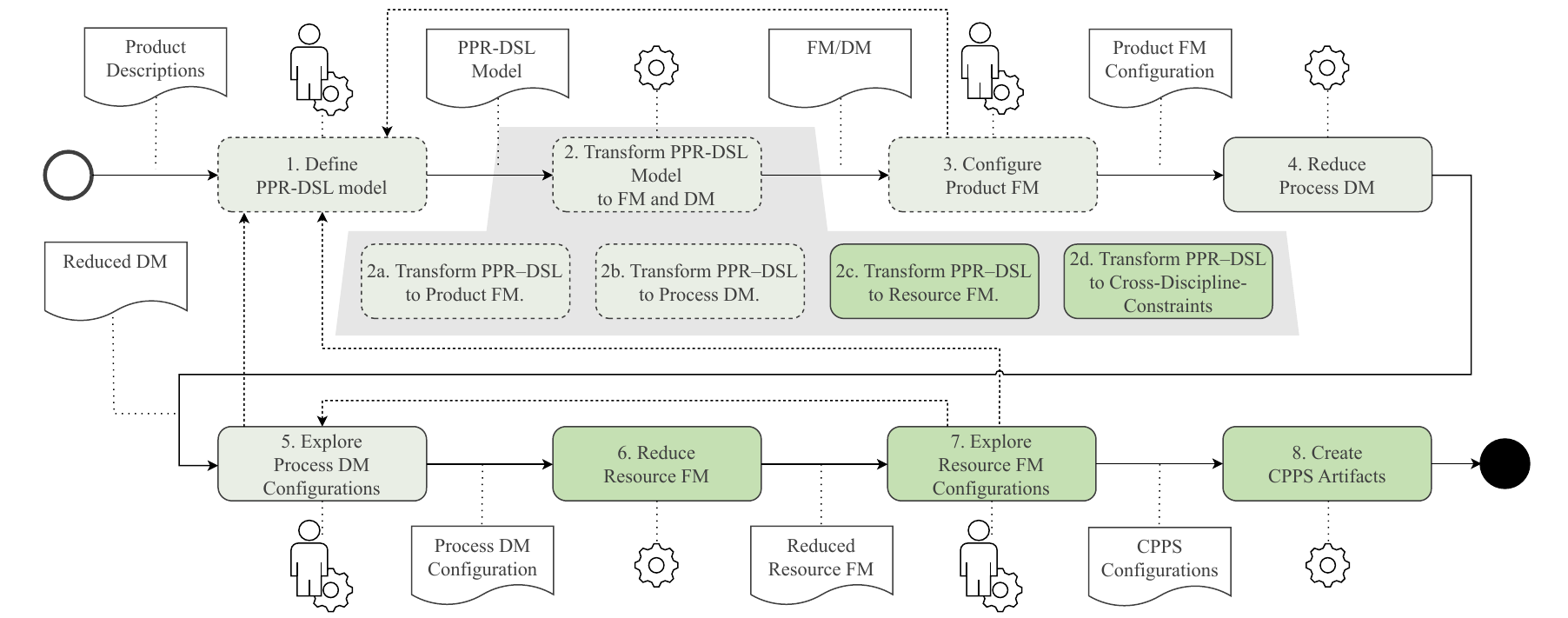}
  \caption{(Human \& automated) \gls{eipse} process steps for exploring production process steps based on a product configuration (updated steps have solid contours, novel steps, additionally a darker color).} 
  \label{fig:process}
\end{figure*}

\paragraph{Overview}
In this article, we contribute the \gls{eipse} approach, which extends the \emph{linear} \gls{ipse} approach with 
\begin{enumerate}[label=(\roman*),nosep]
    \item automated transformations from the \gls{pprdsl} to the \resourceFM{} and the \glspl{cdc} for linking the variability models,
    \item an automated reduction of the \processDM{} configuration based on the \productFM{} configuration and the \emph{tool-supported} process exploration that guides engineers, 
    \item the automated reduction of the \resourceFM{} configuration and its \emph{tool-supported} configuration,
    \item the generation of \gls{cpps} artifacts, such as control code artifacts,
    \item and feedback loops to respect its iterative character.
\end{enumerate}
This way, \gls{eipse} aims to support the disciplines of functional product design, production process engineering, and production resource engineering.
\autoref{fig:process} illustrates the resulting \gls{eipse} process.
Steps with dashed contours in \autoref{fig:process} were carried over \textit{as-is} from the \gls{ipse} approach, steps with solid contours were adapted, and steps with solid contours and in dark green were newly introduced in the \gls{eipse} approach. 
The process consists of ``human'' tasks conducted by engineers with tool support (persona with cog icon) and automated tasks (cog icon). 
\gls{eipse} provides automation support by utilizing the \gls{eipse} tool.
While the \gls{eipse} tool performs the automated tasks entirely, it supports \gls{cpps} engineers during the ``human'' tasks that require access to models.

The \gls{eipse} process starts with the definition of the \gls{ppr} element variants of the desired \gls{cpps} by engineers in the \gls{pprdsl} and ends in creating the artifacts for a configured \gls{cpps} automatically. The \gls{pprdsl} is automatically transformed into three variability models (Step~2a-c) and their respective \glspl{cdc} (Step~2d). 
The variability models are configured by engineers and automatically and subsequently configured for the next configuration step (Steps~3-7). 

In contrast to the \gls{ipse} process, the \gls{eipse} process splits the second step into sub-steps to demonstrate the various variability models that the \gls{eipse} tool creates based on the \gls{pprdsl}. 
It creates an additional \resourceFM{} to represent the production resources (Step~2c) and derives and elicits the \glspl{cdc} among the three variability models (Step~2d). 
The reduction of the \processDM{} configuration (Step~4) is automated.
The exploration and configuration of the \processDM{} (Step~5) is integrated with \textit{state-of-the-art} technology and supported by the \gls{eipse}~tool.
Subsequently, in Step~6, the additionally created \resourceFM{} configuration is automatically reduced based on the decisions taken (i.e., the configuration of the \processDM{}), and next, configured by the \gls{cpps} engineer with the \gls{eipse}~tool. 
Finally, Step~8 creates the control code for the configured production resources to operate the \gls{cpps}.
The following descriptions highlight the added and adapted steps compared to the original \gls{ipse} and explain the \gls{eipse}~tool.

\paragraph{Details}
First, \gls{cpps} engineers model product parts and product variants and identify atomic production steps and production resources to manufacture these products based on analyzing the product descriptions and production requirements in Step~1. 
In this way, they iteratively define a \gls{pprdsl} model (cf.~\autoref{lst:fork}).
In practice, engineers typically select and adapt such atomic process steps and, particularly, production resources from an artifact catalog.

After defining the \gls{pprdsl}, in Step~2, the \gls{eipse} uses transformation technology, such as \travart{}~\citep{FeichtingerTraVarT2021}, to create the variability models automatically.
In contrast to the \gls{ipse} process, which derived a \productFM{} and a \processDM{} only, this second step transforms the \gls{pprdsl} into three variability models with shared \glspl{cdc} (cf. requirement \emph{R3} \& \emph{R5}).  
It creates 
\begin{enumerate}[label=(\roman*), nosep]
    \item a \productFM{} to represent the product variants, their parts, and structure (cf. upper \gls{fm} in~\autoref{fig:fm_shift}),
    \item a \processDM{} to represent the atomic process steps and their dependencies, representing partial behavior of the \gls{cpps}, and product decisions not shown to engineers during the configuration but required for the satisfiability calculation of the model (cf.~\autoref{tab:dm_fork1}),
    \item a \resourceFM{} to represent the hierarchical structure of production resources (cf. lower \gls{fm} in~\autoref{fig:fm_shift}), and
    \item \glspl{cdc} capturing the dependencies between the \productFM{}, the \processDM{}, and the \resourceFM{} in propositional logic (cf. \autoref{lst:cdc-fork}).
\end{enumerate}

The following steps configure and reduce the possible configuration spaces of the variability models. 
The \gls{cpps} engineer starts with configuring the \productFM{} in Step~3, resulting in a valid configuration according to the \gls{fm}. 
For instance, an engineer selects the \emph{Pipe2} and \emph{Lock1} in the in the \productFM{} configuration (cf.~\autoref{fig:fm_shift}).
Based on the configuration, the \gls{eipse} tool in Step~4 automatically reduces the subsequent \processDM{} configuration to exclude decisions that are unnecessary to produce the configured product.
For each selected feature in the \productFM{} configuration, the configuration value for the corresponding decision in the \processDM{} is set to \code{true}.
For instance, for the selected \emph{Pipe2} feature, the \gls{eipse} tool will set the configuration value of the \emph{Pipe2} product decision to \code{true}.
For all product features that are not selected, the configuration value of the respective decision remains \code{false}.
The configuration values in the \processDM{} set in this way affect which process decisions are visible during the configuration, e.g., decision \emph{InsertPipe2} in ~\autoref{tab:dm_fork1}, due to its visibility conditions.

In this reduced \processDM{} configuration, in Step~5, a \gls{cpps} engineer can explore the remaining process decisions iteratively and interactively with the \gls{eipse} tool. 
Based on a constant evaluation, the tool only displays the process steps feasible during the configuration stage according to the visibility conditions. 
Based on the internal constraints of the \processDM, the \gls{eipse} tool sets the subsequent configuration values.
Furthermore, the \gls{eipse} tool stores and visualizes a queue representing the sequence of the currently taken decisions. 
The final sequence of the configured decisions represents the desired production sequence.
This production sequence can be used to define and optimize a valid production process model that is executed on the \gls{cpps}.

In  Step~6, the \gls{eipse} tool automatically reduces the \resourceFM{} configuration (cf. requirement \emph{R3}) by considering the \processDM{} configuration of Step~5.
This results in a partial \resourceFM{} configuration, where the configuration possibilities are a valid subset of production resource configurations for a particular product and process sequence configuration.
For each selected decision in the \processDM{} configuration from Step~5, the \gls{eipse} tool considers the \glspl{cdc} for the production processes and resources.
If a process requires a type of production resource, the configuration value for the corresponding features in the \resourceFM{} is preselected.
For instance, if the \emph{WeldLock1} decision is part of the configured process sequence, the \emph{WeldingRobot} group is selected.

Based on the \resourceFM{}, in Step~7, a \gls{cpps} engineer can configure the desired production resources with the \gls{eipse} tool. 
As the \resourceFM{} is pre-configured by the \processDM{} configuration (Step~5), the configuration of the \resourceFM{} only contains the production resources to be used in the production process.

In the final step, Step~8, engineering and operation artifacts should be generated from the combined configuration of the \productFM, the \processDM, and the \resourceFM.
For instance, IEC~61499~\citep{61499} or AutomationML~\citep{Drath2021} code can be parameterized and generated from reusable artifacts.
This step aims to increase the reusability of artifacts and decrease artifact creation time to increase engineering productivity.
To support this automation, an \gls{eipse} tool requires additional mechanisms for generating \gls{cpps} artifacts.
Therefore, \gls{cpps} engineers need to decide on a \emph{variability mechanism}~\citep{ApelFOSD2013}, such as \emph{Delta Modeling}~\citep{Schaefer2021,Zhang2016}, to implement the shared and the varying \gls{cpps} artifacts. 
In Delta modeling, engineers create elements, such as stubs, templates, and lines of code, that represent the base implementation, in our case, particular \gls{cpps} artifacts such as control code. 
Then, they define Delta models comprising modifications to enrich the base implementation. 
These operations can modify the base implementation or add or remove code (cf. \autoref{lst:delta-model-fork}). 
During artifact generation, a generator parameterizes and combines this base implementation and the variable code parts, i.e., the Deltas, depending on the variability models' configuration.

\paragraph{Summary}
The result of the \gls{eipse} process is a set of configurations as well as engineering and operation artifacts, such as control code for a possible design of a \gls{cpps} or plan for the assembly of a concrete product.
This set of configurations can be used, for instance, to generate a layout of the \gls{cpps}, optimize it, and initiate its operation.

\subsection{\gls{eipse} Toolchain Architecture}
\label{subsec:architecture}
\begin{figure*}[ht]
    \centering
    \includegraphics[width=0.99\textwidth]{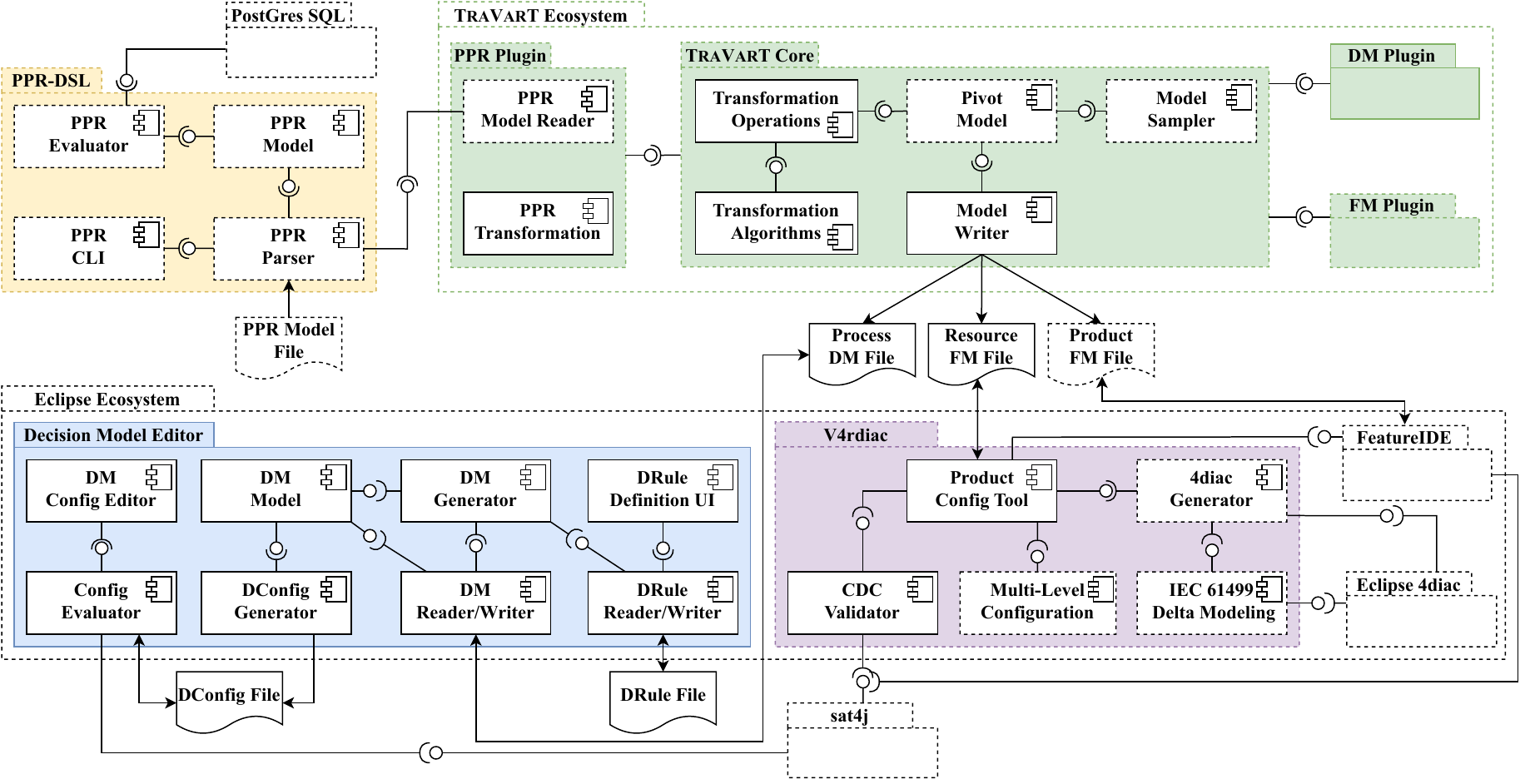}
    \caption{Architecture of the \gls{eipse} toolchain in UML component diagram notation (novel and updated components are depicted with solid contours).}
    \label{fig:ipse_architecture}
\end{figure*}

This section describes our implementation of an \gls{eipse} tool. 
In particular, it presents the architecture of our prototype as well as further implementation details.
\autoref{fig:ipse_architecture} depicts this architecture with its main components, which consist of
\begin{enumerate*}[label=(\roman*)]
    \item the \gls{pprdsl} to define, read, and evaluate \gls{ppr} models (\gls{eipse}, Step~1),
    \item \travart{} to transform \gls{pprdsl} models into \glspl{fm} and \gls{dm} (Step~2),
    \item the FeatureIDE to configure the \productFM{} (Step~3),
    \item the \gls{eipse} \gls{dm} editor to read or define, manipulate, and configure \glspl{dm} (Steps~4 \& 5), and
    \item \gls{v4rdiac} to read, manipulate, and configure models with \glspl{cdc} (Steps~6 \& 7) and generate IEC~61499 code based on Delta models~\citep{Clarke2015_deltaModelling} (Step~8).
\end{enumerate*}
Compared to our previous work, we provide additional automation support with extended \travart{} transformations for production processes and resources in the \gls{pprdsl}, introduce the novel \gls{eipse} \gls{dm} editor, and integrate \gls{v4rdiac}.
To this end, in Figure~\ref{fig:ipse_architecture}, unchanged components are depicted with dashed contours, and novel or adapted components are depicted with solid contours.

\paragraph{\gls{pprdsl} environment.}
The \gls{pprdsl} environment is realized as a standalone Java application (cf.~\autoref{fig:ipse_architecture}, top left in yellow).
Its main components are the \code{\gls{ppr} Model} (based on an extended VDI~3682 model~\citep{vdi_3682}), the \code{\gls{ppr} \gls{cli}}, the \code{\gls{ppr} Parser}, and the \code{\gls{ppr} Evaluator} for constraints.
The parser reads the \code{\gls{ppr} File}, which includes the products, production process steps, production resources, and constraints via the \gls{cli} and builds the \code{\gls{ppr} Model}.
The constraints are mapped onto (recursive) SQL queries and evaluated using a \code{PostgreSQL} database.
This strategy also allows the use of aggregation functions, such as the sum or average of attribute values, over a \gls{ppr} hierarchy and an easy integration with industrial standards.

Currently, the tool support for the \gls{pprdsl} comprises a set of snippets and code completion functions implemented in SublimeText.\footnote{SublimeText: \url{https://www.sublimetext.com/}}
The snippets provide stubs of \gls{ppr} aspects with their attributes (cf. \autoref{lst:fork} lines 1, 14, 31, and 37) that engineers can easily fill in.
The code completion allows the recommendation and autocompletion of keywords and the aspects' IDs and names, such as the resource \textit{Linefeeds} in \autoref{lst:fork} line 19, so engineers can quickly find existing aspects.
However, the text editor does not highlight incorrectly written keywords or whether engineers deleted aspects that are used in other aspect definitions.
To this end, the development of an improved editor that supports features, such as code highlighting and missing aspects, using Eclipse XText\footnote{Eclipse XText: \url{https://www.eclipse.org/Xtext/}} for better integration into the ecosystem, is ongoing.

\paragraph{TRAVART environment.}
\label{sec:travart}
\travart{}~\citep{FeichtingerTraVarT2021} is a plugin-based variability model transformation environment (cf. top right of~\autoref{fig:ipse_architecture}).\footnote{TraVarT: \url{https://github.com/SECPS/TraVarT}}
The \travart{} core plugin is implemented in Java and uses the \gls{uvl}~\citep{SundermannSPLC2021UVL} as the pivot model, building on the current parser implementation.\footnote{UVL Parser: \url{https://github.com/Universal-Variability-Language/uvl-parser}}
For each supported variability model, a plugin needs to be implemented.

Such a plugin must provide functions for reading and writing the supported variability model. 
Furthermore, one has to specify \code{Transformation Operations}, which transform the supported variability model into the pivot model and vice versa.
These operations are usually built upon a mapping table between the supported variability model and the pivot model and then implemented in \code{Transformation Algorithms}~\citep{FeichtingerICSREvolution, FeichtingerMODEVAR2020, FeichtingerSEAA2020}.
\autoref{tab:oneway-mappingtable} shows a mapping between the \gls{pprdsl} and \gls{fm} and \gls{dm} aspects.
Optionally, a plugin can implement a configuration \code{Model Sampler} to enable further testing of the resulting models. 

\begin{table}[ht!]
\centering
\caption{Mapping table of \gls{pprdsl} onto Variability Model Elements}\label{tab:oneway-mappingtable}
\footnotesize
\begin{tabular}{@{}lll@{}}
\toprule
\multicolumn{2}{l}{PPR-DSL}                                                       & \gls{fm} and \gls{dm} elements                                \\ \midrule
\multicolumn{1}{l}{\multirow{3}{*}{\rotatebox[origin=c]{90}{Unit}}} &
  \multicolumn{1}{l}{Product} &
  \multicolumn{1}{l}{Feature with attribute, Decision} \\ \multicolumn{1}{l}{} &
  \multicolumn{1}{l}{Process} &
  \multicolumn{1}{l}{Decision} \\ \multicolumn{1}{l}{}                            & \multicolumn{1}{l}{Resource} & \multicolumn{1}{l}{Feature with attribute}      \\ \midrule
\multicolumn{1}{l}{\multirow{7}{*}{\rotatebox[origin=c]{90}{Properties}}} & \multicolumn{1}{l}{Name}     & \multicolumn{1}{l}{-}                           \\ \multicolumn{1}{l}{}                            & \multicolumn{1}{l}{Abstract} & \multicolumn{1}{l}{Abstract feature}            \\ \multicolumn{1}{l}{} &
  \multicolumn{1}{l}{\multirow{2}{*}{implements}} &
  \multicolumn{1}{l}{\begin{tabular}[c]{@{}l@{}}Feature tree if only one other \\ unit and feature attribute\end{tabular}} \\ \multicolumn{1}{l}{}                            & \multicolumn{1}{l}{}         & \multicolumn{1}{l}{otherwise feature attribute} \\ \multicolumn{1}{l}{}                            & \multicolumn{1}{l}{children} & \multicolumn{1}{l}{Feature tree}                \\ \multicolumn{1}{l}{}                            & \multicolumn{1}{l}{requires} & \multicolumn{1}{l}{Implies constraint}          \\ \multicolumn{1}{l}{}                            & \multicolumn{1}{l}{excludes} & \multicolumn{1}{l}{Excludes constraints}        \\ \midrule
\multicolumn{1}{l}{\multirow{4}{*}{\rotatebox[origin=c]{90}{Constraints}}} &
  \multicolumn{1}{l}{Not} &
  \multicolumn{1}{l}{Not constraint} \\ \multicolumn{1}{l}{}                            & \multicolumn{1}{l}{And}      & \multicolumn{1}{l}{And constraint}              \\ \multicolumn{1}{l}{}                            & \multicolumn{1}{l}{Or}       & \multicolumn{1}{l}{Or constraint}               \\ \multicolumn{1}{l}{}                            & Implies                       & Implies constraint                               \\ \bottomrule
\end{tabular}\end{table} 
Available plugins for \travart{}, including FeatureIDE \glspl{fm}~\citep{FeatureIDE} and DOPLER \glspl{dm}~\citep{DOPLER}, implement transformation operations~\citep{FeichtingerTraVarT2021} that map their concepts to the \gls{uvl}~\citep{SundermannSPLC2021UVL} and vice versa~\citep{FeichtingerSEAA2020}.
For instance, a decision in the DOPLER \gls{dm} is mapped to a feature in the \gls{uvl}~\citep{SundermannSPLC2021UVL}.
Also, a rule in the DOPLER \gls{dm} is mapped to either a feature property (mandatory), the \gls{fm} tree, or a constraint~\citep{FeichtingerSEAA2020}.
In the opposite direction, the hierarchy of the \gls{fm} tree is captured via the visibility conditions of the DOPLER \gls{dm}.
To support the needs of the \gls{eipse} tool, we extended the \code{\gls{dm} Plugin} by a new writer.
This writer creates a file conforming to the \code{\gls{dm} editor}'s syntax, a propositional logic syntax for constraints and visibility conditions (cf. \autoref{sec:dm-editor}).

Moreover, we iteratively extended the existing \code{\gls{pprdsl} Plugin}~\citep{FeichtingerICSREvolution} and transformation operations (cf.~\autoref{tab:oneway-mappingtable}) to transform \gls{pprdsl} processes into a \processDM{} and \gls{pprdsl} resources into a \resourceFM{}. 
Specifically for the \processDM{}, the plugin creates a single Boolean decision for each process in the \gls{pprdsl} with its input products and required preceding processes as visibility conditions.
For instance, the plugin creates a decision \textit{InsertPipe2} in the \processDM{} in \autoref{tab:dm_fork1} from the \textit{InsertPipe2} process in \autoref{lst:fork}.
Furthermore, it creates a visibility condition that fires, if for decision \emph{Pipe}, the value \emph{Pipe2} is selected.
Subsequently, it creates a constraint that selects the abstract decision \emph{InsertPipe}.
Abstract processes are transformed to Boolean decisions with a visibility condition \code{false}, to be implicitly selected by those constraint rules.
Further, the transformations create a feature for each \gls{pprdsl} resource and derive its properties (e.g., whether the feature is abstract or mandatory) from the respective \gls{pprdsl} properties.
For example, for the \emph{WeldingRobot} (cf.~\autoref{lst:fork}), an abstract feature is created in the \resourceFM{}, defining a group of welding robots.
For the \emph{KUKA\_KR\_Agilus}, a concrete feature is generated in the \emph{WeldingRobot} group.
Each group of \gls{pprdsl} resources is converted into an \textit{OR}-group because at least one of the grouped resources will have to be selected if these resources are necessary to produce the configured product. 
Finally, the plugin derives a list of \glspl{cdc} by connecting the features and decisions from the resulting variability artifacts, i.e., \productFM{}, \processDM{}, and \resourceFM{}. 
For instance, the necessary \gls{pprdsl} resources for a given process result in a \gls{cdc} between the respective process decision in the \processDM{} and the respective resource features in the \resourceFM{} (cf. CDC2 in \autoref{lst:cdc-fork}). 

\paragraph{Eclipse Ecosystem}
The Eclipse Ecosystem\footnote{Eclipse Foundation: \url{https://www.eclipse.org}} is a cross-platform, open-source \gls{ide} for (software) engineering in different languages and domains.
The \gls{ide} provides a plugin system that allows the dynamic exchange of components.
Thus, it allows combining different independent software components into a suitable toolchain, in our case, for variability modeling.

\paragraph{FeatureIDE environment.}
The FeatureIDE~\citep{FeatureIDE}\footnote{FeatureIDE: \url{https://github.com/FeatureIDE/FeatureIDE}} is the current de-facto-standard open-source plugin for feature-oriented software development.
It supports several \gls{fm} types, among others, graphical \glspl{fm} modeling and textual variability modeling using UVL~\citep{SundermannSPLC2021UVL}. 
Furthermore, the FeatureIDE allows for configuring \glspl{fm} and validating them through \code{sat4j}.

\paragraph{Eclipse 4diac\texttrademark}
Eclipse 4diac\texttrademark\footnote{Eclipse 4diac\texttrademark: \url{https://www.eclipse.org/4diac}}~\citep{4diacZoitl2010} is an open-source Eclipse-based tool for developing IEC~61499-based~\citep{61499} control software for \glspl{cpps}.

\paragraph{\gls{dm} Editor environment.}
\label{sec:dm-editor}
The \gls{eipse} \gls{dm} Editor (bottom left of \autoref{fig:ipse_architecture}) is inspired by the DOPLER \gls{dm} editor~\citep{DOPLER}. 
The latter was developed for an industry partner and is a closed-source tool.
Instead, our \gls{eipse} \gls{dm} Editor (cf. \autoref{fig:ipse_architecture}) is an open-source Eclipse plugin, compatible with the latest version of FeatureIDE.
Our \gls{dm} Editor offers the following key features:
\begin{enumerate*}[label=(\roman*)]
    \item the import of DOPLER \glspl{dm}, used here to import the \processDM{},
    \item the creation of \code{DRule} files, where each file represents a single decision,
    \item the generation of DOPLER \glspl{dm} from those \code{DRule} files,
\item the generation of \code{DConfig} configuration files for DOPLER \glspl{dm}, and 
    \item the configuration of DOPLER \glspl{dm} via those \code{DConfig} configuration files, used here to explore and configure the \processDM{}.
\end{enumerate*}

\code{DRule} files exhibit a similar structure as the DOPLER \gls{dm} in~\autoref{sec:cpps_vm}, supporting a propositional logic syntax for constraints and visibility conditions.
Currently, we limited the syntax to types that can be translated into common SAT solvers, in this context \code{sat4j}, to enable the DOPLER \gls{dm}'s configuration validation and integration into the toolchain.
On user-triggered generation, the \code{DM Generator} component maps the \code{DRule} files onto an internal \code{DM Model}.
From this \code{DM Model}, the \code{DM Writer} writes a \code{DM File} with the specific syntax.
In our case, this model and the corresponding files contain the decisions for the production process steps that need to be selected to create a suitable process step sequence.
However, this model could also be used for other types of decisions.

The \code{DConfig Generator} component generates a configuration for a \gls{dm} in a \code{DConfig File}. 
It uses the \code{DM Reader} that reads the \gls{dm} model file created either by the \code{DM Generator} or by \travart{}.
The \code{DM Config Editor} reads the generated \code{DConfig File} and presents configuration options in a configuration view.
In this view, users can configure the \gls{dm} model, which is validated in the background by the \code{Config Evaluator}.
The selected sequence of decisions is stored in the \code{DConfig File} for further processing, for instance, to export or visualize the production process sequence.

\paragraph{\gls{v4rdiac}.}

\gls{v4rdiac}~\citep{Fadhlillah22} is a, currently closed source, multidisciplinary variability management approach for \gls{cpps} variability realized as an Eclipse plugin (cf. bottom right of \autoref{fig:ipse_architecture}).
The \gls{eipse} tool uses several \gls{v4rdiac} components for generating customer-specific control software based on the selected products, production processes, and production resources.
Specifically, the \gls{eipse} tool uses the components \code{IEC~61499 Delta Modeling}, \code{Multi-Level Configuration}, \code{CDC Validator}, \code{Product Config Tool}, and the \code{4diac Generator}.

Eclipse 4diac\texttrademark~is used to develop the base implementation of the IEC~61499 control software.
Beyond that, the \code{IEC~61499 Delta Modeling} component is used to implement the Deltas for the control code artifacts (cf. \autoref{lst:delta-model-fork}).
Afterward, a mapping between the base implementation artifacts and the Deltas needs to be established. 
In our case, \gls{cpps} engineers define an additional attribute in the \gls{pprdsl}, which specifies the location of corresponding Deltas, e.g., as URI (cf. \autoref{lst:ppr-fork-delta} Line 8).
The \code{Multi-Level Configuration} is used to define a step-wise configuration in which different variability models can be configured separately.
In our case, the \productFM{} and \processDM{} with their configurations and the \resourceFM{} are loaded into the component and ordered into this sequence.
During the configuration of the \resourceFM{}, which allows the configuration of multiple production resources for a production process, the \code{CDC Validator} ensures that the selected configurations are valid according to the \glspl{cdc}.
The \code{Product Config Tool} component displays the configuration user interface for the \resourceFM{}.
After the \resourceFM{} configuration in the \code{Product Config Tool}, the \gls{eipse} process is finished with a valid configuration.
The \gls{eipse} tool requires an artifact generator that evaluates the mapping between \gls{pprdsl} attributes and \gls{cpps} artifacts to remove, combine, and build \gls{cpps} artifacts given a set of selected products, production processes, and production resources.
The \code{IEC~61499 Delta Modeling} component is linked to Eclipse 4diac\texttrademark~and the \code{4diac Generator} to generate the IEC~61499 control code.
The \code{4diac Generator} then generates the control software for the production resources in IEC~61499.
For a detailed description of Delta modeling for control software, we refer to~\citep{Fadhlillah2023SPLC,FadhlillahETFA2023}. 

The \gls{eipse} tool aims to support a straightforward integration of different \gls{cpps} artifact generators and formats, such as IEC~61499~\citep{61499} or AutomationML~\citep{Drath2021} code.

\section{Evaluation}
\label{sec:evaluation}

This section describes the evaluation of the \gls{eipse} approach and prototype.
\autoref{sec:evaluation_questions} presents the evaluation questions to be answered. 
The subsequent sections (cf.~\autoref{subsec:eval_generalSetup}-\ref{subsec:eval_controlSWgeneration}) first present the general setup of the evaluation activities, followed by describing the concrete setup and the results.

\subsection{Evaluation Questions}
\label{sec:evaluation_questions}

To evaluate the \gls{eipse} approach and prototype, we address the research questions (cf.~\autoref{sec:introduction}) and stated requirements (cf.~\autoref{sec:requirements}) with the following evaluation questions.

\begin{enumerate}[label={\upshape\bfseries EQ\arabic*},noitemsep]
\item{\textit{Is it feasible to apply the \gls{eipse} approach}}  
    \begin{enumerate}
        \item in different real-world case studies?
        \item by engineers from heterogeneous backgrounds?
\end{enumerate}
\end{enumerate}

We postulate that the \gls{eipse} approach facilitates the externalization of knowledge, reduces the effort of \gls{cpps} modeling and configuration through automation, including the exploration of feasible production process sequences, and benefits the reproducibility of the configuration process.
To examine EQ1, we applied our approach to different real-world case studies from industry~\citep{Meixner2021}.
Beyond that, we shadowed~\citep{Shull2007Guide} subjects with different backgrounds to investigate the utility of our approach, i.e., whether it is feasible enough to be used in \gls{cpps} engineering.
Therefore, we measured their efforts as ``time spent'' when applying the \gls{eipse} approach step-wise and configuring a \gls{cpps} as a notable factor of \gls{cpps} engineering optimization~\citep{vdi_3695} and for future investigation. 
Furthermore, we collected feedback from the subjects in post-task discussions.
For both investigations, we take the engineers' hard-to-reproduce and manual approach as a baseline for optimization, where engineers employ mostly implicit domain knowledge to design and configure \glspl{cpps}.

\begin{enumerate}[label={\upshape\bfseries EQ\arabic*}]
\setcounter{enumi}{1}
\item \emph{By how much can using \gls{eipse} reduce the number of decisions needed to configure a production process sequence for a \gls{cpps}?}
\end{enumerate}

The \gls{eipse} approach is grounded on the hypothesis that the configuration of a single product and the formulation of pre- and postconditions for process steps can significantly reduce the configuration space for production process sequences. 
By reducing the configuration space, the users are guided to only configuring necessary process steps, which is essential when configuring commercial and/or industrial software~\citep{Hubaux2012}. 
Therefore, the \gls{eipse} approach uses DOPLER \glspl{dm} due to the concept of visibility conditions that allows for a subsequent unfolding of configuration options in contrast to \glspl{fm}.
We address EQ2 by comparing the entire configuration space of a \processDM{} with the configuration space for the reduced \processDM{} resulting from using \gls{eipse} by utilizing combinatorics.
In particular, we focus on a subsequently created production process sequence for a particular product configuration of the \textit{shift fork} case study~\citep{Meixner2021}.

\begin{enumerate}[label={\upshape\bfseries EQ\arabic*}]
\setcounter{enumi}{2}
\item \emph{Can the \gls{eipse} tool chain generate consistent \gls{cpps} control software code?} 
\end{enumerate}
   
The logical consequence of exploring the process sequences and configuring the production resources for a particular product configuration is generating artifacts that represent various \gls{cpps} aspects. 
In our work, we are currently focused on one type of \gls{cpps} artifact, i.e., IEC~61499~\citep{61499} control software. 
We address this question by preparing multiple valid combinations of selected products, production processes, and production resources. 
We use each valid combination to generate IEC~61499 control software variants using our toolchain. 
Then, we evaluate the consistency of the control software code by verifying whether the elements related to the selected product, production process, and production resources exist in the generated control software. 

\subsection{General Evaluation Setup}
\label{subsec:eval_generalSetup}

For the evaluation, we installed the \gls{eipse} toolchain on one of the author's notebooks.
This is due to the company policies of some of our evaluation subjects from industry, they are not allowed to install any additional software, including our \gls{eipse} toolchain. 
Our setup included 
\begin{enumerate*}[label=(\roman*)]
    \item SublimeText as a text editor to manipulate the \gls{pprdsl} including a ``cheat sheet'' for its syntax (\gls{eipse}, Step~1), 
    \item \travart{} with the \gls{pprdsl} as the library to transform the \gls{pprdsl} models to the required variability models (Step~2),
    \item Eclipse with FeatureIDE (Step~3), the \gls{dm} Editor (Steps~4 \&~5), and \gls{v4rdiac} to manipulate and configure the variability models (cf.~Step~6 \&~7), and to generate the \gls{cpps} artifacts (Step~8) showing them  with 4diac as plugins.
\end{enumerate*}
We used this setup in the sessions to investigate the evaluation question \textbf{EQ1+EQ3}.

Additionally, our evaluation subjects are distributed in diverse locations. 
To solve this limitation, we utilize Zoom's \emph{Screen Sharing} and \emph{Remote Control} features to use the \gls{eipse} toolchain remotely.
In this way, we can use the same machine specification for every subject using the \gls{eipse} toolchain in the evaluation. 
Furthermore, we can record those Zoom sessions for later analysis.

\subsection{Application of eIPSE to Case Studies}
\label{sec:usecase_evaluation}

This evaluation activity addresses \textbf{EQ1 a}  by investigating the applicability of the \gls{eipse} approach to real-world case studies from industry and by measuring the resulting design and configuration space. 

\paragraph{Setup}
In prior work~\citep{Meixner2021}, we introduced four real-world \gls{cpps} case studies: \textit{truck}, \textit{shift fork}, \textit{rocker switch}, and \textit{water filter}, modeled their products in the \gls{pprdsl} and implemented \travart{} transformation operations.
In \citep{MeixnerVamos2022ProcessExploration}, we extended the \textit{shift fork} model with processes and added the corresponding \travart{} transformation.

\begin{figure*}[h]
    \centering
    \includegraphics[width=.75\textwidth]{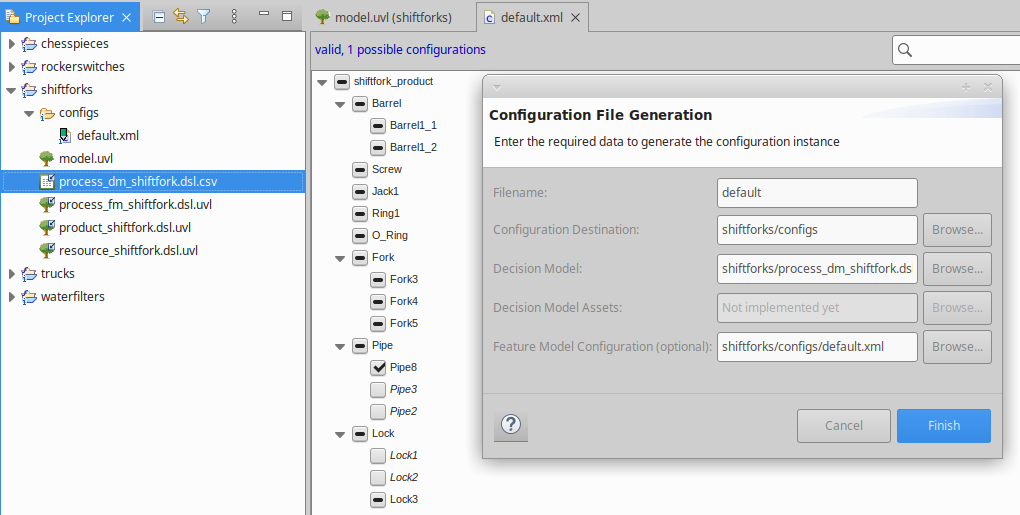}
    \caption{
After configuring the \productFM{} in Step~3 of the \gls{eipse} process (background) for the \textit{shift fork} case study~\citep{Meixner2021}, the reduced \processDM{} configuration is created in Step~4 of the \gls{eipse} process using the \gls{eipse} prototype's wizard (front).}
    \label{fig:first}
\end{figure*}

For the evaluation activity, the author most familiar with the \gls{pprdsl} and the particular \glspl{cpps} modeled the remaining atomic process steps and resources (\gls{eipse}, Step~1) for each case study. 
The author utilized engineering artifacts of the respective \glspl{cpps}.
In the \emph{shift fork} case study, the author most familiar with \travart{} iteratively improved the existing and newly implemented transformation operations, sustaining the research methodology.
These adaptations primarily concerned the hierarchy and grouping of \gls{ppr} aspects but did not change the nature of the \gls{cpps} designs.

\begin{figure*}
    \centering
    \includegraphics[width=.75\textwidth]{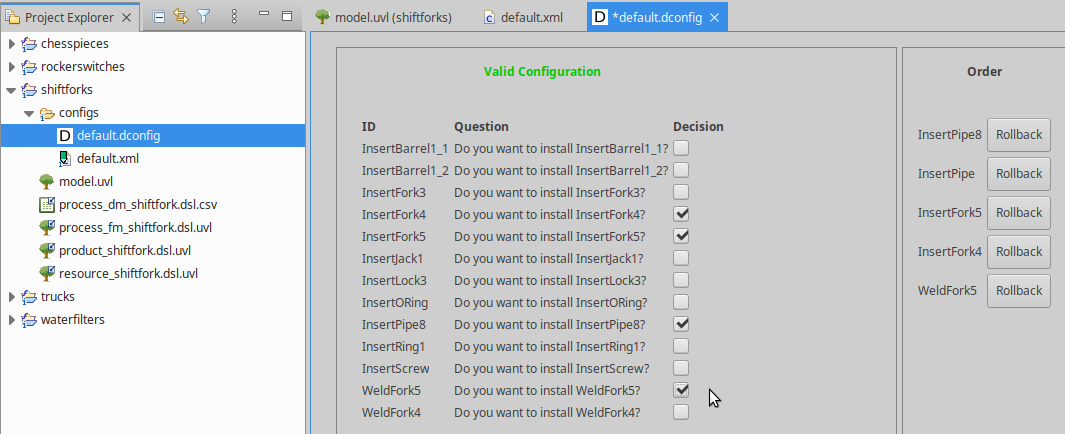}
    \caption{
During the configuration of the production steps in the \processDM{} in Step~5 of the \gls{eipse} process, a suitable production process sequence for the \textit{shift fork} case study~\citep{Meixner2021} is defined. The \gls{eipse} prototype provides a \emph{rollback} option (right-hand side) for systematic process sequence exploration~\citep{MeixnerVamos2022ProcessExploration}.}
    \label{fig:second}
\end{figure*}

\begin{figure}
    \centering
    \includegraphics[width=0.99\columnwidth]{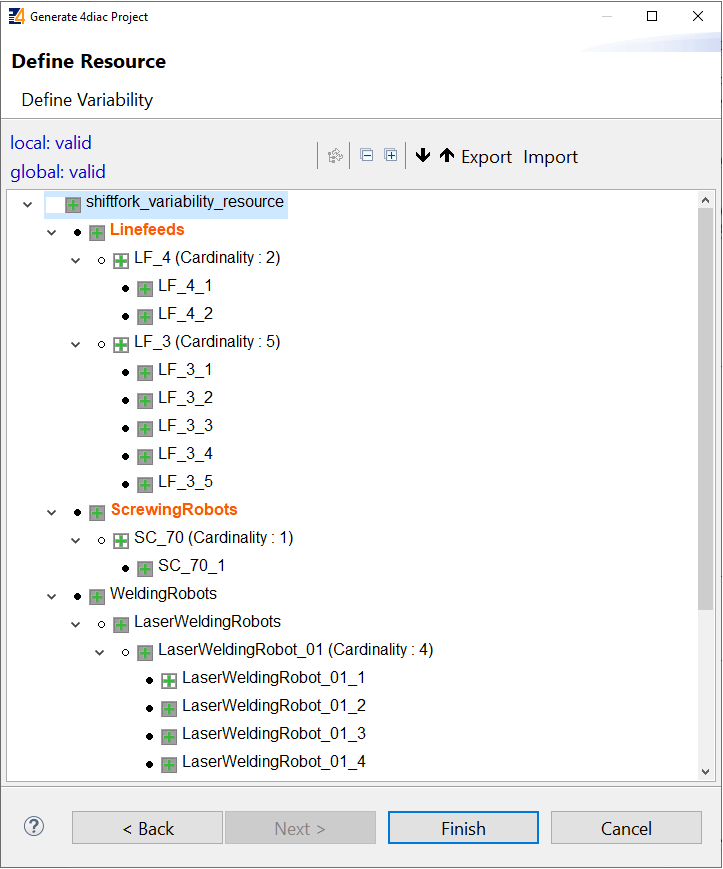}
    \caption{
The configuration of the \resourceFM{} in Step~7 of the \gls{eipse} process for the production resources necessary to execute the configured production process steps of the \processDM{} in Step~5 of the \textit{shift fork} case study~\citep{Meixner2021}.}
    \label{fig:third}
\end{figure}

The two authors alternately applied the remaining steps of the \gls{eipse} approach (Steps~2 to~7) to each of the four case studies. 
\autoref{fig:first}~to~\autoref{fig:third} show 
\begin{enumerate*}[label=(\roman*)]
    \item the configuration view of the \productFM{} in Step~3 (background) and the wizard to create the \processDM{} configuration in Step~4 (front),
    \item the \gls{dm} Editor and a step in configuring the \processDM{} in Step~5, and
    \item the \resourceFM{} configuration in Step~7 for the \emph{shift fork} case study using our \gls{eipse} prototype.
\end{enumerate*}

According to the feedback loops shown in~\autoref{fig:process}, the two authors incorporated changes during the evaluation activity iteratively to correct errors in the \gls{pprdsl} model, such as missing exclusion or grouping constraints and to omit errors in the generated variability models.
To this end, we also used the \gls{eipse} approach to \emph{validate} the \gls{ppr} models of the case studies.
A third author took notes and acted as a referee to minimize bias during the evaluation activity, which could have been introduced by the familiarity of the other two authors with the case studies.
Additionally, during the transformations, we ran automated statistics for each case study to obtain various metrics for the \gls{pprdsl} file and resulting variability models. 
We built on previously defined metrics~\citep{FeichtingerICSREvolution}, such as the size of the models, their constraints, and configuration space.
We summarize the resulting statistics in \autoref{tab:evaluationData} and explain them below.
We used the notes by the third author and the collected statistics to further improve the implementation of our prototype (cf. Section~\ref{sec:methodology}). 

\begin{table*}[th!]
\centering
\begin{scriptsize}
\caption{Statistical data on the \gls{pprdsl} artifact and the generated \productFM, \processDM, \resourceFM, and \glspl{cdc}. }
\label{tab:evaluationData}
\resizebox{\textwidth}{!}{\begin{tabular}{l||ccccc||cccccc||ccc||ccccc||c}
\toprule
\textbf{Case study} &
  \multicolumn{5}{c||}{\textbf{PPR-DSL artifact}} &
  \multicolumn{6}{c||}{\textbf{\productFM}} &
  \multicolumn{3}{c||}{\textbf{\processDM}} &
  \multicolumn{5}{c||}{\textbf{\resourceFM}} &
  \textbf{\glspl{cdc}} \\ \midrule
Metrics &
  \rotatebox[origin=c]{90}{$\#\text{Products}$} &
  \rotatebox[origin=c]{90}{$\#\text{Product}_{\text{comp}}$} &
  \rotatebox[origin=c]{90}{$\#\text{Processes}$} &
  \rotatebox[origin=c]{90}{$\#\text{Resources}$} &
  \rotatebox[origin=c]{90}{$\#\text{Constraints}$} & \rotatebox[origin=c]{90}{$\#\text{Features}$} &
  \rotatebox[origin=c]{90}{$\#\text{Constraints}_{\text{xor}}$} &
  \rotatebox[origin=c]{90}{$\#\text{Constraints}_{\text{or}}$} &
  \rotatebox[origin=c]{90}{$\#\text{Constraints}_{\text{tree}}$} &
  \rotatebox[origin=c]{90}{$\text{Tree height}$} &
  \rotatebox[origin=c]{90}{$\#\text{Configs}$} &
  \rotatebox[origin=c]{90}{$\#\text{Decisions}$} &
  \rotatebox[origin=c]{90}{$\#\text{Constraints}$} &
  \rotatebox[origin=c]{90}{$\#\text{Constraints}_{\text{vis}}$} &
\rotatebox[origin=c]{90}{$\#\text{Features}$} &
  \rotatebox[origin=c]{90}{$\#\text{Constraints}_{\text{xor}}$} &
  \rotatebox[origin=c]{90}{$\#\text{Constraints}_{\text{or}}$} &f
  \rotatebox[origin=c]{90}{$\#\text{Constraints}_{\text{tree}}$} &
  \rotatebox[origin=c]{90}{$\text{Tree Height}$} &
\rotatebox[origin=c]{90}{$\#\text{Rules}$} \\ \midrule
\rowcolor{myLighterGray}
Truck         & 12 & 7  & 13 & 3  & 30  & 8  & 1 & 0 & 0  & 2 & 4 & 20 & 32  & 20  & 5  & 0 & 1 & 0 & 2 & 15 \\
\rowcolor{myDarkerGray}
Shift fork    & 24 & 19 & 36 & 16 & 52  & 20 & 2 & 0 & 10  & 2 & 4 & 55 & 27  & 53  & 17 & 0 & 8 & 0 & 3 & 35 \\
\rowcolor{myLighterGray}
Rocker switch & 46 & 33 & 59 & 13 & 63  & 34 & 0 & 0 & 20 & 2 & 44 & 92 & 26  & 92  & 14 & 0 & 5 & 0 & 3 & 63 \\
\rowcolor{myDarkerGray}
Water filter   & 46 & 37 & 36 & 13 & 108 & 38 & 8 & 0 & 54 & 3 & 10 & 73 & 127 & 73 & 14 & 0 & 5 & 0 & 3 & 43 \\ \bottomrule
\end{tabular}}
\end{scriptsize}
\end{table*} 
\paragraph{Results}
Our evaluation showed that we were able, with the feedback loops, to apply the \gls{eipse} approach in the four selected case studies.
While we had to adapt the \gls{pprdsl} models iteratively throughout the process, we were able to configure reasonable production process sequences and production resources for a particular product configuration.
This indicates that applying \gls{eipse} to industrial \gls{cpps} product lines is feasible.

As a supplementary result, we gathered metrics of the variability models resulting from the transformation, listed in \autoref{tab:evaluationData}.
As a representative of the case studies, we explain the \emph{shift fork} case study in detail.
The first category summarizes the metrics of the \emph{\gls{pprdsl} artifact}. 
For the \emph{shift fork} case study, there are 24 product definitions with 19 product component definitions included, 36 atomic process step definitions, 16 resource definitions, and 52 constraints.
The next set of metrics concerns the generated \emph{\productFM{}}.
We measured 20 features resulting from the set of 19 product components plus the root feature.
The constraints in the \gls{pprdsl} artifact were transformed into 2 \code{xor} groups and 10 cross-tree constraints. 
The reduced complexity compared to the \gls{pprdsl} artifact results from the number of constraints necessary to describe an alternative group in the \gls{pprdsl}.
The 4 possible configurations for the products in the \productFM{} represent exactly the 4 shift fork types produced in the real-world \gls{cpps}.
The same holds for the 4 trucks in the \emph{truck} case study.
However, the \productFM{} is underconstrained, resulting in more possible configurations than the modeled final product types in the \gls{pprdsl} of the \emph{rocker switch} (44 configurations vs. 12 product types) and the \emph{water filter} (10 configurations vs. 8 product types) case studies.
In the generated \emph{\processDM{}}, we measured 55 decisions, consisting of 19 product component decisions, of which 4 were abstract, 15 were concrete, and 36 processes from the \gls{pprdsl} artifact.
Those 19 decisions are used to pre-configure the \processDM{} for the process exploration (cf. Step~4 in \autoref{sec:approach}).
The table shows the large number of decisions (55), rules/constraints (27), and visibility constraints (53) compared to the constraints in the \gls{ppr} model (52).
The generated \emph{\resourceFM{}} contains 17 features derived from the 16 defined resources in the \gls{pprdsl} model plus the root feature.
The features are grouped in \code{or} groups based on the constraints defined in the \gls{pprdsl}.
The last category shows the generated \emph{\glspl{cdc}}, which are 35 derived \gls{cdc} rules for the \emph{shift fork} case study.
The table shows that the \gls{pprdsl} model requires fewer constraints (52) than the variability models combined; 2+10 for the \productFM{}, 27+53 for the \processDM{}, 8 for the \resourceFM{}, and 35 \glspl{cdc}.

\subsection{Application of eIPSE by Different Engineers}
\label{sec:user_evaluation} 
This evaluation activity addresses \textbf{EQ1 b}.
In a user study, we investigate how much effort engineers from heterogeneous backgrounds inexperienced in the \gls{eipse} approach spend for each step of applying \gls{eipse} by shadowing them~\citep{Shull2007Guide}.
We report on their effort, experience, and perceived usefulness.
For a detailed description of the user study, we refer to \autoref{sec:guideline}.

\paragraph{Setup}
For this evaluation activity, we introduce the new \emph{chess piece} case study originating from the TU Wien pilot factory.\footnote{Pilotfabrik TU Wien: \url{https://www.pilotfabrik.at}}
The product line consists of six chess piece types with a body and an aluminum base with either one or two carved reamings.
The body and the base are joined via threaded rods of two lengths. 

Five subjects applied the \gls{eipse} approach to this case study. 
An overview of the subjects and their domain can be found in \autoref{tab:evaluation_effort}.
The first and the second columns of the table state the subject and whether the subjects are engineers (\emph{E}) or researchers (\emph{R}) and their domain.
Subject S1 is an engineer at an industry partner in the field of high-speed \gls{cpps} automation with a background in mechanical engineering (\emph{ME)}.
Two subjects (S2 \& S3) are engineers from an industry partner in the automotive domain with a background in mechanical engineering.
Subject S4 is a senior systems engineer (\emph{SE}) in the \gls{cpps} domain from a research collaboration.
Subject S5 is a computer science researcher with a mechanical engineering background.
All subjects know the principles of the \gls{ppr} concept.

All subjects conducted the evaluation activities in individual Zoom sessions after the evaluation activity for \textbf{EQ1 a} (cf. \autoref{sec:usecase_evaluation}) under the supervision of at least two authors.\footnote{The time frame was limited to roughly two hours for industrial subjects.} 
If the subjects had questions concerning the \gls{pprdsl}, the \gls{eipse} approach, or the toolchain, these authors provided tips and support.
These authors gave hints on how to model aspects of the product line once a subject got stuck and asked for help. 
We also measured the time the subjects took to execute each step of the evaluation activity and \gls{eipse} process.
We aimed to understand better where subjects need more automation support and where we can further improve the \gls{eipse} approach.
Furthermore, we wanted to compare the use of the \gls{eipse} process with a control group that performs the same tasks without the \gls{eipse} process.

The subjects had to model the chess piece types and their parts, reasonable atomic production steps, and production resources in a \gls{pprdsl} model (\gls{eipse}, Step~1). 
In this evaluation, it is not necessary for each subject to model the full version of the chess piece product line. 
We mainly focused on ensuring that the subjects grasp the overall idea of using our \gls{pprdsl} and gain feedback from them. 
Furthermore, we wanted to get an indication of the relation between the tasks for engineering and configuration.
The subjects then use their resulting \gls{pprdsl} model to generate the \productFM{}, the \processDM{}, and the \resourceFM{} using the \travart{} \gls{cli} (Step~2).
Then, the subjects configured the \productFM{} (Step~3) and loaded the configuration into the \gls{dm} Editor to create a reduced \gls{dm} configuration file (Step~4) (cf.~Fig.~\ref{fig:first}).
Afterward, the subjects explored the process sequence and configured the \processDM{} (Step~5).
Lastly, they configured the \resourceFM{} in \gls{v4rdiac}, which reduced the model based on the configured \processDM{} (Step~6 \&~7).
If the subjects thought it was necessary to improve their variability models, they used the feedback loops (cf. Fig.~\ref{fig:ipse_architecture}) to improve their product line.

After the subjects had completed this evaluation task, we asked them for their experience with and feedback on the \gls{eipse} approach as well as its perceived usefulness.
We analyzed the notes taken by comparing them and finding evidence for the usefulness and benefits of the approach, its steps, and potential limitations and improvements.

\paragraph{Results}
\autoref{tab:evaluation_effort} presents the times spent on this evaluation activity.
The third column states the time to introduce the \gls{eipse} approach and the \emph{chess piece} product line.
The fourth and fifth column concern the domain engineering phase, showing the time spent to define the chess piece product line as a \gls{pprdsl} model.
The fifth column shows the time the subjects used to update the product line after an initial configuration.
This update corresponds to the feedback loops of the \gls{eipse} approach (cf.~\autoref{fig:process} dotted arrows).
The last three columns present the times spent during the configuration phase (i.e., application engineering) indicating the efforts for configuring the \productFM{} in the FeatureIDE~\citep{FeatureIDE}, the \processDM{} in the \gls{eipse} \gls{dm} Editor, and the \resourceFM{} in \gls{v4rdiac}~\citep{Fadhlillah22}.
The last row summarizes the rounded average time of each step in minutes.

\begin{table}[]
\centering
\caption{``Time spent'' by engineers with different backgrounds for \gls{eipse} to the \emph{chess piece} case study. E\dots Engineer, R\dots Researcher; ME\dots Mechanical engineering, SE\dots Systems engineering, CS\dots Computer science }
\label{tab:evaluation_effort}
\scriptsize
\begin{tabular}{c|c||l|ll|lll}
\toprule
 \multicolumn{2}{c||}{\multirow{2}{*}{Subject}} & \multicolumn{6}{c}{Activity} \\ \cmidrule{3-8}
            \multicolumn{2}{c||}{} & Intro & \multicolumn{2}{c|}{Engineering} & \multicolumn{3}{c}{Configuration}  \\ \midrule
         \rotatebox[origin=c]{90}{Participant} & \rotatebox[origin=c]{90}{Profession}             & & \rotatebox[origin=c]{90}{Definition}   & \rotatebox[origin=c]{90}{Updates}  & \rotatebox[origin=c]{90}{Product FM} & \rotatebox[origin=c]{90}{Procss DM} & \rotatebox[origin=c]{90}{Resource FM} \\ \midrule
         \rowcolor{myLighterGray}
S1 & E,ME & 10m          & 60m              & 30m             & 3m         & 5m          & 1m          \\  \rowcolor{myDarkerGray}
S2 & E,ME & 13m          & 70m             & 15m             & 2m         & 7m          & 3m          \\ \rowcolor{myLighterGray}
S3 & E,ME & 9m          & 72m              & 3m             & 2m         & 5m          & 4m          \\ \rowcolor{myDarkerGray}
S4 & E,SE & 6m          & 62m              & 4m             & 1m         & 4m          & 1m          \\ \rowcolor{myLighterGray}
S5 & R,ME & 9m          & 63m              & 9m             & 6m         & 9m          & 3m          \\ \midrule
\multicolumn{2}{c||}{Summary (avg)} & 9m & 65m & 12m & 3m & 6m & 2m \\
\bottomrule
\end{tabular}\end{table}

The introduction to \gls{eipse} and the \emph{chess piece} case study took, on average, 9 minutes, [min.~6~mins, max.~13~mins].
Spending on average 65~minutes, [60~mins, 72~mins], the subjects spent most of the time on \textit{defining the product line}.
While most of the subjects did not model the entire product line of the six chess pieces with all the required production processes and resources, they could all grasp the concepts of the approach and continue with the configuration.
\emph{Updating the product line} according to the feedback loops took the subjects on average 12~minutes [3~mins,~30~mins] depending on how much they updated their models.
For the initial \gls{pprdsl} model transformation to the variability models and their configuration, we used the subjects' \gls{pprdsl} models.
However, as the subjects often did not model all products, production process steps, and production resources for timely reasons, after the first investigation of their generated variability models, we switched to a prepared \emph{chess piece} \gls{pprdsl} model to ensure a likewise feedback regarding the configuration with the \gls{eipse} approach.

The \emph{\gls{cpps} configuration} in this approach was fast for the \productFM{} and the \resourceFM{} (both have avg. 3~minutes).
The \processDM{} configuration took the evaluation subjects on average 6~minutes, [4~mins, 9~mins], slightly longer than the configuration of the other variability models.
One reason might be that users novel to the approach do not have enough experience in ordering process steps in a meaningful way.
Additionally, it seems that more domain knowledge is required to decide which process steps seem reasonable to be ordered in a particular way.
Thus, subjects used more time to experiment with different sequences before finalizing and deciding on a specific one.

In post-task discussions, we gathered feedback on the approach from each subject.
Overall, the \gls{eipse} approach was well received.
Impressions were that the approach 
\begin{itemize}[noitemsep]
    \item ``\emph{is very helpful and the toolchain works great}.'' (S2)
    \item ``\emph{makes sense from an engineer's perspective}.'' (S1/S4)
    \item ``\emph{makes the knowledge about the production sequence explicit}'' (S2/S3/S4)
    \item  ``\emph{is straightforward}.'' (S1/S4)
\item ``\emph{allows for a more economic and optimized design of the \gls{cpps}}.'' (S3)
\end{itemize}
Subject S3 stated that ``\emph{the process digitalization is a great idea that can improve reuse of existing configurations}'' and it is ``\emph{easy to understand and use}.''
Several subjects stated that it ``\emph{supports the reproduciblity of process selection}.''
S4 confirmed that the separation of concerns through ``\emph{modeling relations from different discipline perspectives}'' is important.
S2 meant that ``\emph{the often rigid integration structures of large companies might render the approach better suited for small and medium companies}.''

On the constructive side, the subjects also pointed out some issues and suggested several improvements.
Most subjects noted that the approach ``\emph{definitely requires better tool support to use it efficiently},'' such as ``\emph{better tool feedback}'' and ``\emph{a better overview of \gls{ppr} concepts}'' using, for example, ``\emph{low code approaches}.''

Concerning the \gls{pprdsl} for modeling the \emph{chess piece} product line, the subjects stated that  
\begin{itemize}[noitemsep]
    \item it ``\emph{provides the means for better reuse}'' of engineering artifacts (S4) and
    \item was ``\emph{straightforward and easy to use once the syntax is clear}.'' (S2) \item ``\emph{the \gls{pprdsl} is great because it is not as complex as, e.g., SysML}.'' (S2)
\end{itemize}

Nevertheless, three subjects noted that the \gls{pprdsl} ``\emph{has a steep learning curve}.''
Several subjects commented that the \gls{pprdsl} ``\emph{is sometimes redundant and partially confusing},'' for instance, because a ``product'' is the same as a material,\footnote{We note that the VDI~3682 standard uses the term \emph{product} for materials as well as complex composite products.} and that the ``\emph{difference between the requires and children relation is unclear}.''
Such limitations and the ``\emph{several times missing overview}'' can make the \gls{pprdsl} ``\emph{as-is error-prone for larger models}.''
This concerns, for instance, ``\emph{the definition of constraints in the model with a large number of products/processes}.'' While the ``\emph{cheat sheet was of great use},'' the \gls{pprdsl} ``\emph{should be simplified to be usable for engineers}'' and ``\emph{redundancies should be omitted}.''
A suggestion was also to ``\emph{provide more examples and best practices}'' and ``\emph{improve the documentation}'' for the \gls{pprdsl}.
Furthermore, the subjects desired the ``\emph{introduction of parallel processes},'' ``\emph{definition of transport in the \gls{cpps}},'' and ``\emph{use of libraries}.''
We aim to enhance the \gls{pprdsl} and its tool support to address these comments in the future.

Feedback on the transformation to the variability models was positive: ``\emph{extremely fast and happens in the background once syntax errors are fixed}.''
However, the transformation ``\emph{may cause iterative loops of updating the \gls{pprdsl}}.''
We argue that these feedback loops are intended and should be used by engineers to improve their \gls{ppr} models.
We also argue that more experienced users and product designers presumably define better \gls{ppr} models.

Feedback on the configuration was mainly positive:

\begin{itemize}[noitemsep]
    \item it ``\emph{provides an easy configuration}.''  (S1/S4) 
\item ``\emph{having the dependencies explicitly transferred into the configurable models helps to build feasible production sequences}'' (S5)
    \item ``\emph{the experimentation with different process sequences through simulation is a good idea}.'' (S3)
    \item the exploration makes sense particularly with ``\emph{digital twins and asset administration shells for simulation and provides additional value with the modeling of the process relations}.'' (S4)
\end{itemize}

All subjects also provided several remarks to improve our \gls{eipse} approach and toolchain. 
In particular, Subject S5 stated that ``\emph{cost/risk assessment would be nice to further enhance and evaluate the process sequences}.''
Several subjects also stated that ``\emph{the toolchain requires better integration}.'' 
For instance, one subject mentioned that ``\emph{executing the process using the \gls{eipse} toolchain requires a lot of preparatory steps, which could be reduced}.''
Asking for clarification, the subject explained that copy-pasting the necessary files in between the process steps should be enhanced.
We argue that the current state of our toolchain, not the \gls{eipse} process itself, caused this statement.
In future work, we aim to integrate all involved toolchain tools into a single tool environment, such as Eclipse.

\subsection{Reduction of the Process Configuration Space}
\label{sec:reduction_evaluation}

This evaluation activity addresses \textbf{EQ2} by investigating the reduction of the configuration space of a \processDM.

\paragraph{Method}
To measure the reduction of the configuration space, we utilize combinatorics.
Considering the sequence of process steps, the \processDM{} configuration follows a permutation, an arrangement of elements in a specific sequence. 
Additionally, only visible process steps (visibility condition is \code{true}) can be configured in the \processDM{} configuration.
The formula to calculate permutations, where $n$ denotes the total number of visible atomic process steps and $r$ denotes the number of steps occurring only in combination, is defined as:

\begin{equation}
    P(n, r) = \frac{n!}{(n - r)!}
\end{equation}
To this end, we compare the entire configuration space of a \processDM{}, where process steps can be combined arbitrarily, with the configuration space for the reduced \processDM{} resulting from a configured product using \gls{eipse} (cf. Step~3 and Step~4).
As a representative, we examine a subsequently created production process sequence in the \emph{shift fork} case study.

\paragraph{Results}
In this evaluation, we perform a simulation to configure the \emph{shift fork}'s \productFM{} (cf. \autoref{fig:fm_shift} and in \autoref{fig:first}) by selecting the features \emph{Pipe2} and \emph{Lock1} (Step~3).
This selection results in a valid configuration representing a single shift fork product.
Step 4 reduces the configuration options of the \processDM{} to this product where only the investigation of this product's process sequence is progressed.

Consequently, the visibility conditions and configuration values of \emph{Pipe2} and \emph{Lock1} are set \code{true} in the \processDM{} (cf. Tab.~\ref{tab:dm_fork1}).
The \processDM{} configuration is further reduced by setting the visibility conditions of the mandatory process steps to \code{true}. Additionally, any alternative group 
within the \processDM{} is reduced by setting the visibility conditions of only one of the available options to \code{true}, leaving the engineer to select only truly variable process steps. 
For instance, the \textit{InsertLock1} visibility condition is set to \code{true} and, thus, the \textit{InsertLock2} and \textit{InsertLock3} visibility conditions are set to \code{false}.

The overall \processDM{} configurations in the \emph{shift fork} case study comprise 21 process step decisions and the 3 alternative process step group decisions when \emph{considering} the \emph{Pipe2} and \emph{Lock1} product configuration AND the constraints between the decisions. 
In the case of the production process sequence exploration~\citep{MeixnerVamos2022ProcessExploration}, $r=n$ since we can combine all visible production process steps in the particular configuration step.
Furthermore, arbitrarily combining the atomic process steps visible to developers at any time results in $n!$ permutation possibilities for $n$ decisions.
Thus, the entire permutation space comprises $24!=6\cdot10^{23}$ possible process step sequences.

Transforming the pre- and postconditions of the production processes in the \gls{pprdsl} to visibility conditions in the \processDM{} enables the subsequent unfolding of the production process steps for creating a production process sequence.
The shift fork's product configuration allows 11 process steps, e.g., \emph{InsertPipe2} and \textit{InsertLock1}, with no visibility conditions that build the starting point of the \processDM{} configuration.
Furthermore, the shift fork's product configuration allows 4 process steps with visibility conditions related to the 11 previous steps, e.g., \textit{InstallLock1},.
Following the visibility conditions, the \gls{eipse} approach reduces the configuration of the \processDM{} to \emph{five} consecutive steps with 11, 4, 6, 2, and 1 remaining production process steps.
In each step, engineers had to decide in which sequence the currently possible production process steps have to be executed. 
Consequently, the \gls{eipse} approach reduced the set of possible process sequences to a minimum of $11!+4!+6!+2!+1!\approx39.9\cdot10^6$ possible sequences -- a reduction of about $10^{17}$ sequence options.

Even though many sequence configuration options remain, the reduction is significant and helps to reduce the cognitive level of deciding on a valid and feasible sequence.
However, the exploration of optimized production process sequences still demands additional knowledge and training from the engineers, which is a complementary activity to introducing the \gls{eipse} workflow.

\subsection{Generation of Control Software}
\label{subsec:eval_controlSWgeneration}
This evaluation activity addresses \textbf{EQ3} by investigating how to create parts of the control software for a particular \gls{cpps} configuration as an example for a \gls{cpps} artifact. 

\paragraph{Setup}
The author, most familiar with \gls{v4rdiac}, defined the necessary Delta files to generate IEC~61499 control code in 4diac as a \gls{cpps} artifact of the four case studies during Step~8.
In close cooperation with the authors of the first evaluation activity, the link between the processes and resources in the \gls{pprdsl} to the Delta files was realized using a newly introduced \gls{pprdsl} attribute. 

\begin{listing}[ht]
\begin{minted}[linenos,tabsize=2,breaklines,fontsize=\footnotesize,xleftmargin=0.6cm]{xml}
delta DLock1;
uses ShiftForkCaseStudyApp;
{
  <Remove> NetworkElement name=InsertLock1;
  <Remove> NetworkElement name=WeldLock1;
  <Remove> NetworkElement name=E_REND_WeldLock1;
  <Add> FB name=UltrasonicWeldingRobot16 
    type=UltrasonicWeldingRobot_16;
  <Add> EventConnection UltrasonicWeldingRobot16.CNF 
    PopulatedPipe.REQ;
}
\end{minted}
\vspace{-1.25em}
\caption{Snippet of an IEC~61499 Delta model for the \emph{shift fork} case study.}
    \label{lst:delta-model-fork}
\end{listing}

\begin{listing}[t!]
\begin{minted}[linenos,tabsize=2,breaklines,fontsize=\footnotesize,xleftmargin=0.6cm]{bash}
Attribute "deltaFile": { 
  description: "delta file for V4rdiac configuration", 
  defaultValue: "", type: "String"
}

Process "WeldLock1": { name: "WeldLock1",
  implements: [ "WeldLock" ], inputs: [ "Lock1" ], 
  deltaFile: "!DLock1"
}

Resource "WeldingRobot": { name: "WeldingRobot", 
  isAbstract: true }
Resource "LaserWeldingRobot_01":{ name: "LaserWeldingRobot_01", 
  implements: [ "LaserWeldingRobots" ], 
  deltaFile: "DLaserWeldingRobot01"
}
\end{minted}
    \vspace{-1.25em}
    \caption{Excerpt of the \gls{pprdsl} model of the \emph{shift fork} case study with the \texttt{deltaFile} attribute.}
    \label{lst:ppr-fork-delta}
\end{listing}

An example of a Delta model can be seen in \autoref{lst:delta-model-fork}.
Each Delta model may define a unique identifier by using the \code{delta} keyword.
The \code{uses} keyword sets the context for the modification.
A \code{<Remove>} and an \code{<Add>} operation define which element will be removed from or added to the \gls{cpps} artifacts, respectively.

Next, the author introduced a new attribute \code{deltaFile} to the \gls{ppr} models of the four case studies and used it to refer to the particular Delta files (cf. \autoref{lst:ppr-fork-delta}) from the processes and resources (\gls{eipse}, Step~1).
Afterward, the author re-ran the \travart{} transformations (Step~2) to generate the attributes in the variability models.
Then, the author went through the configuration and reduction steps of the \gls{eipse} approach according to the setting of the first evaluation activity (cf.~\autoref{sec:usecase_evaluation}) covering Steps~3 to~7 as input for Step~8.
In Step~8, the author triggered the generation of parts of control software in IEC~61499 in \gls{v4rdiac}.

\paragraph{Results}

\begin{figure}
    \centering
    \includegraphics[width=.9\columnwidth]{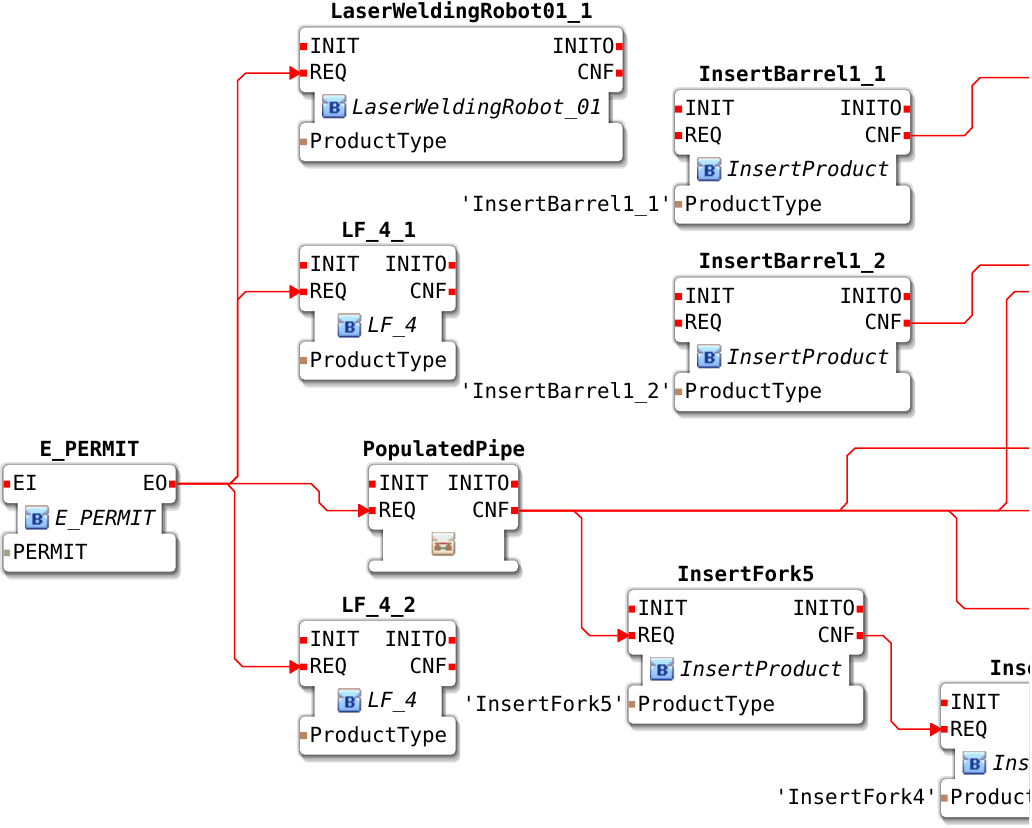}
    \caption{Excerpt of a generated IEC~61499 control software with production process and resource function blocks for a shift fork configuration.}
    \label{fig:fb-network}
\end{figure}

Our toolchain successfully generated an IEC~61499 control software based on a set of selected products, production processes, and production resources. 
Additionally, all elements in the generated control software also reflect the selected products, production processes, and production resources. 
For instance, \autoref{fig:fb-network} shows a generated control software based on selecting production processes, e.g., the \emph{PopulatedPipe}, \emph{InsertFork5}, and \emph{InsertFork4}, and production resources, e.g., \emph{LF\_4\_1} and \emph{LF\_4\_2}, of the shift fork case study. 

\section{Discussion}
\label{sec:discussion}

This section concludes on the evaluation questions from the previous section and answers the research question raised in the introduction (cf. \autoref{sec:introduction}).
Furthermore, it discusses the limitations of the prototype and threats to validity.

\subsection{Observations and Lessons Learned}

This section discusses the observations and lessons learned from the evaluation of the \gls{eipse} approach.

\paragraph{General Comments}
To examine the applicability of \gls{eipse}, we conducted a feasibility study with users experienced in the \gls{eipse} approach on four published case studies~\citep{Meixner2021} and an observational user study with users inexperienced in the \gls{eipse} approach on a novel case study.
We also investigated the reduction of the \processDM{}'s configuration space by \gls{eipse} and experimented with generating control code from such \gls{eipse} configurations.
These studies gave us valuable insight into potential benefits and perceived limitations of the \gls{eipse} approach.

To this end, we go beyond our previous work~\citep{MeixnerVamos2022ProcessExploration}, i.a., by providing feedback on the \gls{eipse} approach of users from different domains.
This feedback indicates that the users perceived the \gls{eipse} approach and toolchain as useful.
In particular, users perceived the \textit{externalization of knowledge}~\citep{Meixner2020DocSym} as valuable (requirements R1, R2, R4), the \textit{separation of concerns for the configuration steps}~\citep{Ananieva2016,Fadhlillah22} as helpful (R3, R5), the \textit{production process sequence exploration} as significant~\citep{Fang2013} (R1), and the \textit{integration of variability models} as beneficial~\citep{Kruger:2017:BSP:3106195.3106217} (R1, R2, R5).
Furthermore, the users provided valuable feedback for approach improvements and future work.

The following paragraphs summarize the observations and lessons learned for each evaluation question.

\paragraph{EQ1a Application of \gls{eipse} to Case Studies}
The primary lesson learned from applying \gls{eipse} to the case studies was that modeling the \gls{ppr} model combined with Steps~2 to~7 led to critical feedback.
Additionally, we experienced that this feedback gained importance with the growing complexity of the product line of products and processes.
For instance, while the \emph{truck} case study contained no defects in the \gls{ppr} model, it required significant improvement for the \emph{rocker switch} and \emph{water filter} case studies.
This improvement mainly concerned missing \code{requires} or \code{excludes} constraints or additional \gls{cdc} rules to limit the configuration options of the manufactured products suitably.
Therefore, it is also possible to configure ``semantically'' incorrect products in the \productFM{}, which can be prevented by more carefully defining the variability in the \gls{pprdsl}.
This confirms research and industry voices~\citep{feichtingerETFA2022IndustryVoices} and further stakeholders from industry regarding the importance of these constraints~\citep{NybergKeynoteSPLC2021}.
To this end, the \gls{eipse} approach helped to reveal flaws, even in the already published \gls{ppr} models~\citep{Meixner2021}.
For instance, in \productFM{}s resulting from such \gls{ppr} models, relevant products could not be configured anymore, the feature groups were incorrect, or the \productFM{} allowed many more configurations than products in the product line.
This also implies that software and \gls{cpps} product lines significantly differ regarding the configured output products, where, in the latter, each manufactured product must contribute to the cost and ``return on investment'' of the planned \gls{cpps}.
Therefore, it requires support for variability model analysis to constrain the models and ensure that unintended products can not be configured.

Second, the \gls{pprdsl}~\citep{MeixnerETFA2021PPRDSL}, \travart{}~\citep{FeichtingerTraVarT2021}, FeatureIDE~\citep{FeatureIDE} \glspl{fm}, DOPLER~\citep{DOPLER} \glspl{dm}, and V4rdiac~\citep{Fadhlillah22} were originally designed as independent tools. 
Thus, we had to align the constraint formats and, respectively, the transformations.

\paragraph{EQ1b Application by Engineers}
The first insight from observing engineers applying the \gls{eipse} approach is that the engineering phase for the \gls{cpps} takes significantly longer (avg. 77m) than the configuration phase (avg. 11m).
This is partly expected because application engineering is supposed to be faster than domain engineering. 
However, the editor support for defining the \gls{pprdsl} may contribute to the time spent on the definition and update tasks and may need improvement.
Still, the numbers confirm that much (mental) effort is invested in the product line definition.
In more detail, we observed that most of the time was spent defining the products and the atomic production process steps rather than the production resources.
The numbers also show that the configuration of the production process sequence in the \processDM{} took longer than the feature models' configuration.
This indicates that creating a meaningful production process sequence is more complex but also requires means to evaluate and simulate the sequences economically.
Furthermore, it implies that the time spent on different tasks requires a \emph{separation of concerns in multidisciplinary engineering} (requirement R3, R5) and the \textit{externalization of knowledge} (R1, R2, R4), which the engineers confirmed.

Secondly, the results show that the automatic creation and reduction of the variability model configuration space is fast and leads to a low time expenditure for the configuration of the models.
This beats the definition of an additional product and potential production process steps as an increment in the \gls{pprdsl}, which is still done manually with our approach.
We argue that the \gls{eipse} approach thus supports the challenge of an evolutionary creation of the product and its production process line.

Thirdly, the feedback of the subjects confirmed the usefulness of the \gls{eipse} approach, in general. 
While the tool support for defining the \gls{pprdsl} and the integration and feedback of the entire toolchain should be improved, the syntax seems to be straightforward for members of the intended user group despite a steep learning curve. 
Furthermore, the variability models could be configured quickly and intuitively because they only required little guidance from the supervising authors. 
However, the feedback explicitly points out \emph{low code approaches} as a potential aim for industry. 

\paragraph{EQ2- Configuration Space Reduction}
Our results confirm our previous hypothesis~\citep{MeixnerVamos2022ProcessExploration} and indicate that reducing the configuration space for decision models improves the guidance for engineers during production process exploration for valid and feasible production process sequences (requirement R1).
We also argue that the \gls{eipse} approach provides better reproducibility of the \gls{cpps} configuration through logging the selection sequence in the \processDM{} configuration.
Nevertheless, it requires additional effort to explicitly model more domain-specific knowledge, such as throughput, cost, or risk, \emph{to evaluate and simulate particular process sequences}.

\paragraph{EQ3 - Generation of Control Artifacts}
The primary lesson learned was that the generated function block networks already present a working version of the control software for the configured \gls{cpps}.
This shows that the integration of the variability models and their configuration works to a large extent (requirements R2, R4).
Nevertheless, the control software engineers still need to connect some process blocks manually.
For instance, the control software engineers must adjust the connections represented in the generated IEC~61499 function block networks to follow the process configuration.

Secondly, the control software engineers need to manually connect the generated production resource instances to the particular processes that use them.
We could improve that by further incorporating the \processDM{} configuration as a basis for automatically generating the Delta models.

\paragraph{Final Remarks}
The evaluation with the (i) application by engineers to existing and a novel case study, (ii) an investigation of the \processDM{} configuration space, and (iii) the successful creation of control code artifacts shows that \gls{eipse} is applicable to realistic cases and demands.
However, the approach also implies an additional overhead.
In particular, this concerns the explicit definition of the \gls{ppr} model in the \gls{pprdsl}, which also took the evaluation subjects the most time in the evaluation.
The additional overhead also concerns the development of the toolchain and its future refinement.
Nevertheless, we argue that the latter is a one-time effort while the former addresses the challenges of \textit{implicit domain knowledge} and \textit{configuration reproducibility} more than counterbalancing the additional effort.
Still, an in-depth investigation of the additional effort implied by the \gls{eipse} approach compared to completely manually finding feasible production process sequences needs to be conducted.

\subsection{Answering the Research Questions}
This section answers our research questions (cf.~\autoref{sec:introduction}).

\begin{enumerate}[nosep, label={\upshape\bfseries RQ\arabic*}]
    \item \emph{How can \gls{cpps} engineers be supported in modeling, exploring, and configuring the combined variability of products, production processes, and production resources, to generate corresponding \gls{cpps} artifacts?} \end{enumerate}

\noindent
To address this research question, this paper introduced the \gls{eipse} approach.
The approach aims at externalizing \gls{cpps} domain knowledge and the underlying variability by utilizing a domain-specific engineering artifact, i.e., the \gls{pprdsl}, combined with two well-established variability models, i.e., \glspl{fm} and \glspl{dm} (cf. requirements \textbf{R1}, \textbf{R2}, and \textbf{R4} in \autoref{sec:requirements}).
Furthermore, it aims to integrate the structural and behavioral variability of \gls{cpps} design aspects.
It also provides defined feedback loops to proactively incorporate changes in requirements or design during \gls{cpps} engineering.

To this end, we go beyond the state-of-the-art~\citep{Ananieva2016,Fang2013,Krueger2019,Meixner2020DocSym,Meixner2019} by providing a framework for externalizing domain expert knowledge, integrating the structural and behavioral variability of \glspl{cpps} and their configuration, including the separation of concerns of different engineering domains.
    
\begin{enumerate}[nosep, label={\upshape\bfseries RQ\arabic*}]
\setcounter{enumi}{1}
    \item \emph{How and to what extent can \gls{cpps} design be automated using variability modeling and \gls{cpps} concepts?} \end{enumerate}

\noindent
To address this research question, this work introduced the semi-automated \gls{eipse} toolchain architecture supporting the modeling and configuration process of a \gls{cpps}' design based on a product configuration with a corresponding prototype.
Therefore, we 
\begin{itemize} [nosep]
    \item adapted the \travart{} transformation operations for production processes according to the new \processDM{} notation (cf.~\autoref{fig:process}, Step~2b and \textbf{R2}),
    \item implemented \travart{} transformation operations for production resources (Step~2c and \textbf{R4}),
    \item implemented transformation operations for \glspl{cdc} (Step~2d and \textbf{R5}),
    \item implemented the \gls{eipse} \gls{dm} Editor including the configuration space reduction of the \processDM{} (Steps~4 and~5), and
    \item adapted \gls{v4rdiac} to integrate the transformed \glspl{cdc} and included a pre-configuration step to read the configurations of the three variability models and generate \gls{cpps} implementation artifacts (Steps~6 to~8) 
\end{itemize}

Thus, we go beyond the state-of-the-art by providing an integrated semi-automated toolchain for modeling and configuring the multidisciplinary structural and behavioral variability of \glspl{cpps}~\citep{Fadhlillah22,Meixner2020DocSym,Meixner2019}.
Additionally, we adapted the DOPLER \gls{dm}~\citep{DOPLER} constraints so that they are solvable via standard SAT solvers, such as sat4j, and provided a \gls{dm} editor not limited to the \gls{cpps} domain.
To this end, we enable the transformation between the \gls{ppr} model and state-of-the-art variability model types~\citep{FeichtingerMODEVAR2020,FeichtingerMODEVAR2021} integrating them into the \emph{de facto} standard tool FeatureIDE~\citep{FeatureIDE} for further dissemination.
Furthermore, we go beyond the state of the practice of the manual approach of engineering \gls{cpps} design also opening the way for better reuse of \gls{cpps} concepts.

Finally, we will discuss our work with industry partners to assess its practical impact.
Therefore, we will investigate which concepts can already be implemented and what needs to be altered or adapted.
Additionally, we will examine which parts of the prototype need to gain a higher technology readiness level for practical use.
We argue that our approach can start a discussion on variability modeling and configuration in \gls{cpps} engineering and that the tools provide a solid foundation for future use.

\subsection{Prototype Limitations}

The \processDM{} and \gls{eipse} \gls{dm} editor currently only support Boolean decisions rather than a broader range of decision types defined by the DOPLER approach~\citep{DOPLER}.
However, unlike the closed source DOPLER approach, our constraint definition syntax is SAT-solvable and thus usable with state-of-the-art software such as FeatureIDE.
Integrating the multiple tools into the \gls{eipse} toolchain still has room for improvement.
For instance, aligning the syntax of the constraint definitions may be investigated.

\subsection{Threats To Validity}

One threat to validity is that \textit{several authors of the paper were involved in steps of the evaluation}.
We tried to mitigate this threat by closely sticking to the provided engineering artifacts for the case studies and, where possible, gathering feedback from the engineers that provided the case studies. 
For the user study of the \gls{eipse} toolchain, we involved five external subjects.

Due to its unstructured nature, another threat concerns the hard-to-measure \textit{effort of the traditional manual approach}.
The \gls{eipse} approach mainly targets the automation of manual undocumented steps for \gls{cpps} engineering in alignment with the VDI~3695~\citep{vdi_3695}.

Additionally, measuring the ``time spent'' for certain tasks might not be the best metric. 
However,  in alignment with the VDI~3695~\citep{vdi_3695}, we argue that the engineering time is a significant factor that should be elicited and optimized.
Furthermore, it at least shows how fast relatively complex tasks can be completed with the \gls{eipse} toolchain and certainly demonstrates that with the approach one would be faster than doing it manually. 
We try to demonstrate users' experiences by presenting their feedback.

Finally, the evaluation of \textit{the approach's feasibility was conducted only for a limited set of case studies} of comparable size.
Furthermore, the evaluation of \emph{the approach's usefulness was conducted only for a single case study} and with \emph{a small number of engineers}.
Therefore, it is unclear how our approach would perform for systems of larger size and complexity.
This may threaten the generalizability of the \gls{eipse} approach.
However, especially \glspl{dm} in literature are smaller than the \glspl{dm} created by our approach~\citep{schmid2011comparison}.
Consequently, our case studies, at least concerning the \glspl{dm}, go beyond the state of the art.
 
\section{Related Work}
\label{sec:related_work}

This section presents approaches related to the \gls{eipse} and work on variability modeling for \gls{cpps}.

\citet{Safdar2021} created a framework for supporting product configuration in the \glspl{cps} domain based on their evaluation of existing variability modeling approaches \citep{Safdar2016}.
The framework mainly uses UML and OCL constraints for expressing \gls{cps} commonality and variability.
The framework is also designed to support automated multi-stage and multi-level product configuration.

\citet{fang2019model} developed a multi-view modeling approach for expressing variability in manufacturing.
Variability is expressed in three different views: (1) software, (2) production process, and (3) plants' topology.
The approach combines a feature meta-model with a topology and process meta-model to define a relation between variability from different views.
Using this meta-model, one can create a topology and process models that are related to a \gls{fm} when expressing variability in the manufacturing domain.
At a later step, one can derive a customer-specific topology and process model that complies with the features selected from the \gls{fm}.

\citet{Fadhlillah22} developed \gls{v4rdiac} as a multidisciplinary variability management approach for \glspl{cpps}.
\gls{v4rdiac} is designed as a generic approach where \gls{cpps} engineers can use any types of variability model to express \gls{cpps} control software.
\gls{cpps} engineers still need to decide which variability model best suits their domain.
We use \gls{v4rdiac} to showcase how our approach can be used to generate \gls{cpps} control software artifacts.

Existing works also use multiple \glspl{fm} to model \gls{cps} system variability from different views or perspectives \citep{Feldmann2015, Kowal2017, Rabiser2018, Canete2021, Geraldi2020} in the machine manufacturing, industrial automation domain, and deployment of IoT application. 
They use cross-tree constraints or cross-model constraints to define the relation of features in the same or from different \glspl{fm}. 
Expressing variability for industrial and complex software using multiple variability models is more beneficial in terms of maintainability and scalability compared to using a single variability model \cite{CzarneckiCool2012, kastner2012, oliinyk2017structuring}.
However, managing multi-view variability modeling, especially using heterogeneous types of variability models, is still a challenge~\citep{Kruger:2017:BSP:3106195.3106217}.

In contrast to existing works, our work expresses \gls{cpps} product, production process, and production resource variability. 
We use the \gls{pprdsl} for expressing \gls{cpps} variability based on a modeling concept that \gls{cpps} engineers are already familiar with.
The \gls{pprdsl} offers a unified syntax for expressing product, production process, and production resource variability as well as dependencies among them.
Additionally, we can transform our \gls{pprdsl} into a Product \gls{fm}, a Process \gls{dm}, and a Resource \gls{fm} to enable reasoning and configuration of a \gls{cpps} using existing product line tools (e.g., FeatureIDE). 
Additionally, production resources in our \gls{pprdsl} can be related to \gls{cpps} artifacts (e.g., IEC~61499 control software). 
Thus, we provide a product configuration mechanism where we can generate customer-specific variants that conform to the selected product, production process, and production resource variants.

\section{Conclusion and Outlook}
\label{sec:conclusion}
\glsresetall

This paper introduced the \gls{eipse} approach to decrease the manual and unstructured efforts while facilitating reproducibility in \gls{cpps} engineering.
On top of our previous work~\citep{FeichtingerVariVolution2020,FeichtingerTraVarT2021,MeixnerVamos2022ProcessExploration,MeixnerETFA2021PPRDSL}, we contributed 
\begin{enumerate*}[label=(\roman*)]
    \item the \gls{eipse} approach with additional steps to transform and configure production resource definitions and control software artifact generation,
    \item an extended prototype which realized the \gls{eipse} approach (including a novel \gls{dm} editor and configuration of SAT solvable \glspl{dm}).
\end{enumerate*}
We provide the corresponding artifacts in additional online material\footref{material} and a demonstration video\footref{demonstration}.
Furthermore, to investigate how the approach performs in practical settings, we:
\begin{enumerate*}[label=(\roman*)]
    \item conducted an evaluation of the feasibility in four published \gls{cpps} case studies~\citep{Meixner2021},
    \item conducted an observational user study with users inexperienced in the \gls{eipse} approach and a novel case study, 
    \item investigated the reduction of the \gls{cpps} configuration space, and
    \item examined the generation of IEC~61499~\citep{61499} control software artifacts.
\end{enumerate*}

In this way, we go beyond the state-of-the-art~\citep{Fadhlillah22,Krueger2019,Meixner2020DocSym,Meixner2019} by providing a framework and semi-automated toolchain for \gls{cpps} variability modeling.
This framework allows \gls{cpps} engineers to externalize better their domain expert knowledge, which comes primarily from experience and undocumented dependencies.
Furthermore, the approach enables engineers to model and configure the multidisciplinary structural and behavioral variability of \glspl{cpps} while separating the concerns of the different engineering disciplines.
Beyond that, the production process sequence exploration fosters reproducibility by recording the exploration steps in the toolchain.
To evaluate the \gls{eipse} approach, we collected the first feedback on the \gls{eipse} approach from users from different domains who perceived the approach and toolchain as useful and recommended improvements for future work.

In future work, we aim to broaden the approach's applicability and perform further evaluation, initiating the next iteration cycle of Design Science. 
On the one hand, currently, the prototypes only support Boolean decisions, which may limit their usability in large industrial settings.  
When integrating advanced solvers, like SMTs, we plan also to support Non-Boolean decisions. 
On the other hand, the \gls{pprdsl} may be improved in terms of editor support and by decoupling the processes from the production resources. 
Describing the overall \gls{cpps} variability may involve heterogeneous multi-view variability models for expressing the variability of different organizational units (e.g., business department, electrical engineering, or signal engineering). 
Thus, we also plan to extend further our \gls{eipse} tool for creating a product configuration tool capable of enacting configuration options from heterogeneous multi-view variability models. 
Given this setup, we plan to conduct a large-scale evaluation with external practitioners from the \gls{cpps} domain to examine the feasibility of the \gls{eipse} approach.

\section*{Acknowledgments}
The financial support by the Christian Doppler Research Association, the Austrian Federal Ministry for Digital and Economic Affairs and the National Foundation for Research, Technology and Development is gratefully acknowledged.
We explicitly want to thank our industry partners for their continuous support.
Partially funded by the Deutsche Forschungsgemeinschaft (DFG, German Research Foundation) – CRC 1608 – 501798263.
The water filter was invented, designed, and produced by Askwar Hilonga at Gongalimodel in Tanzania as a low cost frugal product line for cleaner water (UN SDG6).
We thank Askwar Hilonga for providing and sharing this case study and Yazgül Fidan who initially conducted and analyzed the interviews with Askwar Hilonga.
The truck models and their parts were designed, rendered, and 3D-printed at Czech Technical University in Prague - CIIRC.
We thank V\'{a}clav Jirkovsk\'{y} and Petr Nov\'{a}k for providing and sharing this case study.
The chess piece case study was provided by TU Wien Pilotfabrik and the Center for Digital Production.
We thank Alexander Raschendorfer and Sebastian Kropatschek for providing and sharing this case study.

\appendix

\section{Chess Piece User Study}
\label{sec:guideline}

This section describes the user study for evaluating the \gls{eipse} approach.

\subsection{Introduction}
We report the user study of the evaluation of the \gls{eipse} toolchain.
We investigate the modeling of an industrial product line and the subsequent production process exploration and production resource artifact derivation for a real-world \emph{chess piece} product line designed at TU Wien Pilotfabrik\footnote{TU Wien Pilotfabrik. \url{https://www.pilotfabrik.at/}}
Users assessed should go through the \gls{eipse} process and evaluate the feasibility and usability of the approach and the tools.

We undertook the evaluation reported here as part of a collaboration between different academic and industrial initiatives.
The subjects conducting the user study are, on the one hand, computer scientists and, on the other hand, engineers from companies that design or operate \glspl{cpps}.

\subsection{Rationale}
The user study of the \gls{eipse} evaluation was carried out in the focus of this paper and as part of three of the authors' dissertation projects.
The research scope is to investigate variability modeling for \glspl{cpps} engineering and the required transformation of industrial artifacts to well-established variability models.
The research in this context focuses on the reproducible exploration of production process sequences based on an integrated variability modeling approach.
This variability modeling approach uses different types of variability models, i.e., feature models and decision models, to separate the concerns of the different stakeholders.
There is limited published research on adopting state-of-the-art variability modeling and configuration approaches in the \gls{cpps} engineering industry, and the user study sought to contribute to the body of research in this area.

\subsection{Objective}
The user study took place in an academic setting, which is also the primary audience for the user study. 
The overall objectives were as follows.

\begin{outline}
\1 To perform a \gls{eipse} toolchain evaluation in a setting with subjects from different domains with the focus on \gls{cpps} engineering using a formal evaluation methodology.
\1 To learn from the evaluation about the following:
  \2 Can engineers model the functional view of \gls{cpps} for a small industrial product line with its products, process steps, and resources (domain engineering).
  \2 Can engineers efficiently explore and configure the design space for products, process sequences, and resources.
  \2 Does the integration of feature models and decision models allow for the configuration of a reasonable process sequence and corresponding resources based on a product configuration.
  \2 The time spent to model the functional view of \gls{cpps} for a small industrial product line (domain engineering).
  \2 The time spent to configure a \gls{cpps} design variant (application engineering).
  \2 The perceived usefulness of the \gls{eipse} approach and toolchain for the engineers.
  \2 The lessons learned from using the \gls{eipse} approach and toolchain for industrial product line modeling and configuration.
\1 To learn from the usage of \gls{eipse} approach and toolchain.
\end{outline}

These objectives are decidedly broad and ambitious. 
For conciseness, this chapter focuses on the objectives of learning from the evaluation about the time spent, the perceived usefulness, and the lessons learned.

\subsection{Chess Piece Use Case}

\begin{figure}[ht]
    \centering
    \includegraphics[width=.4\columnwidth]{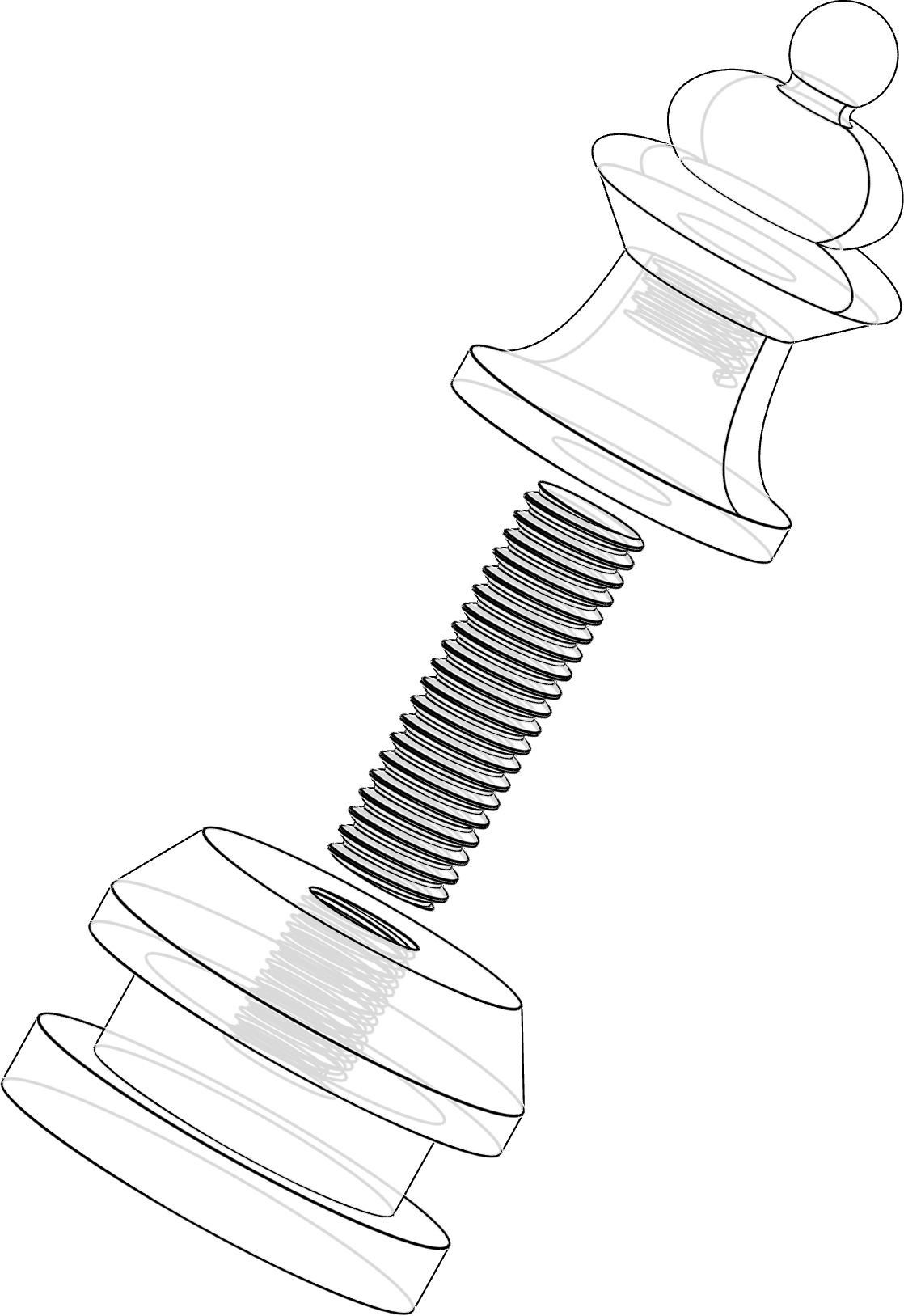}
    \caption{CAD drawing of a pawn chess piece}
    \label{fig:chesspiece}
\end{figure}

For the evaluation, we consider the \emph{chess piece} use case from TU Pilotfabrik.
The use case concerns a \gls{cpps} that manufactures the six chess piece types, i.e., king, queen, bishops, knights, rooks, and pawns.
Figure~\ref{fig:chesspiece} shows a CAD drawing of the pawn chess piece.

Each chess piece consists of a \emph{base}, a \emph{body}, and a threaded \emph{rod} that connects them.

The \textit{base} is produced from \emph{aluminum bars} of 1m length on a \emph{turning machine}.
The \emph{aluminum bar} is \emph{loaded into} the \emph{turning machine} with a \emph{bar loader}.
The \emph{turning machine} \textit{cuts} the \emph{aluminum bar} into the raw \emph{bases} of suitable length for further processing.
These \emph{bases}, which come in two variants, are \emph{turned} on the \emph{turning machine} to get their specific shape.
The king and the queen have a base with \emph{two circumferential reamings} that are \emph{carved into the aluminum}.
The other chess pieces have a base that has \emph{one circumferential reaming}.
After creation, a \emph{laser profiler} measures the bases for turning accuracy.

The \textit{body} of the particular chess pieces is \emph{3D-printed} from \gls{pla} on an industrial \textit{3D printer}.

The \textit{base} and the \textit{body} each have a hole with a thread carved respectively printed in the middle.
The threads each have a diameter for a standardized M6 \textit{rod}.\footnote{ISO metric screw threads: \url{https://w.wiki/_wm23}}
Similarly to the base, the threaded \textit{rod} comes in two variants, one with \textit{20 millimeters} and one with \textit{30 millimeters} in length.

The \emph{base}, the \emph{body}, and the threaded \emph{rod} are assembled in an \emph{assembly station}, where each individual part needs to be loaded into the station.
The parts need to be \emph{screwed} together, which can be done in an arbitrary sequence.

\subsection{Subject Guideline}

Conduct the \gls{eipse} process, as described in Section~\ref{sec:approach}, for the chess piece use case for an imaginative \gls{cpps}.

\paragraph{Chess piece product line modeling}
This task represents \emph{Step 1} of the \gls{eipse} approach in Figure~\ref{fig:process}.
As a user, create a functional \gls{pprdsl} model of the chess piece product line in the Sublime text editor.
Use the provided cheat sheet for the syntax of the \gls{pprdsl}.
The model shall represent three parts, which are
\begin{enumerate*}[label=(\roman*)]
    \item the partial and final products that should be manufactured by the \gls{cpps},
    \item the atomic process steps that create the products with their required input products, predecessors, and resources, as well as
    \item the production resources that can execute a particular process required to manufacture a product.
\end{enumerate*}

\textbf{Products}
For the chess piece use case, create \texttt{partial} products and \texttt{final} products in the \gls{pprdsl}.
Find out which partial products you could group using \texttt{abstract} parent products.
Furthermore, define which of the partial products \texttt{exclude} each other because a similar partial product more or less supplements them.

\textbf{Processes}
As a user, think about how to assemble partial products via specific atomic ``manufacturing'' processes.
Realize these atomic process steps in the \gls{pprdsl} and, similar to the products, group them on their \texttt{abstract} products and \texttt{exclude} process steps that you deem unnecessary.
Furthermore, consider which process steps in a particular assembly process need to be direct predecessors and refer to them as needed in the required section.

\textbf{Resources}
As a user, model the resources similar to products.
For modeling them, you should apply similar rules as for the product variability model.

\paragraph{Model Transformation}
This task represents \emph{Step 2} of the \gls{eipse} approach in Figure~\ref{fig:process}.
Use TRAVART from the \gls{ipse} toolchain from the command line to transform the chess piece \gls{pprdsl} model to the product and resource feature model (\texttt{uvl} file extension) and the process decision model (\texttt{dmodel} file extension).

\paragraph{Iterative Process Exploration}
This task represents \emph{Steps 3 to 5} of the \gls{eipse} approach in Figure~\ref{fig:process}.
Configure a desired product of the chess piece product line, i.e., one of the six chess piece types, using the configurator for feature models in Eclipse (Step 3 in Figure~\ref{fig:process}) by ticking the checkboxes for the features.
Then, generate a reduced \emph{decision model configuration} (\texttt{dconfig} file extension) by right-clicking on the decision model and selecting the configuration file of the previously configured product (\texttt{xml} file ending) (Step 4 in Figure~\ref{fig:process}).
Open the decision model configuration file in the \emph{decision model configurator} and explore feasible production process sequences for the configured chess piece by ticking the decision checkboxes (Step 5 in Figure~\ref{fig:process}). Therefore, you can investigate the process sequence in the right pane of the decision model configurator.

\paragraph{Resource configuration and artifact generation}
This task represents \emph{Steps 6 to 8} of the \gls{eipse} approach in Figure~\ref{fig:process}. 
Use the delta models (\texttt{delta} file extension) that are prepared according to the features or decisions in product \emph{feature model}, process \emph{decision model}, and resource \emph{feature model} that might affect the control source code of the \gls{cpps}.
Additionally, use the prepared delta configuration file (\texttt{deltaconf} file extension) in \gls{v4rdiac} to map the delta models into its corresponding feature or decision.
Then, use \gls{v4rdiac} to load the previously configured product and production processes sequence and configure the desired resources.
You can generate the control source code for the \gls{cpps} according to the selected features or decisions after the configuration is finished.

\subsection{Study Protocol}
No formal study protocol was developed or maintained for the user study.

\subsection{Evaluation Questions}

We formulated the following questions for the evaluation.
\begin{outline}
\1 Is it feasible to apply the \gls{eipse} approach?
  \2 in different scenarios
  \2 by novices from heterogeneous backgrounds
\end{outline}
    
\subsection{Methods of Data Collection}
We used \emph{shadowing} \citep{singer2008software} to investigate the subjects during performing the \gls{eipse} approach on the \emph{chess piece} use case.
Due to the distributed locations of the subjects performing the evaluation, we used Zoom to connect them to the \gls{eipse} toolchain and provided them with remote control. 
At least two of the authors shadowed the subjects during the evaluation sessions.
One author helped the evaluation subject if questions arose, additional explanations were required, or the subjects were stuck in the process.
The other authors present took notes for later investigation.
Furthermore, one author stopped the time for each of the activities and steps of the evaluation process.
Additionally, we recorded the evaluation sessions to replay them later during the internal result analysis.
Afterward, we asked the subjects about the perceived usefulness of the toolchain and the lessons learned during the process.

\subsection{Methods of Data Analysis}
No particular strategy for coding the notes taken was used.
We extracted quotes from the notes that concerns the perceived usefulness and could lead to improvements of the approach and the toolchain.

\subsection{Case Selection Strategy}
The case itself was selected on the basis that its main feature, i.e., the chess pieces, are well known by most people.
Furthermore, the production of the chess pieces seems lucid enough for engineers of different domains to be manageable.
To this end, it should be straightforward to understand how the possible production process could be modeled.
Yet, the case appears to be complex enough to properly investigate the problem of the large configuration space.

\subsection{Data Selection Strategy}
The strategy for selecting data was driven primarily by the activities defined in the subject guideline.

\subsection{Replication Strategy}
There was no strategy for replication on the basis that there was no comparable evaluation previously conducted or even comparable evaluations of other technologies.
We aim that future case studies of the evaluation of the \gls{eipse} approach adopt the here described description for partial replication.

\subsection{Quality Assurance}
To help ensure that data collected were representative of a broad range of stakeholders in the domain of \gls{cpps} engineering, we selected engineers and stakeholders from different companies and domains.

\subsection{Data Collection}
We used \emph{shadowing} \citep{singer2008software} with X participants to investigate the subjects during performing the \gls{eipse} approach on the \emph{chess piece} use case.
Due to the distributed locations of the subjects performing the evaluation, we used Zoom to connect them to the \gls{eipse} toolchain and provided them with remote control. 
At least two of the authors shadowed the subjects during the evaluation sessions.
One author helped the evaluation subject if questions arose, additional explanations were required, or the subjects were stuck in the process.
The other authors present took notes for later investigation.
Furthermore, one author stopped the time for each of the activities and steps of the evaluation process.
Additionally, we recorded the evaluation sessions to replay them later during the internal result analysis.
Afterward, we asked the subjects about the perceived usefulness of the toolchain and the lessons learned during the process.

\printcredits

\bibliographystyle{cas-model2-names}

\bibliography{2023JSS-eIPSE-Manuscript}

\begin{thebibliography}{71}
\expandafter\ifx\csname natexlab\endcsname\relax\def\natexlab#1{#1}\fi
\providecommand{\url}[1]{\texttt{#1}}
\providecommand{\href}[2]{#2}
\providecommand{\path}[1]{#1}
\providecommand{\DOIprefix}{doi:}
\providecommand{\ArXivprefix}{arXiv:}
\providecommand{\URLprefix}{URL: }
\providecommand{\Pubmedprefix}{pmid:}
\providecommand{\doi}[1]{\href{http://dx.doi.org/#1}{\path{#1}}}
\providecommand{\Pubmed}[1]{\href{pmid:#1}{\path{#1}}}
\providecommand{\bibinfo}[2]{#2}
\ifx\xfnm\relax \def\xfnm[#1]{\unskip,\space#1}\fi
\bibitem[{Ananieva et~al.(2016)Ananieva, Kowal, Th{\"{u}}m and
  Schaefer}]{Ananieva2016}
\bibinfo{author}{Ananieva, S.}, \bibinfo{author}{Kowal, M.},
  \bibinfo{author}{Th{\"{u}}m, T.}, \bibinfo{author}{Schaefer, I.},
  \bibinfo{year}{2016}.
\newblock \bibinfo{title}{{Implicit constraints in partial feature models}},
  in: \bibinfo{booktitle}{7th Int. FOSD Workshop, FOSD@SPLASH 2016, Amsterdam,
  Netherlands, October 30, 2016}, pp. \bibinfo{pages}{18--27}.
\bibitem[{Apel et~al.(2013)Apel, Batory, K{\"a}stner and Saake}]{ApelFOSD2013}
\bibinfo{author}{Apel, S.}, \bibinfo{author}{Batory, D.},
  \bibinfo{author}{K{\"a}stner, C.}, \bibinfo{author}{Saake, G.},
  \bibinfo{year}{2013}.
\newblock \bibinfo{title}{{Feature-Oriented Software Development: Concepts and
  Implementation}}.
\newblock \bibinfo{publisher}{Springer}.
\bibitem[{Biffl et~al.(2017)Biffl, Gerhard and
  L{\"u}der}]{biffl2017introduction}
\bibinfo{author}{Biffl, S.}, \bibinfo{author}{Gerhard, D.},
  \bibinfo{author}{L{\"u}der, A.}, \bibinfo{year}{2017}.
\newblock \bibinfo{title}{{Introduction to the Multi-Disciplinary Engineering
  for Cyber-Physical Production Systems}}, in:
  \bibinfo{booktitle}{Multi-Disciplinary Engineering for Cyber-Physical
  Production Systems}. \bibinfo{publisher}{Springer}, pp.
  \bibinfo{pages}{1--24}.
\bibitem[{Campbell et~al.(1990)Campbell, Faulk and
  Weiss}]{campbell1990introduction}
\bibinfo{author}{Campbell, G.H.}, \bibinfo{author}{Faulk, S.R.},
  \bibinfo{author}{Weiss, D.M.}, \bibinfo{year}{1990}.
\newblock \bibinfo{title}{{Introduction To Synthesis}}.
\newblock \bibinfo{type}{Technical Report}. INTRO\_SYNTHESIS\_PROCESS-90019-N,
  Software Productivity Consortium, Herndon, VA, USA.
\bibitem[{Cañete et~al.(2022)Cañete, Amor and Fuentes}]{Canete2021}
\bibinfo{author}{Cañete, A.}, \bibinfo{author}{Amor, M.},
  \bibinfo{author}{Fuentes, L.}, \bibinfo{year}{2022}.
\newblock \bibinfo{title}{Supporting iot applications deployment on edge-based
  infrastructures using multi-layer feature models}.
\newblock \bibinfo{journal}{Journal of Systems and Software}
  \bibinfo{volume}{183}, \bibinfo{pages}{111086}.
\bibitem[{Clarke et~al.(2015)Clarke, Helvensteijn and
  Schaefer}]{Clarke2015_deltaModelling}
\bibinfo{author}{Clarke, D.}, \bibinfo{author}{Helvensteijn, M.},
  \bibinfo{author}{Schaefer, I.}, \bibinfo{year}{2015}.
\newblock \bibinfo{title}{Abstract delta modelling}.
\newblock \bibinfo{journal}{Mathematical Structures in Computer Science}
  \bibinfo{volume}{25}, \bibinfo{pages}{482–527}.
\bibitem[{Clements and Northrop(2002)}]{clements2002software}
\bibinfo{author}{Clements, P.}, \bibinfo{author}{Northrop, L.},
  \bibinfo{year}{2002}.
\newblock \bibinfo{title}{Software product lines}.
\newblock \bibinfo{publisher}{Addison-Wesley Boston}.
\bibitem[{Czarnecki et~al.(2012)Czarnecki, Grünbacher, Rabiser, Schmid and
  W{\k{a}}sowski}]{CzarneckiCool2012}
\bibinfo{author}{Czarnecki, K.}, \bibinfo{author}{Grünbacher, P.},
  \bibinfo{author}{Rabiser, R.}, \bibinfo{author}{Schmid, K.},
  \bibinfo{author}{W{\k{a}}sowski, A.}, \bibinfo{year}{2012}.
\newblock \bibinfo{title}{{Cool Features and Tough Decisions: A Comparison of
  Variability Modeling Approaches}}, in: \bibinfo{booktitle}{6th International
  Workshop on Variability Modeling of Software-Intensive Systems},
  \bibinfo{publisher}{ACM}. pp. \bibinfo{pages}{173--182}.
\bibitem[{Dhungana et~al.(2011)Dhungana, Grünbacher and Rabiser}]{DOPLER}
\bibinfo{author}{Dhungana, D.}, \bibinfo{author}{Grünbacher, P.},
  \bibinfo{author}{Rabiser, R.}, \bibinfo{year}{2011}.
\newblock \bibinfo{title}{{The DOPLER Meta-Tool for Decision-Oriented
  Variability Modeling: A Multiple Case Study}}.
\newblock \bibinfo{journal}{Automated Software Engineering}
  \bibinfo{volume}{18}, \bibinfo{pages}{77--114}.
\bibitem[{Drath(2009)}]{drath2009datenaustausch}
\bibinfo{author}{Drath, R.}, \bibinfo{year}{2009}.
\newblock \bibinfo{title}{{Datenaustausch in der Anlagenplanung mit
  AutomationML: Integration von CAEX, PLCopen XML und COLLADA}}.
\newblock \bibinfo{publisher}{Springer-Verlag}.
\bibitem[{Drath(2021)}]{Drath2021}
\bibinfo{editor}{Drath, R.} (Ed.), \bibinfo{year}{2021}.
\newblock \bibinfo{title}{AutomationML - A Practical Guide}.
\newblock \bibinfo{publisher}{De Gruyter Oldenbourg}.
\bibitem[{Drath et~al.(2008)Drath, Luder, Peschke and Hundt}]{Drath2008}
\bibinfo{author}{Drath, R.}, \bibinfo{author}{Luder, A.},
  \bibinfo{author}{Peschke, J.}, \bibinfo{author}{Hundt, L.},
  \bibinfo{year}{2008}.
\newblock \bibinfo{title}{Automationml - the glue for seamless automation
  engineering}, in: \bibinfo{booktitle}{2008 IEEE International Conference on
  Emerging Technologies and Factory Automation}, pp. \bibinfo{pages}{616--623}.
\bibitem[{Fadhlillah et~al.(2022a)Fadhlillah, Feichtinger, Bauer, Kutsia and
  Rabiser}]{FadhlillahSPLC22}
\bibinfo{author}{Fadhlillah, H.S.}, \bibinfo{author}{Feichtinger, K.},
  \bibinfo{author}{Bauer, P.}, \bibinfo{author}{Kutsia, E.},
  \bibinfo{author}{Rabiser, R.}, \bibinfo{year}{2022}a.
\newblock \bibinfo{title}{V4rdiac: Tooling for multidisciplinary delta-oriented
  variability management in cyber-physical production systems},
  \bibinfo{publisher}{Association for Computing Machinery},
  \bibinfo{address}{New York, NY, USA}. p. \bibinfo{pages}{34–37}.
\bibitem[{Fadhlillah et~al.(2022b)Fadhlillah, Feichtinger, Meixner,
  Sonnleithner, Rabiser and Zoitl}]{Fadhlillah22}
\bibinfo{author}{Fadhlillah, H.S.}, \bibinfo{author}{Feichtinger, K.},
  \bibinfo{author}{Meixner, K.}, \bibinfo{author}{Sonnleithner, L.},
  \bibinfo{author}{Rabiser, R.}, \bibinfo{author}{Zoitl, A.},
  \bibinfo{year}{2022}b.
\newblock \bibinfo{title}{Towards multidisciplinary delta-oriented variability
  management in cyber-physical production systems}, in:
  \bibinfo{booktitle}{Proc. of the 16th International Working Conference on
  Variability Modelling of Software-Intensive Systems (VaMoS)},
  \bibinfo{publisher}{ACM}.
\bibitem[{Fadhlillah et~al.(2023a)Fadhlillah, Fern\'{a}ndez, Rabiser and
  Zoitl}]{Fadhlillah2023SPLC}
\bibinfo{author}{Fadhlillah, H.S.}, \bibinfo{author}{Fern\'{a}ndez, A.M.G.},
  \bibinfo{author}{Rabiser, R.}, \bibinfo{author}{Zoitl, A.},
  \bibinfo{year}{2023}a.
\newblock \bibinfo{title}{Managing cyber-physical production systems
  variability using v4rdiac: Industrial experiences}, in:
  \bibinfo{booktitle}{Proceedings of the 27th ACM International Systems and
  Software Product Line Conference - Volume A}, \bibinfo{publisher}{Association
  for Computing Machinery}, \bibinfo{address}{New York, NY, USA}. p.
  \bibinfo{pages}{223–233}.
\bibitem[{Fadhlillah et~al.(2023b)Fadhlillah, Sharma, Gutierrez~Fernandez,
  Rabiser and Zoitl}]{FadhlillahETFA2023}
\bibinfo{author}{Fadhlillah, H.S.}, \bibinfo{author}{Sharma, S.},
  \bibinfo{author}{Gutierrez~Fernandez, A.M.}, \bibinfo{author}{Rabiser, R.},
  \bibinfo{author}{Zoitl, A.}, \bibinfo{year}{2023}b.
\newblock \bibinfo{title}{Delta modeling in iec 61499: Expressing control
  software variability in cyber-physical production systems}, in:
  \bibinfo{booktitle}{2023 IEEE 28th International Conference on Emerging
  Technologies and Factory Automation (ETFA)}, pp. \bibinfo{pages}{1--8}.
\newblock \DOIprefix\doi{10.1109/ETFA54631.2023.10275693}.
\bibitem[{Fang(2019)}]{fang2019model}
\bibinfo{author}{Fang, M.}, \bibinfo{year}{2019}.
\newblock \bibinfo{title}{Model-Based Software Derivation for Industrial
  Automation Management Systems}.
\newblock Ph.D. thesis. Technische Universität Kaiserslautern.
\bibitem[{Fang et~al.(2013)Fang, Leyh, Elsner and D{\"{o}}rr}]{Fang2013}
\bibinfo{author}{Fang, M.}, \bibinfo{author}{Leyh, G.},
  \bibinfo{author}{Elsner, C.}, \bibinfo{author}{D{\"{o}}rr, J.},
  \bibinfo{year}{2013}.
\newblock \bibinfo{title}{Challenges in managing behavior variability of
  production control software}, in: \bibinfo{booktitle}{PLEASE@ICSE},
  \bibinfo{publisher}{{IEEE} Computer Society}. pp. \bibinfo{pages}{21--24}.
\bibitem[{Feichtinger et~al.(2022a)Feichtinger, Meixner, Biffl and
  Rabiser}]{FeichtingerICSREvolution}
\bibinfo{author}{Feichtinger, K.}, \bibinfo{author}{Meixner, K.},
  \bibinfo{author}{Biffl, S.}, \bibinfo{author}{Rabiser, R.},
  \bibinfo{year}{2022}a.
\newblock \bibinfo{title}{Evolution support for custom variability artifacts
  using feature models: A study in the cyber-physical production systems
  domain}, in: \bibinfo{editor}{Perrouin, G.}, \bibinfo{editor}{Moha, N.},
  \bibinfo{editor}{Seriai, A.D.} (Eds.), \bibinfo{booktitle}{Reuse and Software
  Quality}, \bibinfo{publisher}{Springer International Publishing},
  \bibinfo{address}{Cham}. pp. \bibinfo{pages}{79--84}.
\bibitem[{Feichtinger et~al.(2020)Feichtinger, Meixner, Rabiser and
  Biffl}]{FeichtingerVariVolution2020}
\bibinfo{author}{Feichtinger, K.}, \bibinfo{author}{Meixner, K.},
  \bibinfo{author}{Rabiser, R.}, \bibinfo{author}{Biffl, S.},
  \bibinfo{year}{2020}.
\newblock \bibinfo{title}{{Variability Transformation from Industrial
  Engineering Artifacts: An Example in the Cyber-Physical Production Systems
  Domain}}, in: \bibinfo{booktitle}{3rd International Workshop on Variability
  and Evolution of Software-Intensive Systems (VariVolution), {SPLC} '20: 24th
  {ACM} International Systems and Software Product Line Conference, Volume
  {B}}, \bibinfo{publisher}{{ACM}}. pp. \bibinfo{pages}{65--73}.
\bibitem[{Feichtinger et~al.(2022b)Feichtinger, Meixner, Rinker, Koren,
  Eichelberger, Heinemann, Holtmann, Konersmann, Michael, Neumann, Pfeiffer,
  Rabiser, Riebisch and Schmid}]{feichtingerETFA2022IndustryVoices}
\bibinfo{author}{Feichtinger, K.}, \bibinfo{author}{Meixner, K.},
  \bibinfo{author}{Rinker, F.}, \bibinfo{author}{Koren, I.},
  \bibinfo{author}{Eichelberger, H.}, \bibinfo{author}{Heinemann, T.},
  \bibinfo{author}{Holtmann, J.}, \bibinfo{author}{Konersmann, M.},
  \bibinfo{author}{Michael, J.}, \bibinfo{author}{Neumann, E.M.},
  \bibinfo{author}{Pfeiffer, J.}, \bibinfo{author}{Rabiser, R.},
  \bibinfo{author}{Riebisch, M.}, \bibinfo{author}{Schmid, K.},
  \bibinfo{year}{2022}b.
\newblock \bibinfo{title}{Industry voices on software engineering challenges in
  cyber-physical production systems engineering}, in: \bibinfo{booktitle}{2022
  IEEE 27th International Conference on Emerging Technologies and Factory
  Automation (ETFA)}, pp. \bibinfo{pages}{1--8}.
\bibitem[{Feichtinger and Rabiser(2020a)}]{FeichtingerMODEVAR2020}
\bibinfo{author}{Feichtinger, K.}, \bibinfo{author}{Rabiser, R.},
  \bibinfo{year}{2020}a.
\newblock \bibinfo{title}{Towards transforming variability models: Usage
  scenarios, required capabilities and challenges}, in:
  \bibinfo{booktitle}{24th ACM International Systems and Software Product Line
  Conference - Volume B}, \bibinfo{publisher}{ACM}, \bibinfo{address}{New York,
  NY, USA}. p. \bibinfo{pages}{44–51}.
\bibitem[{Feichtinger and Rabiser(2020b)}]{FeichtingerSEAA2020}
\bibinfo{author}{Feichtinger, K.}, \bibinfo{author}{Rabiser, R.},
  \bibinfo{year}{2020}b.
\newblock \bibinfo{title}{Variability model transformations: Towards unifying
  variability modeling}, in: \bibinfo{booktitle}{46th Euromicro Conference on
  Software Engineering and Advanced Applications}, \bibinfo{publisher}{IEEE},
  \bibinfo{address}{Portoroz, Slovenia}.
\bibitem[{Feichtinger and Rabiser(2021)}]{FeichtingerMODEVAR2021}
\bibinfo{author}{Feichtinger, K.}, \bibinfo{author}{Rabiser, R.},
  \bibinfo{year}{2021}.
\newblock \bibinfo{title}{How flexible must a transformation approach for
  variability models and custom variability representations be?}, in:
  \bibinfo{booktitle}{4rd International Workshop on Languages for Modelling
  Variability (MODEVAR), co-located with SPLC 2021}, \bibinfo{publisher}{ACM}.
  pp. \bibinfo{pages}{69--72}.
\bibitem[{Feichtinger et~al.(2021)Feichtinger, St\"{o}bich, Romano and
  Rabiser}]{FeichtingerTraVarT2021}
\bibinfo{author}{Feichtinger, K.}, \bibinfo{author}{St\"{o}bich, J.},
  \bibinfo{author}{Romano, D.}, \bibinfo{author}{Rabiser, R.},
  \bibinfo{year}{2021}.
\newblock \bibinfo{title}{Travart: An approach for transforming variability
  models}, in: \bibinfo{booktitle}{15th International Working Conference on
  Variability Modelling of Software-Intensive Systems},
  \bibinfo{publisher}{ACM}. pp. \bibinfo{pages}{8:1--8:10}.
\bibitem[{Feldmann et~al.(2015)Feldmann, Legat and Vogel-Heuser}]{Feldmann2015}
\bibinfo{author}{Feldmann, S.}, \bibinfo{author}{Legat, C.},
  \bibinfo{author}{Vogel-Heuser, B.}, \bibinfo{year}{2015}.
\newblock \bibinfo{title}{{Engineering support in the machine manufacturing
  domain through interdisciplinary product lines: An applicability analysis}}.
\newblock \bibinfo{journal}{IFAC-PapersOnLine} \bibinfo{volume}{28},
  \bibinfo{pages}{211--218}.
\bibitem[{Galster et~al.(2013)Galster, Weyns, Tofan, Michalik and
  Avgeriou}]{galster2013variability}
\bibinfo{author}{Galster, M.}, \bibinfo{author}{Weyns, D.},
  \bibinfo{author}{Tofan, D.}, \bibinfo{author}{Michalik, B.},
  \bibinfo{author}{Avgeriou, P.}, \bibinfo{year}{2013}.
\newblock \bibinfo{title}{Variability in software systems-a systematic
  literature review}.
\newblock \bibinfo{journal}{IEEE Transactions on Software Engineering}
  \bibinfo{volume}{40}, \bibinfo{pages}{282--306}.
\bibitem[{Geraldi et~al.(2020)Geraldi, Reinehr and Malucelli}]{Geraldi2020}
\bibinfo{author}{Geraldi, R.T.}, \bibinfo{author}{Reinehr, S.},
  \bibinfo{author}{Malucelli, A.}, \bibinfo{year}{2020}.
\newblock \bibinfo{title}{Software product line applied to the internet of
  things: A systematic literature review}.
\newblock \bibinfo{journal}{Information and Software Technology}
  \bibinfo{volume}{124}, \bibinfo{pages}{106293}.
\bibitem[{Gunes et~al.(2014)Gunes, Peter, Givargis and Vahid}]{gunes2014survey}
\bibinfo{author}{Gunes, V.}, \bibinfo{author}{Peter, S.},
  \bibinfo{author}{Givargis, T.}, \bibinfo{author}{Vahid, F.},
  \bibinfo{year}{2014}.
\newblock \bibinfo{title}{A survey on concepts, applications, and challenges in
  cyber-physical systems.}
\newblock \bibinfo{journal}{KSII Transactions on Internet \& Information
  Systems} \bibinfo{volume}{8}.
\bibitem[{Hevner(2007)}]{hevner2007three}
\bibinfo{author}{Hevner, A.R.}, \bibinfo{year}{2007}.
\newblock \bibinfo{title}{A three cycle view of design science research}.
\newblock \bibinfo{journal}{Scandinavian journal of information systems}
  \bibinfo{volume}{19}, \bibinfo{pages}{4}.
\bibitem[{Hevner et~al.(2008)Hevner, March, Park and Ram}]{hevner2008design}
\bibinfo{author}{Hevner, A.R.}, \bibinfo{author}{March, S.T.},
  \bibinfo{author}{Park, J.}, \bibinfo{author}{Ram, S.}, \bibinfo{year}{2008}.
\newblock \bibinfo{title}{Design science in information systems research}.
\newblock \bibinfo{journal}{Management Information Systems Quarterly}
  \bibinfo{volume}{28}, \bibinfo{pages}{6}.
\bibitem[{Hubaux et~al.(2012)Hubaux, Xiong and Czarnecki}]{Hubaux2012}
\bibinfo{author}{Hubaux, A.}, \bibinfo{author}{Xiong, Y.},
  \bibinfo{author}{Czarnecki, K.}, \bibinfo{year}{2012}.
\newblock \bibinfo{title}{A user survey of configuration challenges in linux
  and ecos}, in: \bibinfo{booktitle}{Proceedings of the 6th International
  Workshop on Variability Modeling of Software-Intensive Systems},
  \bibinfo{publisher}{Association for Computing Machinery},
  \bibinfo{address}{New York, NY, USA}. p. \bibinfo{pages}{149–155}.
\bibitem[{{International Electrotechnical Commission (IEC)}(2014)}]{62714}
\bibinfo{author}{{International Electrotechnical Commission (IEC)}},
  \bibinfo{year}{2014}.
\newblock \bibinfo{title}{{IEC 62714-1, Engineering data exchange format for
  use in industrial automation systems engineering - Automation markup language
  - Part 1: Architecture and general requirements}}.
\bibitem[{{International Electrotechnical Commission (IEC),
  TC65/WG6}(2012)}]{61499}
\bibinfo{author}{{International Electrotechnical Commission (IEC), TC65/WG6}},
  \bibinfo{year}{2012}.
\newblock \bibinfo{title}{{IEC 61499-1, Function Blocks - part 1:~Architecture:
  Edition 2.0}}.
\bibitem[{Jazdi et~al.(2010)Jazdi, Maga, Göhner, Ehben, Tetzner and
  Löwen}]{Jazdi2010}
\bibinfo{author}{Jazdi, N.}, \bibinfo{author}{Maga, C.},
  \bibinfo{author}{Göhner, P.}, \bibinfo{author}{Ehben, T.},
  \bibinfo{author}{Tetzner, T.}, \bibinfo{author}{Löwen, U.},
  \bibinfo{year}{2010}.
\newblock \bibinfo{title}{{Mehr Systematik für den Anlagenbau und das
  industrielle Lösungsgeschäft - Gesteigerte Effizienz durch Domain
  Engineering}} \bibinfo{volume}{58}, \bibinfo{pages}{524--532}.
\newblock \DOIprefix\doi{doi:10.1524/auto.2010.0867}.
\bibitem[{Järvenpää et~al.(2019)Järvenpää, Siltala, Hylli and
  Lanz}]{Jaervenpaeae2019}
\bibinfo{author}{Järvenpää, E.}, \bibinfo{author}{Siltala, N.},
  \bibinfo{author}{Hylli, O.}, \bibinfo{author}{Lanz, M.},
  \bibinfo{year}{2019}.
\newblock \bibinfo{title}{Implementation of capability matchmaking software
  facilitating faster production system design and reconfiguration planning}.
\newblock \bibinfo{journal}{Journal of Manufacturing Systems}
  \bibinfo{volume}{53}, \bibinfo{pages}{261--270}.
\bibitem[{Kang et~al.(1990)Kang, Cohen, Hess, Novak and
  Peterson}]{kang1990feature}
\bibinfo{author}{Kang, K.C.}, \bibinfo{author}{Cohen, S.G.},
  \bibinfo{author}{Hess, J.A.}, \bibinfo{author}{Novak, W.E.},
  \bibinfo{author}{Peterson, A.S.}, \bibinfo{year}{1990}.
\newblock \bibinfo{title}{{Feature-oriented domain analysis (FODA) feasibility
  study}}.
\newblock \bibinfo{type}{Technical Report}. Carnegie-Mellon Univ., Pittsburgh,
  Pa, Software Engineering Inst.
\bibitem[{K\"{a}stner et~al.(2012)K\"{a}stner, Ostermann and
  Erdweg}]{kastner2012}
\bibinfo{author}{K\"{a}stner, C.}, \bibinfo{author}{Ostermann, K.},
  \bibinfo{author}{Erdweg, S.}, \bibinfo{year}{2012}.
\newblock \bibinfo{title}{A variability-aware module system}, in:
  \bibinfo{booktitle}{Proc. of the ACM Int'l Conf. on Object Oriented
  Programming Systems Languages and Applications}, \bibinfo{publisher}{ACM},
  \bibinfo{address}{New York, NY, USA}. p. \bibinfo{pages}{773–792}.
\bibitem[{Kowal et~al.(2017)Kowal, Ananieva, Th{\"{u}}m and
  Schaefer}]{Kowal2017}
\bibinfo{author}{Kowal, M.}, \bibinfo{author}{Ananieva, S.},
  \bibinfo{author}{Th{\"{u}}m, T.}, \bibinfo{author}{Schaefer, I.},
  \bibinfo{year}{2017}.
\newblock \bibinfo{title}{{Supporting the Development of Interdisciplinary
  Product Lines in the Manufacturing Domain}}.
\newblock \bibinfo{journal}{IFAC-PapersOnLine} \bibinfo{volume}{50},
  \bibinfo{pages}{4336--4341}.
\bibitem[{Kr{\"u}ger et~al.(2019)Kr{\"u}ger, Mukelabai, Gu, Shen, Hebig and
  Berger}]{Krueger2019}
\bibinfo{author}{Kr{\"u}ger, J.}, \bibinfo{author}{Mukelabai, M.},
  \bibinfo{author}{Gu, W.}, \bibinfo{author}{Shen, H.}, \bibinfo{author}{Hebig,
  R.}, \bibinfo{author}{Berger, T.}, \bibinfo{year}{2019}.
\newblock \bibinfo{title}{{Where is my feature and what is it about? A case
  study on recovering feature facets}}.
\newblock \bibinfo{journal}{Journal of Systems and Software}
  \bibinfo{volume}{152}, \bibinfo{pages}{239--253}.
\bibitem[{Kr{\"{u}}ger et~al.(2017)Kr{\"{u}}ger, Nielebock, Krieter, Diedrich,
  Leich, Saake, Zug and Ortmeier}]{Kruger:2017:BSP:3106195.3106217}
\bibinfo{author}{Kr{\"{u}}ger, J.}, \bibinfo{author}{Nielebock, S.},
  \bibinfo{author}{Krieter, S.}, \bibinfo{author}{Diedrich, C.},
  \bibinfo{author}{Leich, T.}, \bibinfo{author}{Saake, G.},
  \bibinfo{author}{Zug, S.}, \bibinfo{author}{Ortmeier, F.},
  \bibinfo{year}{2017}.
\newblock \bibinfo{title}{{Beyond Software Product Lines: Variability Modeling
  in Cyber-Physical Systems}}, in: \bibinfo{booktitle}{21st International
  Systems and Software Product Line Conference, {SPLC} 2017, Volume A, Sevilla,
  Spain, September 25-29, 2017}, \bibinfo{publisher}{ACM},
  \bibinfo{address}{New York, NY, USA}. pp. \bibinfo{pages}{237--241}.
\bibitem[{Lee(1989)}]{lee1989disassembly}
\bibinfo{author}{Lee, S.}, \bibinfo{year}{1989}.
\newblock \bibinfo{title}{Disassembly planning based on subassembly
  extraction}, in: \bibinfo{booktitle}{Third ORSA/TIMS Conference on Flexible
  Manufacturing System}, pp. \bibinfo{pages}{383--388}.
\bibitem[{Meinicke et~al.(2017)Meinicke, Th{\"u}m, Schr{\"o}ter, Benduhn, Leich
  and Saake}]{FeatureIDE}
\bibinfo{author}{Meinicke, J.}, \bibinfo{author}{Th{\"u}m, T.},
  \bibinfo{author}{Schr{\"o}ter, R.}, \bibinfo{author}{Benduhn, F.},
  \bibinfo{author}{Leich, T.}, \bibinfo{author}{Saake, G.},
  \bibinfo{year}{2017}.
\newblock \bibinfo{title}{Mastering Software Variability with FeatureIDE}.
\newblock \bibinfo{publisher}{Springer}.
\bibitem[{Meixner(2020)}]{Meixner2020DocSym}
\bibinfo{author}{Meixner, K.}, \bibinfo{year}{2020}.
\newblock \bibinfo{title}{{Integrating Variability Modeling of Products,
  Processes, and Resources in Cyber-Physical Production Systems Engineering}},
  in: \bibinfo{booktitle}{24th ACM International Systems and Software Product
  Line Conference - Volume B}, \bibinfo{publisher}{{ACM}}. pp.
  \bibinfo{pages}{96--103}.
\bibitem[{Meixner et~al.(2021a)Meixner, Feichtinger, Rabiser and
  Biffl}]{Meixner2021}
\bibinfo{author}{Meixner, K.}, \bibinfo{author}{Feichtinger, K.},
  \bibinfo{author}{Rabiser, R.}, \bibinfo{author}{Biffl, S.},
  \bibinfo{year}{2021}a.
\newblock \bibinfo{title}{A reusable set of real-world product line case
  studies for comparing variability models in research and practice}, in:
  \bibinfo{booktitle}{25th International Systems and Software Product Line
  Conference - Volume B}, \bibinfo{publisher}{{ACM}}. pp.
  \bibinfo{pages}{105--112}.
\bibitem[{Meixner et~al.(2022)Meixner, Feichtinger, Rabiser and
  Biffl}]{MeixnerVamos2022ProcessExploration}
\bibinfo{author}{Meixner, K.}, \bibinfo{author}{Feichtinger, K.},
  \bibinfo{author}{Rabiser, R.}, \bibinfo{author}{Biffl, S.},
  \bibinfo{year}{2022}.
\newblock \bibinfo{title}{{Efficient Production Process Variability
  Exploration}}, in: \bibinfo{editor}{Arcaini, P.}, \bibinfo{editor}{Devroey,
  X.}, \bibinfo{editor}{Fantechi, A.} (Eds.), \bibinfo{booktitle}{VaMoS '22:
  16th International Working Conference on Variability Modelling of
  Software-Intensive Systems, Florence, Italy, February 23 - 25, 2022},
  \bibinfo{publisher}{{ACM}}. pp. \bibinfo{pages}{14:1--14:9}.
\bibitem[{Meixner et~al.(2019)Meixner, Rabiser and Biffl}]{Meixner2019}
\bibinfo{author}{Meixner, K.}, \bibinfo{author}{Rabiser, R.},
  \bibinfo{author}{Biffl, S.}, \bibinfo{year}{2019}.
\newblock \bibinfo{title}{Towards modeling variability of products, processes
  and resources in cyber-physical production systems engineering}, in:
  \bibinfo{booktitle}{23rd International Systems and Software Product Line
  Conference - Volume B}, \bibinfo{publisher}{{ACM}}. pp.
  \bibinfo{pages}{68:1--68:8}.
\bibitem[{Meixner et~al.(2020)Meixner, Rabiser and Biffl}]{Meixner2020}
\bibinfo{author}{Meixner, K.}, \bibinfo{author}{Rabiser, R.},
  \bibinfo{author}{Biffl, S.}, \bibinfo{year}{2020}.
\newblock \bibinfo{title}{{Feature Identification for Engineering Model
  Variants in Cyber-Physical Production Systems Engineering}}, in:
  \bibinfo{booktitle}{14th International Working Conference on Variability
  Modelling of Software-Intensive Systems}, \bibinfo{publisher}{{ACM}}. pp.
  \bibinfo{pages}{18:1--18:5}.
\bibitem[{Meixner et~al.(2021b)Meixner, Rinker, Marcher, Decker and
  Biffl}]{MeixnerETFA2021PPRDSL}
\bibinfo{author}{Meixner, K.}, \bibinfo{author}{Rinker, F.},
  \bibinfo{author}{Marcher, H.}, \bibinfo{author}{Decker, J.},
  \bibinfo{author}{Biffl, S.}, \bibinfo{year}{2021}b.
\newblock \bibinfo{title}{{A Domain-Specific Language for
  Product-Process-Resource Modeling}}, in: \bibinfo{booktitle}{26th {IEEE}
  International Conference on Emerging Technologies and Factory Automation,
  {ETFA} 2021, Vasteras, Sweden, September 7-10, 2021},
  \bibinfo{publisher}{{IEEE}}. pp. \bibinfo{pages}{1--8}.
\bibitem[{Monostori(2014)}]{MONOSTORI20149}
\bibinfo{author}{Monostori, L.}, \bibinfo{year}{2014}.
\newblock \bibinfo{title}{{Cyber-physical Production Systems: Roots,
  Expectations and R\&D Challenges}}.
\newblock \bibinfo{journal}{Procedia CIRP} \bibinfo{volume}{17},
  \bibinfo{pages}{9--13}.
\bibitem[{Nyberg(2021)}]{NybergKeynoteSPLC2021}
\bibinfo{author}{Nyberg, M.}, \bibinfo{year}{2021}.
\newblock \bibinfo{title}{Generating safety cases for large-scale industrial
  product lines}.
\newblock \URLprefix \url{https://splc2021.net/program/keynotes}.
  \bibinfo{note}{keynote at 25th ACM International Systems and Software Product
  Line Conference}.
\bibitem[{Oliinyk et~al.(2017)Oliinyk, Petersen, Schoelzke, Becker and
  Schneickert}]{oliinyk2017structuring}
\bibinfo{author}{Oliinyk, O.}, \bibinfo{author}{Petersen, K.},
  \bibinfo{author}{Schoelzke, M.}, \bibinfo{author}{Becker, M.},
  \bibinfo{author}{Schneickert, S.}, \bibinfo{year}{2017}.
\newblock \bibinfo{title}{Structuring automotive product lines and feature
  models: an exploratory study at opel}.
\newblock \bibinfo{journal}{Requirements Engineering} \bibinfo{volume}{22},
  \bibinfo{pages}{105--135}.
\bibitem[{Paetzold(2017)}]{Paetzold2017}
\bibinfo{author}{Paetzold, K.}, \bibinfo{year}{2017}.
\newblock \bibinfo{title}{Product and systems engineering/ca* tool chains}, in:
  \bibinfo{booktitle}{Multi-Disciplinary Engineering for Cyber-Physical
  Production Systems}. \bibinfo{publisher}{Springer}, pp.
  \bibinfo{pages}{27--62}.
\bibitem[{Rabiser et~al.(2018)Rabiser, Pr{\"{a}}hofer, Gr{\"{u}}nbacher,
  Petruzelka, Eder, Angerer, Kromoser and Grimmer}]{Rabiser2018}
\bibinfo{author}{Rabiser, D.}, \bibinfo{author}{Pr{\"{a}}hofer, H.},
  \bibinfo{author}{Gr{\"{u}}nbacher, P.}, \bibinfo{author}{Petruzelka, M.},
  \bibinfo{author}{Eder, K.}, \bibinfo{author}{Angerer, F.},
  \bibinfo{author}{Kromoser, M.}, \bibinfo{author}{Grimmer, A.},
  \bibinfo{year}{2018}.
\newblock \bibinfo{title}{Multi-purpose, multi-level feature modeling of
  large-scale industrial software systems}.
\newblock \bibinfo{journal}{Software and Systems Modeling}
  \bibinfo{volume}{17}, \bibinfo{pages}{913--938}.
\bibitem[{Runeson et~al.(2012)Runeson, Host, Rainer and
  Regnell}]{runeson2012case}
\bibinfo{author}{Runeson, P.}, \bibinfo{author}{Host, M.},
  \bibinfo{author}{Rainer, A.}, \bibinfo{author}{Regnell, B.},
  \bibinfo{year}{2012}.
\newblock \bibinfo{title}{Case study research in software engineering:
  Guidelines and examples}.
\newblock \bibinfo{publisher}{John Wiley \& Sons}.
\bibitem[{Safdar et~al.(2021)Safdar, Lu, Yue, Ali and Nie}]{Safdar2021}
\bibinfo{author}{Safdar, S.A.}, \bibinfo{author}{Lu, H.}, \bibinfo{author}{Yue,
  T.}, \bibinfo{author}{Ali, S.}, \bibinfo{author}{Nie, K.},
  \bibinfo{year}{2021}.
\newblock \bibinfo{title}{{A framework for automated multi-stage and multi-step
  product configuration of cyber-physical systems}} \bibinfo{volume}{20},
  \bibinfo{pages}{211--265}.
\bibitem[{Safdar et~al.(2016)Safdar, Yue, Ali and Lu}]{Safdar2016}
\bibinfo{author}{Safdar, S.A.}, \bibinfo{author}{Yue, T.},
  \bibinfo{author}{Ali, S.}, \bibinfo{author}{Lu, H.}, \bibinfo{year}{2016}.
\newblock \bibinfo{title}{{Evaluating variability modeling techniques for
  supporting cyber-physical system product line engineering}}.
\newblock \bibinfo{journal}{Lecture Notes in Computer Science (including
  subseries Lecture Notes in Artificial Intelligence and Lecture Notes in
  Bioinformatics)} \bibinfo{volume}{9959 LNCS}, \bibinfo{pages}{1--19}.
\bibitem[{Sch\"{a}fer et~al.(2021)Sch\"{a}fer, Becker, Andres, Kistenfeger and
  Rohlf}]{Schaefer2021}
\bibinfo{author}{Sch\"{a}fer, A.}, \bibinfo{author}{Becker, M.},
  \bibinfo{author}{Andres, M.}, \bibinfo{author}{Kistenfeger, T.},
  \bibinfo{author}{Rohlf, F.}, \bibinfo{year}{2021}.
\newblock \bibinfo{title}{Variability realization in model-based system
  engineering using software product line techniques: An industrial
  perspective}, in: \bibinfo{booktitle}{Proceedings of the 25th ACM
  International Systems and Software Product Line Conference - Volume A},
  \bibinfo{publisher}{Association for Computing Machinery},
  \bibinfo{address}{New York, NY, USA}. p. \bibinfo{pages}{25–34}.
\bibitem[{Schleipen et~al.(2015)Schleipen, L{\"u}der, Sauer, Flatt and
  Jasperneite}]{schleipen2015requirements}
\bibinfo{author}{Schleipen, M.}, \bibinfo{author}{L{\"u}der, A.},
  \bibinfo{author}{Sauer, O.}, \bibinfo{author}{Flatt, H.},
  \bibinfo{author}{Jasperneite, J.}, \bibinfo{year}{2015}.
\newblock \bibinfo{title}{Requirements and concept for plug-and-work}.
\newblock \bibinfo{journal}{at-Automatisierungstechnik} \bibinfo{volume}{63},
  \bibinfo{pages}{801--820}.
\bibitem[{Schmid and John(2004)}]{Schmid2004}
\bibinfo{author}{Schmid, K.}, \bibinfo{author}{John, I.}, \bibinfo{year}{2004}.
\newblock \bibinfo{title}{A customizable approach to full lifecycle variability
  management}.
\newblock \bibinfo{journal}{Sci. Comput. Program.} \bibinfo{volume}{53},
  \bibinfo{pages}{259--284}.
\bibitem[{Schmid et~al.(2011)Schmid, Rabiser and
  Gr{\"u}nbacher}]{schmid2011comparison}
\bibinfo{author}{Schmid, K.}, \bibinfo{author}{Rabiser, R.},
  \bibinfo{author}{Gr{\"u}nbacher, P.}, \bibinfo{year}{2011}.
\newblock \bibinfo{title}{A comparison of decision modeling approaches in
  product lines}, in: \bibinfo{booktitle}{5th International Workshop on
  Variability Modelling of Software-Intensive Systems},
  \bibinfo{organization}{ACM}. pp. \bibinfo{pages}{119--126}.
\bibitem[{Shull et~al.(2007)Shull, Singer and Sj{\o}berg}]{Shull2007Guide}
\bibinfo{author}{Shull, F.}, \bibinfo{author}{Singer, J.},
  \bibinfo{author}{Sj{\o}berg, D.I.}, \bibinfo{year}{2007}.
\newblock \bibinfo{title}{{Guide to Advanced Empirical Software Engineering}}.
\newblock \bibinfo{publisher}{Springer}.
\bibitem[{Singer et~al.(2008)Singer, Sim and Lethbridge}]{singer2008software}
\bibinfo{author}{Singer, J.}, \bibinfo{author}{Sim, S.E.},
  \bibinfo{author}{Lethbridge, T.C.}, \bibinfo{year}{2008}.
\newblock \bibinfo{title}{Software engineering data collection for field
  studies}.
\newblock \bibinfo{journal}{Guide to advanced empirical software engineering} ,
  \bibinfo{pages}{9--34}.
\bibitem[{Sundermann et~al.(2021)Sundermann, Feichtinger, Engelhardt, Rabiser
  and Th\"{u}m}]{SundermannSPLC2021UVL}
\bibinfo{author}{Sundermann, C.}, \bibinfo{author}{Feichtinger, K.},
  \bibinfo{author}{Engelhardt, D.}, \bibinfo{author}{Rabiser, R.},
  \bibinfo{author}{Th\"{u}m, T.}, \bibinfo{year}{2021}.
\newblock \bibinfo{title}{Yet another textual variability language? a community
  effort towards a unified language}, in: \bibinfo{booktitle}{Proceedings of
  the 25th ACM International Systems and Software Product Line Conference -
  Volume A}, \bibinfo{publisher}{Association for Computing Machinery},
  \bibinfo{address}{New York, NY, USA}. p. \bibinfo{pages}{136–147}.
\bibitem[{Tolio et~al.(2010)Tolio, Ceglarek, ElMaraghy, Fischer, Hu,
  Laperrière, Newman and Váncza}]{Tolio2010}
\bibinfo{author}{Tolio, T.}, \bibinfo{author}{Ceglarek, D.},
  \bibinfo{author}{ElMaraghy, H.A.}, \bibinfo{author}{Fischer, A.},
  \bibinfo{author}{Hu, S.J.}, \bibinfo{author}{Laperrière, L.},
  \bibinfo{author}{Newman, S.T.}, \bibinfo{author}{Váncza, J.},
  \bibinfo{year}{2010}.
\newblock \bibinfo{title}{Species—co-evolution of products, processes and
  production systems}.
\newblock \bibinfo{journal}{CIRP Annals} \bibinfo{volume}{59},
  \bibinfo{pages}{672--693}.
\bibitem[{{VDI/VDE 3682}()}]{vdi_3682}
{VDI/VDE 3682}, \bibinfo{year}{2005}.
\newblock \bibinfo{title}{{VDI/VDE 3682: Formalised Process Descriptions}}.
\newblock \bibinfo{howpublished}{Beuth Verlag}.
\bibitem[{{VDI/VDE 3695}()}]{vdi_3695}
{VDI/VDE 3695}, \bibinfo{year}{20010-2013}.
\newblock \bibinfo{title}{{VDI/VDE 3695}: {E}ngineering of industrial plants.}
\newblock \bibinfo{howpublished}{Beuth Verlag, 5 parts}.
\bibitem[{Wieringa(2014)}]{wieringa2014design}
\bibinfo{author}{Wieringa, R.J.}, \bibinfo{year}{2014}.
\newblock \bibinfo{title}{Design science methodology for information systems
  and software engineering}.
\newblock \bibinfo{publisher}{Springer}.
\bibitem[{Wohlin et~al.(2012)Wohlin, Runeson, H{\"o}st, Ohlsson, Regnell and
  Wessl{\'e}n}]{wohlinExperimentationInSoftwareEngineering2012}
\bibinfo{author}{Wohlin, C.}, \bibinfo{author}{Runeson, P.},
  \bibinfo{author}{H{\"o}st, M.}, \bibinfo{author}{Ohlsson, M.C.},
  \bibinfo{author}{Regnell, B.}, \bibinfo{author}{Wessl{\'e}n, A.},
  \bibinfo{year}{2012}.
\newblock \bibinfo{title}{Experimentation in software engineering}.
\newblock \bibinfo{publisher}{Springer Science \& Business Media}.
\bibitem[{Zhang et~al.(2016)Zhang, Duszynski and Becker}]{Zhang2016}
\bibinfo{author}{Zhang, B.}, \bibinfo{author}{Duszynski, S.},
  \bibinfo{author}{Becker, M.}, \bibinfo{year}{2016}.
\newblock \bibinfo{title}{Variability mechanisms and lessons learned in
  practice}, in: \bibinfo{booktitle}{Proceedings of the 1st International
  Workshop on Variability and Complexity in Software Design},
  \bibinfo{publisher}{Association for Computing Machinery},
  \bibinfo{address}{New York, NY, USA}. p. \bibinfo{pages}{14–20}.
\bibitem[{Zoitl et~al.(2010)Zoitl, Strasser and Valentini}]{4diacZoitl2010}
\bibinfo{author}{Zoitl, A.}, \bibinfo{author}{Strasser, T.},
  \bibinfo{author}{Valentini, A.}, \bibinfo{year}{2010}.
\newblock \bibinfo{title}{Open source initiatives as basis for the
  establishment of new technologies in industrial automation: 4diac a case
  study}, in: \bibinfo{booktitle}{2010 IEEE Int'l Symp. on Industrial
  Electronics}, \bibinfo{organization}{IEEE}. pp. \bibinfo{pages}{3817--3819}.

\end{thebibliography}

\end{document}